%% file: paper.tex
\documentclass[12pt,a4paper]{article}
\pdfoutput=1 %
\usepackage[utf8]{inputenc}
\usepackage{fullpage}
\usepackage{cite}
\usepackage{caption}
\usepackage{subcaption}
\usepackage[english]{babel}
\usepackage{amsmath}
\usepackage{physics}
\usepackage{amsthm}
\usepackage{amsfonts}
\usepackage{algorithm}
\usepackage[noend]{algpseudocode}
\usepackage{amssymb}
\usepackage{bbold}
\usepackage{mathrsfs}
\usepackage{mathbbol}
\usepackage[utf8]{inputenc}
\usepackage{mathtools}
\usepackage[toc,page]{appendix}
\usepackage{subcaption}
\usepackage{graphicx}
\usepackage{bbding}
\usepackage{esvect}
\usepackage{float}  
\usepackage{commath}
\usepackage{enumitem}
\usepackage[affil-it]{authblk}
\usepackage{siunitx}
\usepackage{braket}
\usepackage{phaistos}
\usepackage{multicol}
\usepackage{overpic}
\usepackage{tabularx}
\usepackage{ulem}

\usepackage{layout}
\usepackage{printlen}
\usepackage[grid=false, gridunit=pt, gridcolor=blue!40, subgridcolor=blue!20]{eso-pic}
\usepackage{booktabs, array, ragged2e, ltablex, ltxtable}

\usepackage{comment}
\usepackage[colorlinks=true
  ,urlcolor=blue
  ,anchorcolor=blue
  ,citecolor=blue
  ,filecolor=blue
  ,linkcolor=blue
  ,menucolor=blue
  ,linktocpage=true
  ,pdfproducer=medialab
  ,pdfa=true
]{hyperref}
\usepackage{xcolor}
\usepackage{jheppub}
\usepackage{tikz}
\usetikzlibrary{calc} 
\pgfdeclarelayer{background}
\pgfdeclarelayer{foreground}
\pgfsetlayers{background,foreground}
\usetikzlibrary{shapes.geometric, arrows}
\usetikzlibrary{shapes, arrows.meta, positioning, calc, fit}

\usepackage{environ}
\NewEnviron{myequation}{%
\begin{equation*}
\scalebox{1.7}{$\BODY$}
\end{equation*}
}

\usepackage[usestackEOL]{stackengine}

\newtheorem{theorem}{Theorem}
\newtheorem{corollary}{Corollary}

\usepackage{bbm}

\DeclareMathOperator*{\argmin}{arg\,min}
\newcommand{\Hil}{\mathcal{H}}

\def\ket#1{{|{#1}\rangle}} 

\title{Navigating the Quantum Resource Landscape of Entropy Vector Space Using Machine Learning and Optimization}

\author[a, \dagger]{Nothando Khumalo\note[$\dagger$]{These authors contributed equally},}
\author[b, \dagger]{Aman Mehta,}
\author[a, \dagger \ast]{William Munizzi,}
\author[a,b]{Prineha Narang}

\affiliation[a]{Division of Physical Sciences, College of Letters and Science, University of California, Los Angeles}
\affiliation[b]{Department of Electrical and Computer Engineering, University of California, Los Angeles}

\emailAdd{nkhumalo@g.ucla.edu}
\emailAdd{mehtaaman@ucla.edu}
\emailAdd{wmunizzi17@ucla.edu}
\emailAdd{prineha@ucla.edu}

\abstract{We present a machine learning framework to study the dynamics of entropy vectors and quantum resources, including entanglement and magic, focusing on violations of entropy inequalities. Using a reinforcement learning agent formulated as a Markov decision process, we identify quantum circuits that optimally navigate the entropy vector space to generate violations of Ingleton's inequality. We complement this approach with a classical optimization algorithm to produce arbitrary numbers of Ingleton-violating states, with tunable degrees of violation, and empirically determine the maximal attainable violation for Ingleton's inequality. Our analysis reveals characteristic patterns of quantum resources that accompany Ingleton violation. A comprehensive statistical analysis shows that Ingleton-violating states occupy sharply-defined, isolated regions of the Hilbert space, and are extremely rare. Together, these results establish a unified computational toolkit for studying entropy vector dynamics, tracking quantum resource evolution, and engineering circuits with controlled information-theoretic features.}

\begin{document}
\maketitle
\flushbottom

\section{Introduction}

Quantum entropy is more than a measure of uncertainty, but a useful quantity for understanding the constraints governing entangled subsystems of a composite quantum state. Since Shannon's pioneering work on information theory~\cite{Shannon:1948}, information inequalities have played a central role in delineating the intrinsic limitations of information storage and processing across both classical and quantum regimes~\cite{Dembo:1991,Blachman:1965,Cover:1983,Matus:2007,Cadney:2012}. In the quantum domain, the existence of entanglement introduces an additional layer of structure, imposing constraints on the strength of quantum correlations and how they can be shared across the Hilbert space~\cite{Li:2007,Dougherty:2005,Hayashi:2007,Hayashibook:2007}. Measures like entanglement entropy~\cite{VidalWerner2002,Horodecki2009} serve to quantify the strength of bipartite quantum correlations, however, the pattern of entanglement distributed among multiple parties is further restricted by entropy inequalities that define the permissible ways quantum information can be shared~\cite{Coffman2000,Linden2002}. A universal inequality, obeyed by all quantum states, is the subadditivity of quantum entropy~\cite{LiebRuskai1973,Lieb1973a,Lieb1973b, He:2023aif}, though a hierarchy~\cite{Linden2013,ZhangYeung1997,Lashkari:2014kda,ZhangYeung1998,Matus2007,DoughertyFreilingZeger2007,LindenWinter2005,Hern_ndez_Cuenca_2024} of additional entropy inequalities exist which are satisfied only by states with precise information-theoretic properties and entanglement structure.

For a composite Hilbert space $\Hil = \bigotimes_i \Hil_i$ with fixed factorization, calculating the entanglement shared across each Hilbert space partition defines the \textit{entropy vector} for a state $\ket{\psi} \in \Hil$, providing a complete characterization of its bipartite entanglement structure. Entropy vectors naturally induce a classification on states in $\Hil$, since every state has an associated entropy vector, and different states which share the same entropy vector exhibit an equivalent entanglement pattern. Entropy inequalities restrict entropy vectors to convex subspaces of the ambient entropy vector space, defining distinct \textit{entropy cones} associated to classes of quantum states that obey those inequalities. While the restriction of entropy vectors to specific entropy cones has been extensively studied~\cite{Bao2015,Bao2020a,Linden2013,Czech2021,HernandezCuenca2019,Bao:2020mqq}, less is known about the mechanisms through which quantum states transition between cones under allowed dynamics~\cite{Keeler2022,Keeler:2023xcx,Keeler:2023shl,Munizzi:2023ihc,Munizzi:2024huw,Munizzi:2025suf}, the nesting relations between different entropy cones~\cite{Bao:2020mqq,He:2023rox,Czech:2023xed}, and how transitions between cones reflect a deeper resource trade-off. Furthermore, the high dimensionality and non-intuitive geometry of the entropy vector space raise a fundamental challenge: how can one systematically search this space to identify cone boundaries, extremal rays, and inequality-violating regions?

The high dimensionality, and often non-intuitive geometry, of entropy vector space makes analytic exploration inherently challenging, particularly when attempting to characterize the regions between distinct entropy cones or identifying states whose entropy vectors satisfy, saturate, or fail a specific set of inequalities in a precise manner. Machine-learning and convex optimization offer powerful and complementary techniques for addressing these challenges and systematically navigating the entropy vector landscape. For quantum states evolving under unitary operations, reinforcement learning offers a data-driven approach for mapping how specific quantum circuits steer a state's entropy vector through this space, and between different entropy cones. Likewise, the problem of finding states that satisfy, saturate, or violate a chosen set of entropy inequalities is intuitively framed as an optimization problem, with the inequalities themselves comprising a natural set of cost functions. The combination of these computational techniques enables the efficient and scalable exploration of entropy vector geometry, revealing structural features of entropy vector space, as well as the associated quantum resource landscape, that remain largely intractable through analytic methods alone. 

The simplest non-trivial linear rank inequality, originally formulated in the context of classical network theory, is Ingleton’s inequality~\cite{Ingleton:1971}, a $4$-party entropy inequality that determines whether a network admits a linear coding solution~\cite{Fong2008Ingleton,debeaudrap:2015,Kobayashi:2011}. Systems satisfying Ingleton’s inequality can be efficiently described using linear network coding, a framework that maximizes transmission rates when operations are restricted to linear transformations of the input data. In the quantum setting, Ingleton's inequality constrains the entanglement structure of states with specific quantum resource profiles and information processing capabilities. For example, all stabilizer states, i.e. those quantum states that are efficiently classically simulable, satisfy Ingleton's inequality~\cite{Linden2013,Walter2016}. Likewise, all holographic quantum states~\cite{Bao2015}, states which are dual to a smooth classical geometry in AdS/CFT, also satisfy Ingleton's inequality. In fact, Ingleton-violating states occupy a very small region of the Hilbert space, and violating Ingleton's inequality requires a precise configuration of entanglement and non-stabilizerness, i.e. quantum magic. Relatedly, we may ask which combinations of entanglement and magic are necessary to drive Ingleton violation and what resource signatures characterize states near or beyond this entropy boundary? These questions naturally motivate exploring how inequality violation impacts practical quantum tasks, since the same resources that enable violation often govern a system’s computational and communicational power. Beyond entropy cones, the improved understanding of Ingleton violation informs the design of quantum algorithms, and distributed computing architectures, with tailored information-theoretic capabilities.

In this work, we introduce a machine learning protocol for probing entropy vector dynamics, and examining the role of quantum resources, e.g. entanglement, magic, and non-local magic, in the evaluation of entropy inequalities. Using this framework, we identify arbitrary numbers of Ingleton-violating states, of which only a few were known prior to this work~\cite{Linden2013,Fong2008Ingleton}, and construct explicit quantum circuits to prepare these states and probe the boundary of the Ingleton entropy cone. We analyze how entanglement and magic evolve as a state transitions from inside to outside the Ingleton entropy cone, revealing the resource signatures associated with Ingleton violation. Complementing this approach, we develop a classical optimization procedure for generating families of Ingleton-violating states, with tunable degrees of violation. Using our results, we infer the maximal attainable violation of Ingleton's inequality, and characterize the quantum resource profiles of states at maximal violation. Finally, we perform a comprehensive statistical analysis of both the stability of maximally-violating solutions and the distribution of Ingleton-violating states across the Hilbert space. Together, these methods establish a robust computational toolkit for investigating entropy vector dynamics, and elucidating how specific resources determine the information-theoretic capabilities of quantum states.

The structure of this paper is organized as follows. In Section~\ref{ReviewSection}, we review entropy cones, quantum magic, and necessary elements of reinforcement learning. In Section~\ref{ReinforcementLearningSection} we present our machine learning algorithm for studying inequality violation, and the dynamics of entropy vectors and quantum resources under unitary evolution. In Section~\ref{IngletonCircuitSection} we apply our protocol to generate circuits which prepare Ingleton-violating states, and analyze the evolution of entanglement and magic near the boundary of, and outside, the Ingleton entropy cone. Section~\ref{StatsSection} introduces a classical optimization protocol for generating broad classes of Ingleton-violating states, with arbitrary violation amounts, and analyzes the complementary behavior of entanglement and magic as Ingleton violation is maximized. Section~\ref{DiscussionSection} discusses the impact of this work, and suggests future directions for study beyond this manuscript. A set of packages for implementing the computational techniques introduced, as well as associated data sets are publicly available at~\cite{Khumalo_Quantum_Resource_Dynamics}

\section{Review}\label{ReviewSection}

\subsection{Entropy Vectors and Entropy Cones}

Given a state $\rho$ in a Hilbert space $\mathcal{H}$, and assuming a factorization on $\mathcal{H}$, we compute the entanglement entropy of $\rho$ according to the von Neumann entropy $S_{\rho}$, defined
\begin{equation}\label{vnEntropy}
   S_{\rho} \equiv -\Tr \left( \rho \log_d \rho \right),
\end{equation}
where $d$ is the dimension of $\mathcal{H}$. In this work information is measured in \textit{bits}, and entropies are calculated using $\log_2$.

When $\rho$ is pure $\rho^2 = \rho$, and thereby $S_{\rho} =0$. However, for $\rho$ a multipartite system there can exist non-zero entanglement entropy between complementary subsystems of $\rho$. For $\rho$ a $n$-party quantum system, and $I$ an $\ell$-party subsystem of $\rho$, the entanglement entropy of $I$ with its $\left(n-\ell \right)$ complement $\bar{I}$ is computed as
\begin{equation}\label{EntanglementEntropy}
   S_{I} = -\Tr(\rho_I \log_2 \rho_I),
\end{equation}
where $\rho_I$ is the reduced density matrix for $I$, obtained by tracing out $\bar{I}$ in $\rho$.

Given an $n$-partite factorization of $\Hil$, there are $2^n-1$ unique entanglement entropies that can be calculated using Eq.\ \eqref{EntanglementEntropy}. Each $S_I$ is computed and arranged into an ordered tuple, known as the entropy vector for $\rho$. For example, a tripartite pure state $\ket{\psi}$ has an entropy vector $\vec{S}_{\psi}$, with the following form
\begin{equation}\label{EntropyVector}
    S_{\psi} = \left(S_A,\ S_B,\ S_O,\ S_{AB},\ S_{AO},\ S_{BO},\ S_{ABO},  \right).
\end{equation} 
where $O$ is customarily used to denote the $n^{th}$ subsystem, which acts as a purifier for the other subsystems. When the overall state $\psi$ is pure, subsystem entanglement entropy obeys the complementarity relations $S_I = S_{\bar{I}}$.

A state's entropy vector encodes a complete description of the state's bipartite entanglement, and can be used to classify states according to their entanglement structure \cite{Temple:2024,Bao2015,Linden:2023,Zhang:1998,Bao:2024}. Entanglement entropies of quantum states are constrained to obey specific relations, known as entropy inequalities. The most well known inequality is subadditivity, which reads
\begin{equation}\label{SA}
    S_A + S_B \geq S_{AB},
\end{equation}
and is satisfied by all quantum states. Moreover all tripartite quantum states are strongly subadditive, and thereby also satisfy the inequality
\begin{equation}\label{SSA}
    S_{AB} + S_{BC} \geq S_{B} + S_{ABC},
\end{equation}
Other inequalities, however, are only satisfied by quantum states with specific features. For example, the monogamy of mutual information (MMI) inequality \cite{Hayden:2013}
\begin{equation}\label{MMI}
    S_{AB} + S_{AC} + S_{BC} \geq S_{A} + S_{B} + S_{C} + S_{ABC},
\end{equation}
is satisfied \cite{Hayden:2013} by all quantum states which admit a smooth classical dual geometry through the AdS/CFT correspondence. Furthermore, MMI constrains the correlations that a single system can simultaneously share with any pair of systems, thus impacting the quantum information processing capabilities of the entangled system.

The set of all entropy inequalities obeyed by a particular class of states defines the entropy cone for those states. For linear inequalities, the entropy cone comprises a convex polyhedral subspace of the ambient entropy vector space. Quantum states which satisfy the requisite set of inequalities possess entropy vectors which are confined to the interior of the entropy cone. While certain entropy cones, e.g. the holographic entropy cone, have been studied in depth, very little is understood about the relative nesting of different entropy cones, as well as the dynamics of entropy vectors as a state transitions between distinct entropy cones.

In this work, we focus on Ingleton's inequality, which serves as a constraint on  both the holographic and stabilizer entropy cones. Ingleton's inequality, which reads
\begin{equation}\label{Ingleton}
S_{AB} + S_{AC} + S_{AD} + S_{BC} + S_{BD} \geq S_{A} + S_{B} + S_{ABC} + S_{ABD} + S_{CD},
\end{equation}
constrains the possible entanglement arrangements that can be prepared using linear encoding. Consequently, all stabilizer states satisfy Ingleton's inequality, and violating this inequality requires some amount of non-classicality in the quantum system.

\subsection{The Clifford Group and Stabilizer States}

The Clifford group is a multiplicative matrix group, generated by the Hadamard, phase, and $\mathrm{CNOT}$ operations. In a matrix representation, the Clifford generators may be written
\begin{equation}
    H\equiv \frac{1}{\sqrt{2}}\begin{bmatrix}1&1\\1&-1\end{bmatrix}, \quad P\equiv \begin{bmatrix}1&0\\0&i\end{bmatrix}, \quad     C_{1,2} \equiv \begin{bmatrix}
            1 & 0 & 0 & 0\\
            0 & 1 & 0 & 0\\
	    0 & 0 & 0 & 1\\
	    0 & 0 & 1 & 0
            \end{bmatrix},
\end{equation}
where the first index on $CNOT$ denotes the control qubit, and the second denotes the target qubit. 

The quantum $T$ gate is a single-qubit operation that induces a $\pi/4$ rotation, and thereby can reach single-qubit states beyond the stabilizer octahedron. Accordingly, the $T$ can generate magic in a quantum system, and is represented as the matrix
\begin{equation}
    T \equiv \begin{bmatrix}
            1 & 0 \\
            0 & e^{i\pi/4} 
            \end{bmatrix}.
\end{equation}
Since $T$ and $H$ are sufficient to generate arbitrary single-qubit rotations, the group $\langle H_i, P_i, C_{i,j}, T_i\rangle$ is a universal gate set.

\subsection{Quantum Magic and Non-Local Magic}

One important measure of non-classicality is quantum magic, which describes the difficulty of simulating a quantum system using classical resources~\cite{Veitch2013, Howard:2017}. Magic has been shown to be a necessary resource for quantum algorithms to achieve advantage over classical computation, yet precisely quantifying and characterizing the magic in a quantum system is an active area of contemporary research~\cite{tirrito:2024, haug:2025, Andreadakis:2025mfw, Dallas:2025zxs, Tarabunga:2025wym}.

Stabilizer Renyi entropy~\cite{Leone:2021rzd} quantifies the ``non-stabilizerness'', i.e. magic, of a state. For an $n$-qubit pure state $\ket{\psi}$, we define the Stabilizer Renyi entropy of order $\alpha$ as
\begin{equation}\label{eq:SRE1}
    M_{\alpha}\left(\psi\right) \equiv \frac{1}{1-\alpha} \ln \left(\sum_{P \in \mathcal{P}_n} \frac{\bra{\psi} P \ket{\psi}^{2\alpha}}{2^n} \right) - \ln \left(2^n\right),
\end{equation}
where $\mathcal{P}_n$ denotes the set of $n$-qubit Pauli strings. The calculation of SRE considers the sum over a normalized stabilizer probability distribution, computed from Pauli expectation values. For $\ket{\psi}$ a stabilizer state, this distribution is sharply peaked, and therefore $M_{\alpha}\left(\psi\right) = 0$ \cite{Turkeshi:2023lqu}. SRE constitutes a good magic measure from the perspective of resource theory, as it satisfies the following properties
\begin{enumerate}
    \item \textbf{Faithfulness:} $M_{\alpha}\left(\ket{\psi}\right) = 0$ if and only if $\ket{\psi}$ is a stabilizer state, otherwise $M_{\alpha}\left(\ket{\psi}\right) > 0$.
    \item \textbf{Clifford Invariance:} $M_{\alpha}\left(C\ket{\psi}\right) = M_{\alpha}\left(\ket{\psi}\right)$ for all $C$ in the set of Clifford unitaries.
    \item \textbf{Additivity:} $M_{\alpha}\left(\ket{\psi} \otimes \ket{\phi} \right) = M_{\alpha}\left(\ket{\psi}\right) + M_{\alpha}\left( \ket{\phi}\right)$.
\end{enumerate}

Magic can also exist non-locally within a quantum system, where it is closely tied to the entanglement structure of the state. Given any magic measure $M$, and bipartite Hilbert space $\Hil = \Hil_A \otimes \Hil_B$, the non-local magic $M^{NL}$ of $\ket{\psi} \in \Hil$ is defined
\begin{equation}\label{eq:NLMagic}
    M^{NL} \equiv \min_{U_A \otimes U_B} M\left(U_A \otimes U_B \ket{\psi} \right), 
\end{equation}
where $U_A$ and $U_B$ are arbitrary unitaries which act locally on either partition of $\Hil$. Much like entanglement, non-local magic reflects the correlations in magic, as a resource, shared across a Hilbert space partition. 

The calculation of non-local magic, as given in Eq.\ \eqref{eq:NLMagic}, requires an optimization over all local unitaries and thereby is costly to perform. Accordingly, numerous estimates of non-local magic have been developed~\cite{Cao:2024nrx,haug:2025}, which can efficiently approximate and bound the non-local magic of a quantum system. One such non-local magic estimate is the capacity of entanglement $C_E$, defined~\cite{DeBoer:2018kvc} for a state $\rho$ as
\begin{equation}\label{eq:EntCap1}
        C_E\left(\rho \right) \equiv \Tr \left(\rho \left( -\log_d \rho\right)^2\right) - \left(-\Tr \left( \rho \log_d \rho \right)\right)^2.
\end{equation}
Equivalently, $C_E\left(\rho \right)$ can be interpreted%
\footnote{Some readers may be more familiar with the concept of $C_E\left(\rho \right)$ as the second cumulant of the modular Hamiltonian $K$, that is $C_E\left(\rho \right) = \langle K^2 \rangle - \langle K \rangle^2$.}
as the variance of entanglement entropy of $\rho$. Moreover, $C_E\left(\rho \right)$ quantifies the non-flatness of the entanglement spectrum of $\rho$, and thereby provides a rigorous lower bound for the non-local magic
\begin{equation}\label{eq:EntCap2}
        C_E\left(\rho \right) \leq M^{NL} \left(\rho \right).
\end{equation}
%


\subsection{Machine Learning}

In this work we implement a reinforcement learning algorithm to generate states, and construct quantum circuits, that violate select entropy inequalities. Reinforcement learning (RL), a branch of machine learning, trains an agent to learn its environment, e.g. entropy vector space, motivated solely by scalar reward signals~\cite{vanOtterlo2012}. The agent observes the current state of the system, and selects subsequent actions according to a policy designed to maximize future rewards. More specifically, we rely on an RL algorithm known as \textit{Q-learning}, where the long-term reward of each action is described by its associated \textit{q-value}, a numerical prediction of overall reward~\cite{Watkins:1992}. The updated q-value $\mathcal{Q}'(S,A)$, corresponding to action $A$ taken on state $S$, is calculated using Bellman's equation
\begin{equation}
    \mathcal{Q}'(S, A) = (1 - \alpha)\cdot \mathcal{Q}(S, A) + \alpha\left[R(S, A) + \gamma \cdot \text{argmax}_{A'}\left\{\mathcal{Q}(S', A')\right\}\right],
\end{equation}
where $\mathcal{Q}(S, A)$ is the previous q-value estimate, $\alpha$ denotes the learning rate, $\gamma$ the discount factor, and $R(S,A)$ the reward associated with taking action $A$ on state $S$. A larger learning rate prioritizes actions with immediate rewards, while a higher discount value favors future rewards. In order to balance exploration and reward we employ an $\epsilon$-greedy policy,
\begin{equation}
    A_n = 
    \begin{cases}
        \text{argmax}_{\{A\}} \left\{\mathcal{Q}(S_n, A_n) \right \} :1- \epsilon\\
        A_{Rand.} : \epsilon
    \end{cases}
\end{equation}
where the action $A_n$, taken at the $n^{th}$ iteration, is either taken to be the action of largest $\mathcal{Q}(S_n, A_n)$, with probability $1- \epsilon$, or an entirely random action $A_{Rand.}$, with probability $\epsilon$. The value of $\epsilon$ varies between $0$ and $1$, with a high $\epsilon$ value favoring higher exploration. Optimizing the values $\epsilon, \alpha,$ and $\gamma$ is critical for an efficient exploration of the landscape and optimal agent performance. 

\section{A Reinforcement Learning Algorithm for Entropy Vector Dynamics}\label{ReinforcementLearningSection}

\begin{figure}[h]
    \centering
    \includegraphics[width=0.8\linewidth]{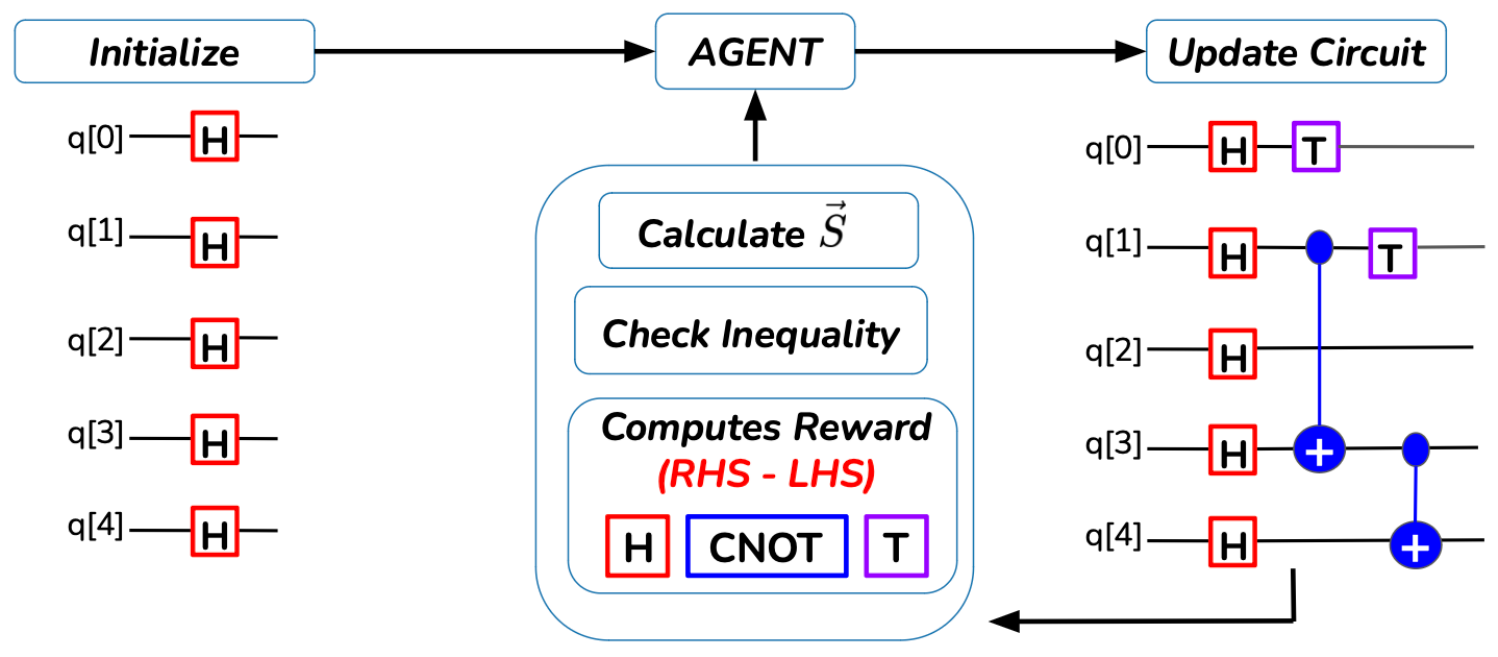}
    \caption{Workflow illustrating reinforcement learning protocol for generating states that fail a chosen entropy inequality, using a select gate set, and analyzing the entropy vector dynamics across violating circuits. After initialization, the circuit is updated with a gate chosen from the Q-learning. The agent calculates the entropy vector after each gate, and checks for violation of the inequality. A reward is computed based on the inequality evaluation, the Q-table is updated, and the process repeats until violation is reached.}    
    \label{fig:ql_workflow}
\end{figure}

We now introduce a reinforcement learning algorithm to analyze the evolution of entropy vectors under quantum circuits, and evaluate the specified violation of entropy inequalities. We emphasize that quantum gates do not act on entropy vectors directly, but rather on the underlying states, whose transformations determine the entropy vector evolution. In this work, we focus on the monogamy of mutual information (MMI) and Ingleton inequalities, given their quantum information-theoretic implications. Furthermore, we often consider quantum circuits generated by elementary quantum gates, e.g. Hadamard, CNOT, and T gates. However, this prescription is fully general and can be applied to analyze the evaluation of generic entropy inequalities, under arbitrary quantum evolution. The protocol proceeds as follows.
\begin{enumerate}\label{Protocol}
    \item A register of $n$ qubits is prepared in an initial state $\ket{\psi_0}$, e.g. the all-zero state $\ket{\psi_0} = \ket{0}^{\otimes n}$.
    
    \item A set of gates $G = \{g_1,\ g_2,\ ...,\ g_i\}$ is chosen to evolve $\ket{\psi_0}$. After each gate is applied, the entropy vector $\vec{S_\psi}$, of the current state $\ket{\psi}$, is computed. Additional properties, e.g. magic, are also computed and analyzed as in Sections \ref{ResourceEvolutionSection} and \ref{ConvergenceSubsection}.

    \item An entropy inequality is chosen, with the objective of driving $\ket{\psi}$ towards violation of this inequality. A reward function is defined using the \textit{inequality difference}, i.e. the right-hand%
    \footnote{Our convention is to state entropy inequalities such that the left-hand side is greater than the right-hand side at satisfaction.} %
    side of the inequality minus the left-hand side 
    \begin{equation}\label{RewardFunction}
        R(\vec{S}_{\psi}) = RHS - LHS
    \end{equation}
    such that the reward increases as the system approaches inequality violation.
    
    \item A reinforcement agent is employed which evaluates $R(\vec{S}_{\psi})$, and determines the next gate $g_j$, according to a Q-learning table, such that $R(S_{\vec{\rho}})$ is maximized.
    
    \item When violation of the inequality is achieved, by the desired amount, the protocol terminates and the current state $\ket{\psi}$ is returned.
\end{enumerate}
The protocol outlined above inputs a specified gate set and a target entropy inequality to explore, and determines both a quantum state that violates the chosen inequality by a desired margin as well as a circuit that prepares the violating state from the initial input state. Throughout the evolution, the state's entropy vector, along with any additional quantum properties of interest, is computed and evaluated against the target inequality. The protocol returns a detailed characterization of the entropy vector dynamics, and other relevant properties, across the violating circuit. Figure~\ref{fig:ql_workflow} provides a sample schematic for the reinforcement learning protocol, evaluating an example inequality on a $5$-qubit system evolving under the action of $H$, $CNOT$, and $T$ gates.

The ordered difference (RHS-LHS) between the two sides of an entropy inequality, or family of inequalities, defines a natural reward function for the reinforcement agent with a larger reward given to actions which push the system closer towards violation. Interpreted geometrically, the reward consistently drives an entropy vector toward the boundary, and ultimately out of, the entropy cone associated with the chosen family of inequalities. Coupling the reward function with an appropriately chosen learning rate $\alpha$, exploration probability $\epsilon$, and discount rate $\gamma$ is crucial to ensure the agent efficiently converges to a violating state. In our implementation, we determine suitable learning parameters to be $\alpha = 0.8$, $\epsilon = 0.2$, and $\gamma = 0.5$. Arrival to these values was informed by known relations between Clifford operators~\cite{Keeler:2023xcx}, as well as by empirical inspection of Q-tables after many training iterations.

The pseudocode below gives the implementation of our reinforcement learning protocol.
\begin{algorithm}[h]
\caption{Q-learning agent for identifying violations using quantum circuits.}
\begin{algorithmic}
\Require Learning rate $\alpha$, exploration rate $\epsilon$, discount factor $\gamma$
\State Initialize Q-table $Q(s,a)$ arbitrarily
\While{goal state is not reached}
    \State Initialize $s$ (starting state)
    \Repeat
        \State Choose action $a \in A(s)$ using $\epsilon$-greedy policy based on $Q$
        \State Take action $a$, observe reward $r$ and next state $s'$
        \State Update $Q(s,a) \gets Q(s,a) + \alpha \left[ r + \gamma \max_{a'} Q(s',a') - Q(s,a) \right]$
        \State $s \gets s'$
    \Until{goal state is reached}
\EndWhile
\end{algorithmic}
\end{algorithm}
Two benefits of Q-learning are that the training process is not time-intensive and the learning process is transparent. In Q-learning, the agent learns in real-time by taking actions and updating the q-values for each of the possible actions. For our problem, the possible actions are each of the universal quantum gates acting on all combinations of the qubits. Actions that bring the agent closer to a quantum state that violates the inequality have higher q-values. These values are updated after each step is taken. The agent is training as it moves through the state space towards the violator state. Since the entanglement entropy is calculated as part of the learning process before each quantum gate is chosen, we are able to track the entanglement dynamics after each quantum gate is performed. This transparency demonstrates the dynamics of the entropy vector and permits analysis of the entropy vector on the path to violation. 

\paragraph{Example}

\begin{figure}[h]
    \centering
    \begin{minipage}{0.49\linewidth}
        \centering
        \includegraphics[width=\linewidth]{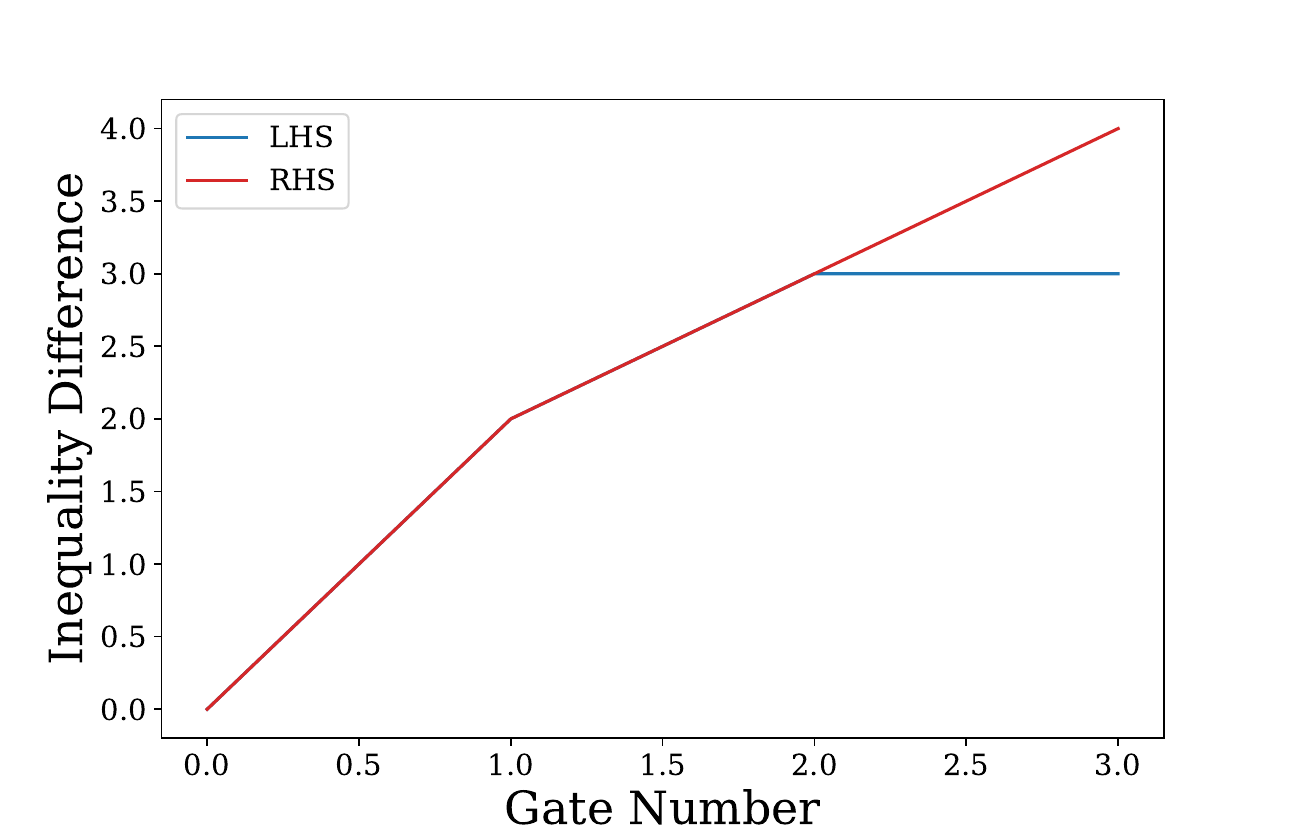}
    \end{minipage}
    \hfill
    \begin{minipage}{0.49\linewidth}
        \centering
        \includegraphics[width=\linewidth]{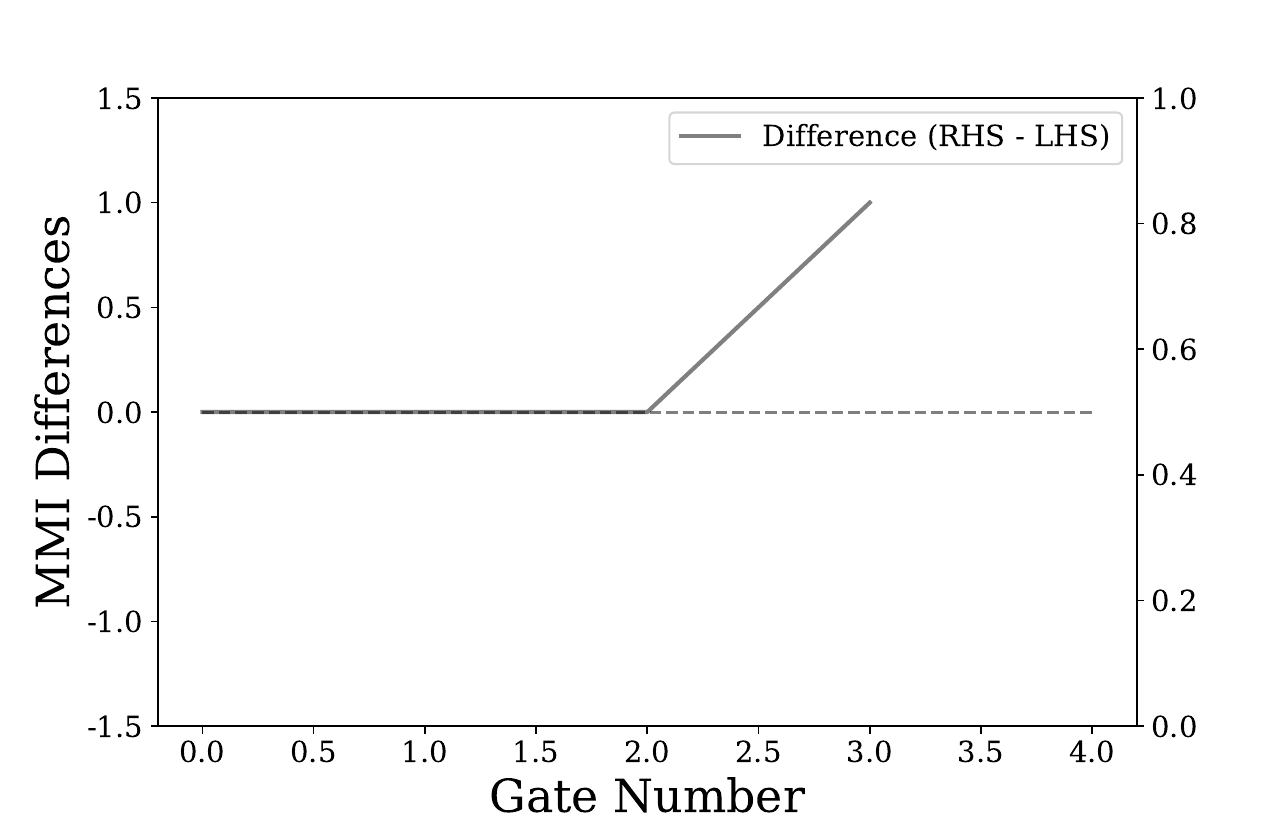}
    \end{minipage}
    \caption{Left image shows the evolution of the left-hand and right-hand sides of MMI, in Eq.\ \eqref{MMI}, as the system evolves from satisfaction to failure. After the $4$ gates in Figure~\ref{fig:mmi-violation-circuit} are applied to $\ket{0}^{\otimes4}$, the system is in state $\ket{GHZ_4}$, and violates MMI. The right image shows the MMI difference, computed by Eq.\ \eqref{MMIDiff}, which becomes positive at violation.}
    \label{fig:mmi-violation}
\end{figure}

For clarity, we now provide an explicit example studying the violation of MMI, defined in Eq.\ \eqref{MMI}, on a $4$-qubit system using a subgroup of Clifford operators. We begin by initializing a quantum register of four qubits into the state $\ket{\psi_0} \equiv \ket{0}^{\otimes 4}$. We then choose the set of gates $G$, where 
\begin{equation}
    G = \{H_i,\ C_{i,j}\}, \quad i,j \in [1,4],
\end{equation}
to evolve $\ket{\psi_0}$. Following Eq.\ \eqref{MMI}, the reward function for the reinforcement agent is given by
\begin{equation}\label{MMIDiff}
    R_{MMI}(S_{\vec{\rho}}) = \left(S_{A} + S_{B} + S_{C} + S_{ABC}\right) - \left(S_{AB} + S_{AC} + S_{BC}\right).
\end{equation}
The agent computes the entropy vector for the evolving state after, each gate is applied, and evaluates the next gate based on the MMI difference. Figure~\ref{fig:mmi-violation} displays the evolution of the MMI difference, in Eq.\ \eqref{MMIDiff}, until violation is achieved.

\begin{figure}[h]
        \centering
        \includegraphics[width=0.45\linewidth]{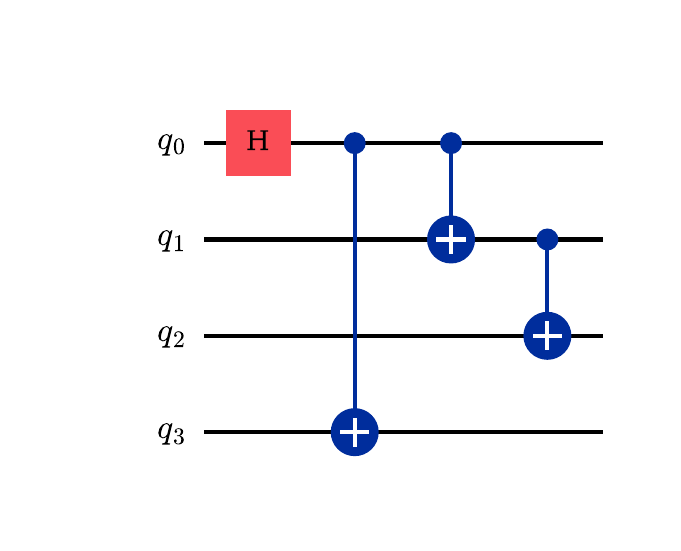}
    \caption{Circuit which prepares the state $\ket{GHZ_4}$ starting from $\ket{0}^{\otimes 4}$. This circuit is constructed using the $H$ and $CNOT$ gates, according to the reinforcement learning protocol described in Figure~\ref{fig:ql_workflow}. The final state $\ket{GHZ_4}$ violates MMI, as in Eq.\ \eqref{MMI}.}
    \label{fig:mmi-violation-circuit}
\end{figure}

The circuit shown in Figure \ref{fig:mmi-violation-circuit}, starting from the initial state $\ket{\psi_0} = \ket{0}^{\otimes 4}$, prepares the state $\ket{GHZ_4}$, defined
\begin{equation}\label{FinalMMIState}
    \ket{GHZ_4} = \frac{1}{\sqrt{2}}\big(|0000\rangle +|1111\rangle \big).
\end{equation}
The state $\ket{GHZ_4}$ has the entropy vector $\vec{S}_{\ket{GHZ_4}}$, which reads
\begin{equation}\label{FinalMMIVector}
    \vec{S}_{\ket{GHZ_4}} = \left(1,\ 1,\ 1,\ 1,\ 1,\ 1,\ 1 \right).
\end{equation}
The vector $\vec{S}_{\ket{GHZ_4}}$ explicitly violates the instance of MMI given by Eq.\ \eqref{MMI}, and thus the algorithm terminates and returns both the state $\ket{GHZ_4}$ and the circuit in Figure~\ref{fig:mmi-violation-circuit} to prepare it.

The $4$-qubit violation of MMI illustrated in Figure \ref{fig:mmi-violation}, while a simple example, highlights the capability for the proposed reinforcement learning approach to identify entropy inequality-violating states, and construct circuits to generate entanglement arrangements with desired information-theoretic properties. The eventual breakdown of MMI illustrates the clear transition from a state with entanglement structure satisfying MMI, to a state whose entropy vector lies outside the MMI entropy cone. Furthermore, since MMI defines several facets of the holographic entropy cone, its violation signals departure from the holographic entropy cone as well. Learning the sequence of gates which drive this transition, the reinforcement agent determines a controlled trajectory through entropy vector space, offering a constructive approach to probing the boundaries and nested structure of distinct entropy cones.

In this section, we introduced a reinforcement learning algorithm designed to probe the dynamics of entropy vectors, under evolution by quantum circuits, and to systematically drive a system toward violation of a specified entropy inequality. The protocol employed a Q-learning agent that selected gate sequences to apply which maximized a reward function, derived from the inequality difference, guiding the state's entropy vector toward, and ultimately beyond, the boundary of the corresponding entropy cone. We demonstrated the effectiveness of this protocol using a simple $4$-qubit example, in which the agent determined a state that violates MMI and constructed a circuit comprised of Hadamard and CNOT gates to prepare that state. In the following sections we apply this protocol to investigate violations of the Ingleton inequality, and analyze the associated entanglement and magic characteristics responsible for this violation.

\section{Violating Ingleton's Inequality}\label{IngletonCircuitSection}

Constructing quantum circuits that prepare states which violate Ingleton's inequality enables a detailed analysis of how bipartite entanglement, and more broadly a set of quantum resources, distribute and reorganize to produce violation of entropy inequalities. Using the reinforcement learning protocol introduced in Section~\ref{ReinforcementLearningSection}, we train an agent, modeled as a Markov decision process, to navigate the Hilbert space from the computational basis to states which violate Ingleton's inequality. We begin by proving that Ingleton violation in qubit systems requires a minimum of six qubits, establishing a lower bound on system size. Using a universal gate set comprised of Hadamard, $T$, and $CNOT$ gates, we then construct explicit circuits that prepare Ingleton-violating states on $6$ qubits. Tracking the state's entropy vector through each circuit reveals how subsystem entanglement grows and redistributes across different subsystems, breaking qubit exchange symmetry and leading to Ingleton violation. Since all stabilizer states satisfy Ingleton's inequality~\cite{Linden2013}, we analyze the emergence of non-stabilizerness, e.g. magic and non-local magic, along the path to violation. Finally, starting from a state that saturates Ingleton's inequality, we perturb around the edges of the Ingleton entropy cone to produce new violating states, and characterize their associated resource profiles.

\subsection{A Minimum Qubit Requirement for Ingleton Violation}\label{MinQubitSection}

Ingleton's inequality, as defined in Eq.\ \eqref{Ingleton}, is a $4$-party entropy inequality, and therefore minimally requires a $4$-party quantum system to violate it. However, since we are interested in realizing Ingleton violation in a pure state, comprised of qubits evolving under quantum circuits, we first determine the minimum number of qubits needed to realize an entanglement structure which violates Ingleton. In this section we prove that Ingleton's inequality cannot be violated in any $5$-qubit pure state, and minimally requires a state of $6$ qubits to occur. The primary argument of this proof relies on the fact that all $n$-qubit entropy vectors, for $n \leq 5$, can be realized by a min-cut protocol on graphs or hypergraphs, and thereby lie within the hypergraph entropy cone. The hypergraph entropy cone is strictly contained within the stabilizer entropy cone~\cite{Bao2020}, of which Ingleton's inequality constitutes facets, and therefore all $5$-qubit entropy vectors satisfy Ingleton's inequality.

Before presenting the proof that Ingleton violation minimally requires six qubits, we first review the necessary graph-theoretic framework used to represent entanglement entropy an entropy vectors. Consider a graph $G=(V,E)$, defined by a set of vertices $V$ and a set of weighted edges $E$. The set $V$ is partitioned%
\footnote{While the difference between internal and external vertices does not impact this particular proof, we retain the distinction as it is important for related graph constructions of entanglement.} %
into subsets $V_{Int.}$, corresponding to \textit{internal} vertices, and $V_{Ext.}$ corresponding to \textit{external} vertices, such that $V = V_{Int.} \cup V_{Ext.}$. For an $n$-party state $\psi$, with purifier, each single-party subsystem $k$
is assigned an external vertex $v_k \in V_{Ext.}$. A graph cut $C(G)$ is defined as a partitioning of $V$ into two disjoint subsets $\mathcal{V}$ and $\overline{\mathcal{V}}$. Given such a cut, the cut \textit{weight} $W(C)$ is the sum of all edge weights $\{w_i\}$ for edges $\{e_i\} \in E$ that connect vertices in $\mathcal{V}$ to those in $\overline{\mathcal{V}}$. For a subsystem $I \subseteq \ket{\psi}$, comprised of single-party systems $\{k\}$, the entanglement entropy $S_I$ is then defined~\cite{Bao2015} as the total edge weight of the minimum-weight cut that partitions $V$ such that $\{v_{k}\} \in \mathcal{V}$ and its complement $\{v_{\bar{k}}\} \subseteq V_{Ext.}$ lies in $\overline{\mathcal{V}}$.

\paragraph{Example}

\begin{figure}[h]   
    \begin{center}
        \begin{overpic}[width=0.3\linewidth]{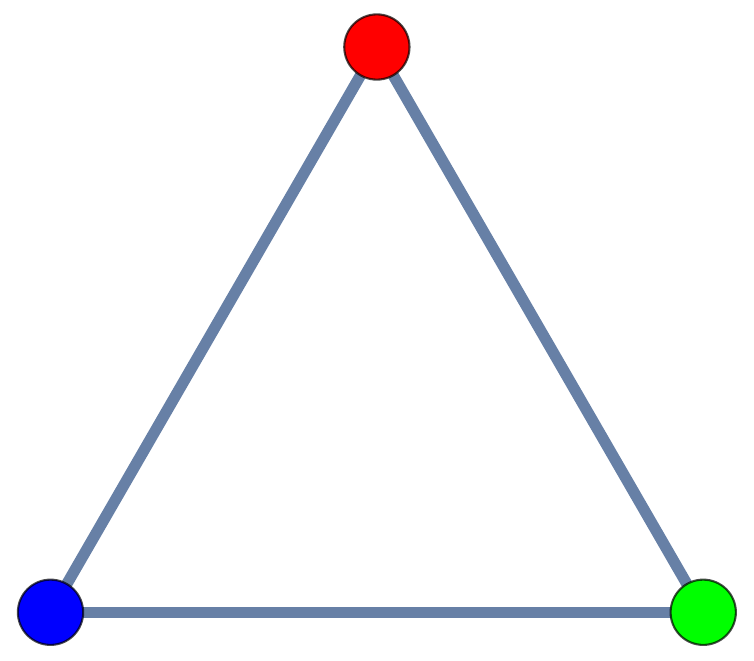}
            \put (37,83) {$A$}
            \put (-8,0) {$B$}
            \put (100,0) {$C$}
            \put (17,45) {$w_1$}
            \put (72,45) {$w_2$}
            \put (45,-2) {$w_3$}
        \end{overpic}
    \caption{Example of an undirected graph that realizes all possible entropy vectors for $3$-qubit quantum states, using a min-cut prescription. Each vertex corresponds to an individual qubit, and each edge carries a weight $w_i$. The entanglement entropy $S_i$, of the $i^{th}$ qubit, corresponds to the weight of cutting both edges connected to it, e.g. $S_A = w_1+w_2$.}
    \label{fig:undirectedGraph-3node}
    \end{center}
\end{figure}
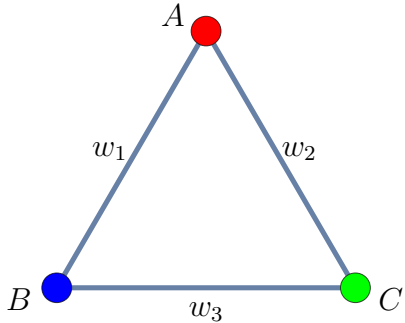

As a simple example let $\ket{\psi}$ denote a generic $3$-qubit pure state, with entropy vector $\vec{S}(\ket{\psi}) = \left(S_A,\, S_B,\ S_C\right)$ consisting of $2^{3-1}-1 = 3$ entanglement entropies. Consider a weighted, undirected triangular graph consisting of three vertices and three edges, as shown in Figure \ref{fig:undirectedGraph-3node}. We assign each single-qubit region, $A,\ B,$ and $C$, to a vertex in this graph, and allow the variables $w_1,\ w_2,$ and $w_3$ to denote arbitrary edge weights.
Computing $\vec{S}(\ket{\psi})$ corresponds to calculating the edge weights of the min-cuts which determine $S_A, S_B,$ and $S_C$. We observe that separating vertex $A$ from its complement vertex set $B \cup C$ requires cutting edges $w_1$ and $w_2$, and therefore $S_A = w_1 + w_2$. By a similar argument we have $S_B = w_1+w_3$ and $S_C = w_2+w_3$. We can represent the equations defining each entanglement entropy for our $3$-qubit system as the following linear system
\begin{equation}\label{UndirectedLinear}
    \begin{pmatrix}
    1 & 1 & 0 & S_A\\
    1 & 0 & 1 & S_B\\
    0 & 1 & 0 & S_C\\
    \end{pmatrix}.
\end{equation}
The system in Eq.\ \eqref{UndirectedLinear} contains three variables, each corresponding to an entry of the generic $3$-qubit pure state entropy vector, and three independent equations. Consequently, for any set of entropies $\{S_A,\ S_B,\ S_C\}$, it is possible to assign weights $\{w_i\}$ such that $\vec{S}(\ket{\psi})$ is realized as a min-cut on the graph shown in Figure~\ref{fig:undirectedGraph-3node}. We refer to states whose entropy vectors can be realized as the weights of a min-cut on an undirected graph as \textit{undirected graph realizable}. A notable example of such states are holographic states, which possess entropy vectors satisfying this property.

At four qubits, the structure of undirected graphs is not rich enough to capture all pure state entropy vectors. A simple counterexample is the $4$-qubit $GHZ$ state with entropy vector $\vec{S} = (1,1,1,1,1,1,1)$, which cannot be realized by the above min-cut protocol on undirected graphs and which violates MMI as shown in Section \ref{ReinforcementLearningSection}. However, this $GHZ$ entropy vector can be realized from the weights of a graph min-cut if we promote the undirected graphs to hypergraphs~\cite{Bao2020a}. In a hypergraph $H = (V,E)$, we allow the edges $\{e_i\} \in E$ to simultaneously connect any number of vertices. Accordingly, removing a single hyperedge disconnects all vertices connected by that hyperedge.

To prove that six qubits are minimally required to violate the Ingleton inequality, we will demonstrate that all $5$-qubit entropy vectors are realizable via a min-cut protocol on hypergraphs%
\footnote{Any hypergraph can be realized as a directed graph, and an analogous construction exists within a directed graph framework.}
.

\begin{theorem}\label{MinQubitTheorem}
    Every $n$-qubit pure state entropy vector, for $n\leq 5$, admits a hypergraph representation, where each entropy component is the sum of edge weights across a corresponding minimum-weight cut.
\end{theorem}
    
\begin{figure}[h]
    \begin{center}
        \begin{overpic}[width=1\linewidth]{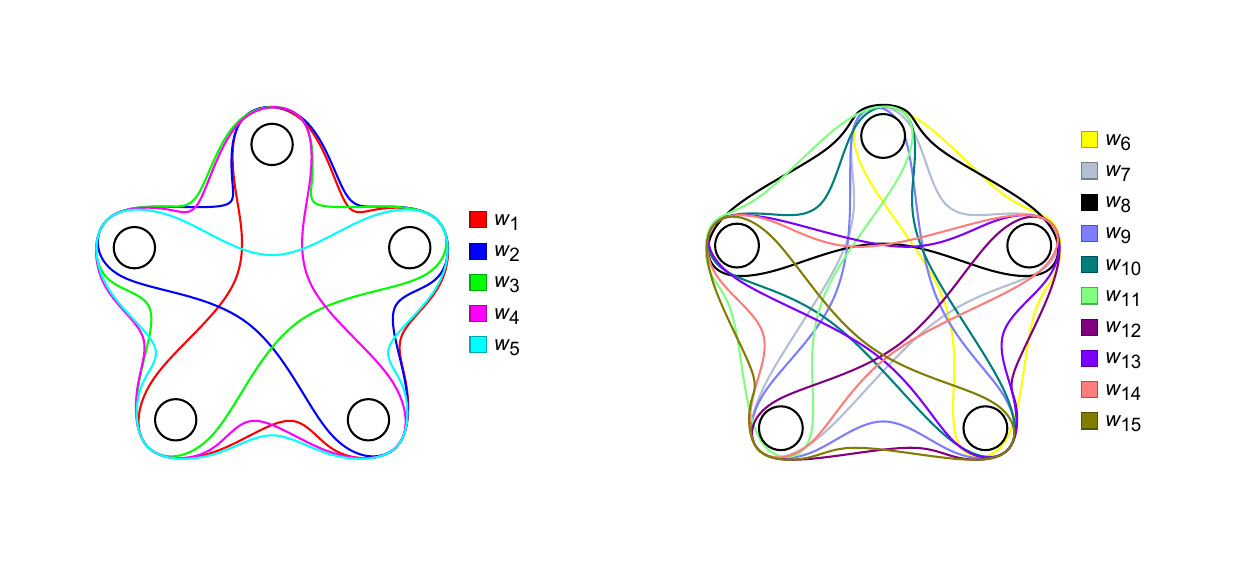}
            \put (47,23) {\Large{$\bigcup$}}
            \put (20.75,33.25) {$A$}
            \put (70,34) {$A$}
            \put (32,24.75) {$B$}
            \put (82,25) {$B$}
            \put (28.6,10.9) {$C$}
            \put (78.5,10.3) {$C$}
            \put (13,10.9) {$D$}
            \put (62,10.3) {$D$}
            \put (9.75,24.75) {$E$}
            \put (58.25,25) {$E$}
        \end{overpic}
    \caption{Hypergraph possessing a min-cut protocol that realizes any $5$-qubit pure state entropy vector. The full graph is shown as the union of two subgraphs, with all $4$-edges shown to the left and all $3$-edges to the right. Eq.\ \eqref{HypergraphLinear} demonstrates how these $15$ edges weights combine to produce any $5$-qubit pure state entropy vector.}
    \label{fig:undirectedGraph-5node}
    \end{center}
\end{figure}
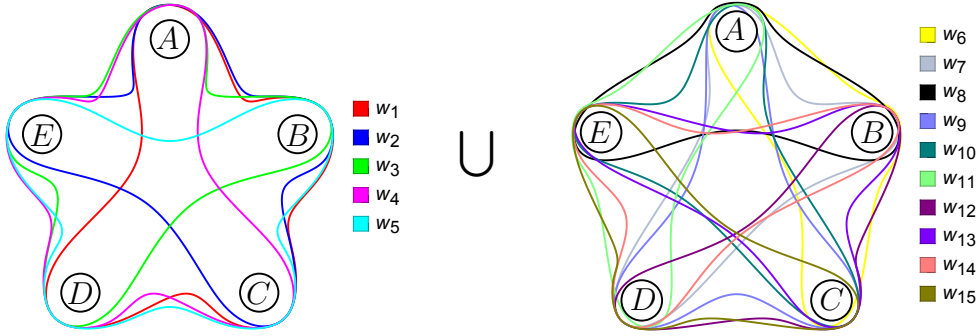

\textit{Proof.} Let $\ket{\psi_{ABCDE}}$ be an arbitrary $5$-qubit pure state. The entropy vector  $\vec{S}\left(\psi_{ABCDE}\right)$ is then composed of $15$ entanglement entropies
\begin{footnotesize}
\begin{equation}\label{FiveQubitEV}
    \vec{S}\left(\psi_{ABCDE}\right) = \left(S_A,S_B,S_C,S_D,S_E,S_{AB},S_{AC},S_{AD},S_{AE},S_{BC},S_{BD},S_{BE},S_{CD},S_{CE},S_{DE} \right).
\end{equation}
\end{footnotesize}
We seek to identify a hypergraph whose min-cut weights realize $\vec{S}\left(\ket{\psi_{ABCDE}}\right)$, for any $\ket{\psi_{ABCDE}}$. Consider the hypergraph $H$, consisting of $5$ vertices and $15$ hyperedges, shown in Figure \ref{fig:undirectedGraph-5node}. For clarity, we present $H$ as the graph union of the two subgraphs shown in the figure.

We first identify the $5C4=5$ possible $4$-edges in $H$, shown to the left of Figure \ref{fig:undirectedGraph-5node}. Next, we identify the $5C3=10$ possible $3$-edges in $H$, shown to the right of Figure \ref{fig:undirectedGraph-5node}. The full graph $H$ is then defined as the graph union of the two graphs shown in Figure \ref{fig:undirectedGraph-5node}. As before, we compute each entropy of $\vec{S}\left(\ket{\psi_{ABCDE}}\right)$ by determining the min-cut required to separate $\{v_k\} \subseteq V$ from $\{v_k\} \subseteq V$, for each $k \in \mathcal{P}(\{A,\ B,\
 C,\ D,\ E\})\setminus \emptyset$. For example, separating vertex $A$ from vertices $B,\ C,\ D,$ and $E$ requires cutting four $4$-edges and six $3$-edge, with $S_A$ the sum of weights for all edges cut
\begin{equation}\label{Min-Cut}
    S_A = w_1 + w_2 + w_4 + w_5 + w_6 + w_9 + w_{10} + w_{11} + w_{13} + w_{14}.
\end{equation}
Repeating the min-cut procedure in Eq.\ \eqref{Min-Cut} for all other $v_k$, we generate the following linear system for all $S_k$ entries of the entropy vector.
\vspace{-1em}
\begin{equation}\label{HypergraphLinear}
\left(\begin{array}{cccccccccccccccc}
    1 & 1 & 0 & 1 & 1 & 1 & 0 & 0 & 1 & 1 & 1 & 0 & 1 & 1 & 0 & S_{A}\\
    1 & 1 & 1 & 0 & 1 & 1 & 1 & 0 & 0 & 1 & 0 & 1 & 0 & 1 & 1 & S_{B}\\
    1 & 0 & 1 & 1 & 1 & 1 & 1 & 1 & 0 & 0 & 1 & 0 & 1 & 0 & 1 & S_{C}\\
    1 & 1 & 1 & 1 & 0 & 0 & 1 & 1 & 1 & 0 & 1 & 1 & 0 & 1 & 0 & S_{D}\\
    0 & 1 & 1 & 1 & 1 & 0 & 0 & 1 & 1 & 1 & 0 & 1 & 1 & 0 & 0 & S_{E}\\
    1 & 1 & 1 & 1 & 1 & 1 & 1 & 0 & 1 & 1 & 1 & 1 & 1 & 1 & 1 & S_{AB}\\
    1 & 1 & 1 & 1 & 1 & 1 & 1 & 1 & 1 & 1 & 1 & 0 & 1 & 1 & 1 & S_{AC}\\
    1 & 1 & 1 & 1 & 1 & 1 & 1 & 1 & 1 & 1 & 1 & 1 & 1 & 1 & 0 & S_{AD}\\
    1 & 1 & 1 & 1 & 1 & 1 & 0 & 1 & 1 & 1 & 1 & 1 & 1 & 1 & 1 & S_{AE}\\
    1 & 1 & 1 & 1 & 1 & 1 & 1 & 1 & 0 & 1 & 1 & 1 & 1 & 1 & 1 & S_{BC}\\
    1 & 1 & 1 & 1 & 1 & 1 & 1 & 1 & 1 & 1 & 1 & 1 & 0 & 1 & 1 & S_{BD}\\
    1 & 1 & 1 & 1 & 1 & 1 & 1 & 1 & 1 & 1 & 0 & 1 & 1 & 1 & 1 & S_{BE}\\
    1 & 1 & 1 & 1 & 1 & 1 & 1 & 1 & 1 & 0 & 1 & 1 & 1 & 1 & 1 & S_{CD}\\
    1 & 1 & 1 & 1 & 1 & 1 & 1 & 1 & 1 & 1 & 1 & 1 & 1 & 0 & 1 & S_{CE}\\
    1 & 1 & 1 & 1 & 1 & 0 & 1 & 1 & 1 & 1 & 1 & 1 & 1 & 1 & 1 & S_{DE}\\
    \end{array}\right)
\end{equation}

The linear system in Eq.\ \eqref{HypergraphLinear} consists of an invertible matrix, with $15$ independent equations, and an augmented column of $15$ free variables. Therefore, for any $\vec{S}\left(\psi_{ABCDE}\right)$ in Eq.\ \eqref{FiveQubitEV} we can solve Eq.\ \eqref{HypergraphLinear} exactly. Solving Eq.\ \eqref{HypergraphLinear} yields the following relations among all $15$ entanglement entropies (for compactness we denote the entanglement entropy of a subsystem by its label alone, e.g. $S_A \to A$)
\begin{scriptsize}
\begin{equation}\label{Generic5QubitEntropies}
    \begin{split}
        w_1 &= \frac{1}{7} (-3 A+3 AB+3 AC+3 AD+3 AE+4 B-4 BC-4 BD-4 BE+4 C-4 CD-CE+4 D-4 DE+4 E)\\
        w_2 &=\frac{1}{7} (3 A+4 AB-3 AC-3 AD-3 AE-4 B+4 BC+4 BD+4 BE+3 C-3 CD-6 CE+3 D-3 DE+3 E) \\
        w_3 &=\frac{1}{7} (3 A-3 AB+4 AC-3 AD-3 AE+3 B+4 BC-3 BD-3 BE-4 C+4 CD+CE+3 D-3 DE+3 E) \\
        w_4 &=\frac{1}{7} (3 A-3 AB-3 AC+4 AD-3 AE+3 B-3 BC+4 BD-3 BE+3 C+4 CD-6 CE-4 D+4 DE+3 E) \\
        w_5 &=\frac{1}{7} (3 A-3 AB-3 AC-3 AD+4 AE+3 B-3 BC-3 BD+4 BE+3 C-3 CD+CE+3 D+4 DE-4 E) \\
        w_6 &=\frac{1}{7} (-A+AB+AC+AD-6 AE-B+BC+BD+BE-C+CD+2 CE-D+DE-E) \\
        w_7 &=\frac{1}{7} (-A+AB-6 AC+AD+AE-B+BC+BD+BE-C+CD+2 CE-D+DE-E) \\
        w_8 &=\frac{1}{7} (-A+AB+AC+AD+AE-B+BC+BD+BE-C-6 CD+2 CE-D+DE-E) \\
        w_9 &=\frac{1}{7} (-A+AB+AC+AD+AE-B+BC-6 BD+BE-C+CD+2 CE-D+DE-E) \\
        w_{10} &=\frac{1}{7} (-A+AB+AC+AD+AE-B+BC+BD-6 BE-C+CD+2 CE-D+DE-E) \\
        w_{11} &=\frac{1}{7} (-A+AB+AC-6 AD+AE-B+BC+BD+BE-C+CD+2 CE-D+DE-E) \\
        w_{12} &=\frac{1}{7} (-A+AB+AC+AD+AE-B-6 BC+BD+BE-C+CD+2 CE-D+DE-E) \\
        w_{13} &=\frac{1}{7} (-A+AB+AC+AD+AE-B+BC+BD+BE-C+CD+2 CE-D-6 DE-E) \\
        w_{14} &=\frac{1}{7} (-A-6 AB+AC+AD+AE-B+BC+BD+BE-C+CD+2 CE-D+DE-E) \\
        w_{15} &=\frac{1}{7} (-A+AB+AC+AD+AE-B+BC+BD+BE-C+CD-5 CE-D+DE-E) \\
    \end{split}
\end{equation}
\end{scriptsize}

Since Eq.\ \eqref{Generic5QubitEntropies} assigns a weight to each edge in Figure~\ref{fig:undirectedGraph-5node}, expressed as some linear combination of arbitrary entanglement entropies, the set of all entropy vectors for $5$-qubit pure states is hypergraph realizable.\qed

One immediate consequence of Theorem \ref{MinQubitTheorem} is that no $5$-qubit pure state can fail Ingleton's inequality, which we detail in the corollary below.
\begin{corollary}\label{HypergraphCorollary}
    Every $n$-qubit pure state for $n \leq 5$ satisfies or saturates, but does not fail, Ingleton's inequality.
\end{corollary}
\textit{Proof.} The proof of Corollary \ref{HypergraphCorollary} follows directly from the fact that the hypergraph entropy cone, i.e. the convex hull of all hypergraph realizable entropy vectors, is strictly contained~\cite{Bao2020} within the stabilizer entropy cone, beginning at $n = 4$ qubits. Recall that all stabilizer states always satisfy, or saturate, Ingleton's inequality~\cite{Linden2013}, at any $n$, and thereby Ingleton's inequality comprises a collection of facets for the $n$-party stabilizer entropy cone. Since all hypergraph entropy vectors are strictly contained within the stabilizer entropy cone, no hypergraph realizable entropy vector may fail Ingleton's inequality.\qed

In this section, we proved that no $5$-qubit pure state possesses an entropy vector that violates Ingleton’s inequality. We reviewed a graph-theoretic framework for computing entanglement entropies as edge weights of a min-cut on a mathematical graph. We showed that any generic $5$-qubit pure state entropy vector can be represented using this min-cut protocol on a hypergraph. Since all $5$-qubit pure state entropy vectors are hypergraph realizable, and the hypergraph entropy cone is strictly contained within the stabilizer entropy cone, whose facets include Ingleton’s inequality, no $5$-qubit pure state can violate Ingleton’s inequality. Consequently, our search for Ingleton-violating states begins in the next section with systems of six qubits.

\subsection{Quantum Resource Evolution in Ingleton-Violating Circuits}\label{ResourceEvolutionSection}

In this section, we analyze the structure and dynamics of entanglement and magic in quantum circuits that yield violation of Ingleton's inequality. Starting with a known $4$-party Ingleton-violating state, we prepare this state on a $6$-qubit quantum register using a universal gate set composed of $H$, $T$, and $CNOT$ gates. This construction by elementary gates enables a step-by-step analysis of how successive unitary operations drive the system from within the Ingleton entropy cone to its exterior. As the circuit evolves, we compute the state's entropy vector and quantum magic, revealing how each resource transforms as Ingleton's inequality is violated. Finally we demonstrate that the emergence of some minimal amount of non-local magic, across different Hilbert space bipartitions, accompanies failure of Ingleton's inequality.

We begin by considering the Ingleton-violating state~\cite{Linden2013}, shown in Eq.\ \eqref{eq:lindenstate}. Let $\rho_{ABCD}$ be the $4$-party mixed state defined
\begin{equation}\label{eq:lindenstate}
    \rho_{ABCD} \equiv \frac{1}{2} \ket{GHZ_4}\bra{GHZ_4} + \frac{1}{4} \ket{0101} \bra{0101} + \frac{1}{4} \ket{1001} \bra{1001},
\end{equation}
where $\ket{GHZ_4}$ denotes%
\footnote{State $\rho_{ABCD}$ is equivalent to the state given in Eq. (6) of \cite{Linden2013}, with two qubits swapped.} %
the $4$-qubit $GHZ$ state $(\ket{0000} + \ket{1111})/\sqrt{2}$. In order to construct a quantum circuit that prepares $\rho_{ABCD}$ on a system of qubits, we employ a purification scheme with two ancilla qubits $q_0$ and $q_1$. Let $R = [q_0,\ q_1]$ denote the $2$-qubit ancilla register, such that the $6$-qubit%
\footnote{Following Theorem~\ref{MinQubitTheorem}, two ancilla qubits are minimally needed to purify $\rho_{ABCD}$.} %
pure state $\ket{\psi_{ABCDR}}$ is defined
\begin{equation}\label{PureState}
    \ket{\psi_{ABCDR}} \equiv \sqrt{\frac{1}{2}} \ket{GHZ_4} \otimes\ket{00}_R + \sqrt{\frac{1}{4}} \ket{0101} \otimes \ket{01}_R + \sqrt{\frac{1}{4}} \ket{1001} \otimes \ket{10}_R
\end{equation}
The Ingleton-violating subsystem $\rho_{ABCD}$ is obtained from $\ket{\psi_{ABCDR}}$ by performing a partial trace over qubits $q_0$ and $q_1$,
\begin{equation}
    \rho_{ABCD} = \Tr_{q_0q_1}\left(\ket{\psi_{ABCDR}}\bra{\psi_{ABCDR}}\right).
\end{equation}

\begin{figure}[h]
    \centering
    \includegraphics[width=0.6\linewidth]{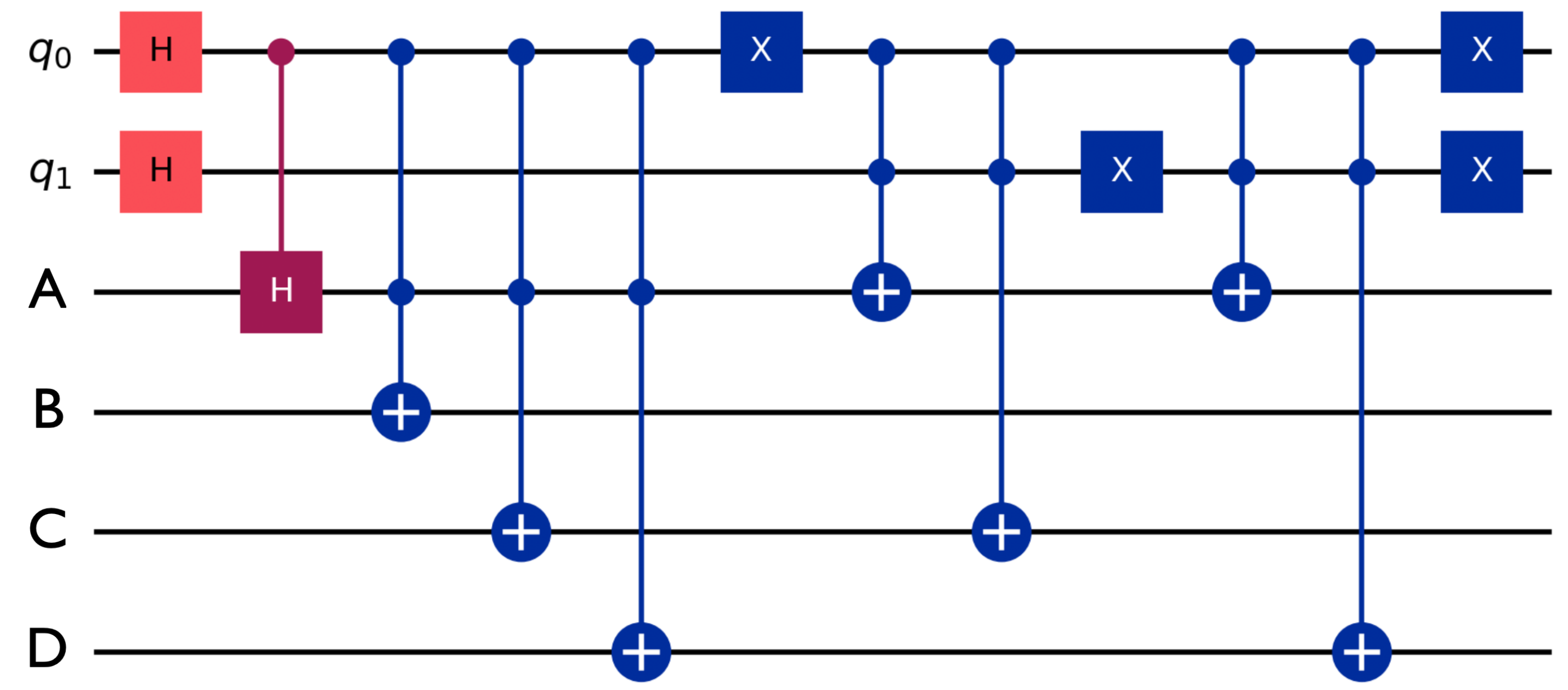}
    \caption{Circuit that prepares the Ingleton-violating state $\ket{\psi_{ABCDR}}$, in Eq.\ \eqref{PureState}, starting from the computational basis. The circuit is constructed using $H,\ CH,\ X,$ and $CCX$ gates, with a decomposition into $H,\ T,$ and $CNOT$ given in Figure~\ref{fig:lindencircuit}. Ingleton's inequality is failed on subsystem $\rho_{ABCD}$, obtained by tracing out qubits $q_0$ and $q_1$.}
    \label{fig:contractedCircuit}
\end{figure}

Preparing the state $\ket{\psi_{ABCDR}}$ in Eq.\ \eqref{PureState} using a quantum circuit allows for the evolution of quantum resources to be studied as the state is driven towards Ingleton violation. Examining the structure of Eq.\ \eqref{PureState}, we derive a circuit, shown in Figure~\ref{fig:contractedCircuit}, which prepares $\ket{\psi_{ABCDR}}$ from the measurement basis state $\ket{\psi_0} \equiv \ket{0}^{\otimes 6}$.

\begin{figure}[h]
    \centering
    \includegraphics[width=0.8\linewidth]{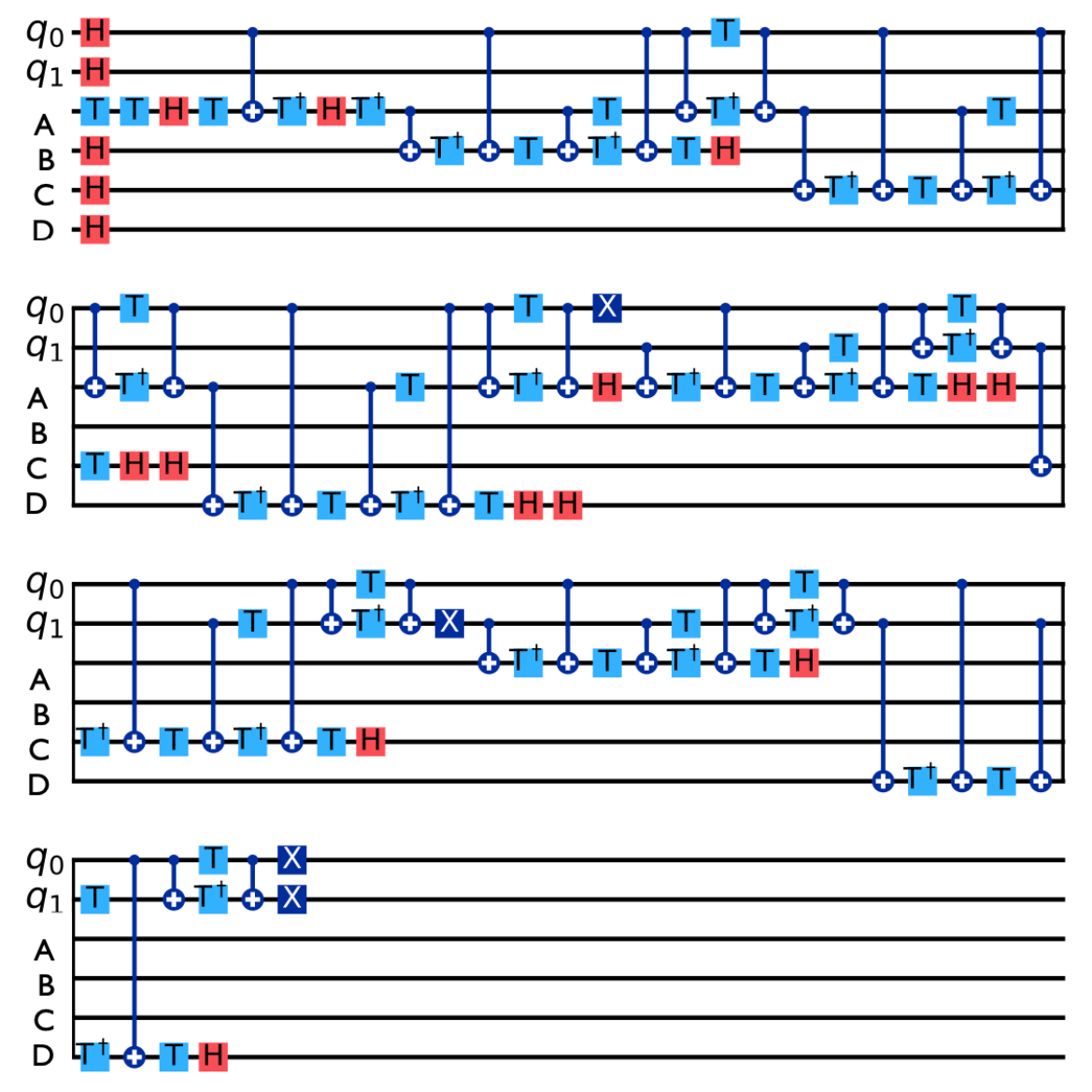}
    \caption{Circuit preparing the Ingleton-violating state $\ket{\psi_{ABCDR}}$, in Eq.\ \eqref{PureState}. This circuit is equivalent to that in Figure~\ref{fig:contractedCircuit}, decomposed into $H,\ T,$ and $CNOT$ gates. This decomposition, into Cliffords and an elementary magic gate, enables a fine-grained analysis of quantum resource evolution as a state leaves the Ingleton entropy cone.}
    \label{fig:lindencircuit}
\end{figure}

We now decompose the Ingleton-violating circuit in Figure~\ref{fig:contractedCircuit} into $H$, $T$, and $CNOT$ quantum gates, shown in Figure~\ref{fig:lindencircuit}, which serves a dual purpose. First, while $\{H_i,\ T_i,\ C_{i,j}\}$ forms a universal quantum gate set%
\footnote{Together, the gates $H$ and $T$ generate arbitrary single-qubit rotations, while $CNOT$ enables maximal entanglement. Consequently, the set $\{H_i,\ T_i,\ C_{i,j}\}$ can approximate any unitary to arbitrary precision with a finite depth circuit.} %
, the finite and small generating set of elementary gates enables a fine-grained analysis of the state's quantum resources, e.g. the entropy vector and magic, as the state leaves the Ingleton entropy cone. Moreover, the set $\{H_i,\ T_i,\ C_{i,j} \}$ consists of two Clifford operations and a single non-Clifford operation. Since $T$ gate count serves as a proxy for non-stabilizerness, the number of $T$ gates can be directly compared to magic monotones, such as stabilizer Renyi entropy, for a comparative analysis of magic evolution. The circuit in Figure~\ref{fig:lindencircuit} prepares $\ket{\psi_{ABCDR}}$ beginning from the from the all-zero state $\ket{\psi_0} = \ket{0}^{\otimes 6}$.
The circuit in Figure~\ref{fig:lindencircuit} prepares $\ket{\psi_{ABCDR}}$ using $119$ gates selected from the set $\{H_i,\ T_i,\ C_{i,j}\}$. In Section~\ref{EdgeOfConeSection}, the gate set $\{H_i,\ T_i,\ C_{i,j} \}$ will comprise the set of actions provided to our reinforcement agent as we probe the exterior of the Ingleton entropy cone. 

\begin{figure} [h]
    \centering
    \begin{overpic}[width=0.75\linewidth]{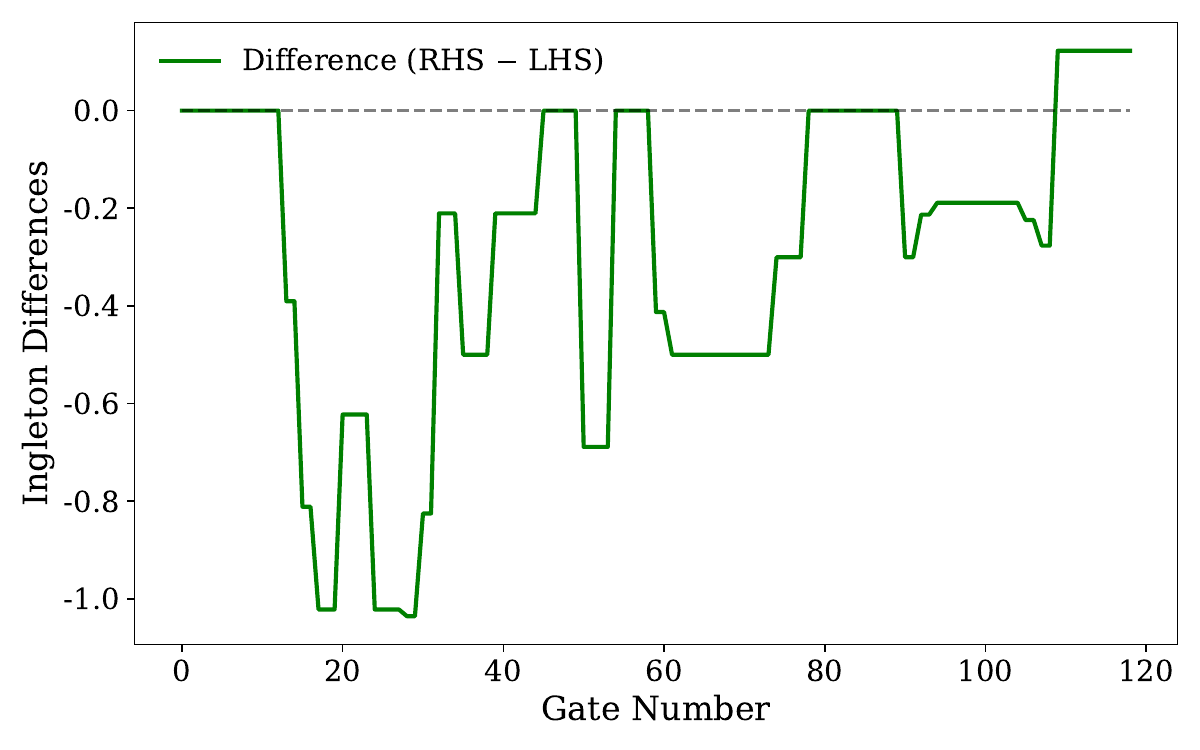} 
    \end{overpic}
    \caption{Evolution of the Ingleton difference, given by Eq.\ \eqref{IngletonDifference}, over the circuit shown in Figure~\ref{fig:lindencircuit} that prepares $\ket{\psi_{ABCDR}}$. Regions when the difference is negative indicate satisfaction of Ingleton's inequality, while positive regions indicate violation. The dashed line corresponds to saturation of the inequality.}
    \label{fig:lindendiff}
\end{figure}

After each gate in Figure~\ref{fig:lindencircuit} is applied, the entropy vector $\vec{S}_{\rho_{ABCD}}$ is computed and Ingleton's inequality is evaluated. We define the Ingleton difference as the right-hand side of the inequality in Eq.\ \eqref{Ingleton} minus the left-hand side, which we use to quantify proximity to violation. For the specific instance of Ingleton's inequality in Eq.\ \eqref{Ingleton}, this difference takes the form
\begin{equation}\label{IngletonDifference}
\left(S_{A} + S_{B} + S_{ABC} + S_{ABD} + S_{CD}\right) - \left(S_{AB} + S_{AC} + S_{AD} + S_{BC} + S_{BD}\right).
\end{equation}
A negative value of Eq.\ \eqref{IngletonDifference} indicates that Ingleton's inequality is satisfied, zero corresponds to saturation, and positive values signify failure. Figure~\ref{fig:lindendiff} illustrates the evolution of this Ingleton difference for $\vec{S}_{\rho_{ABCD}}$, computed after each gate is applied in the circuit.
Figure \ref{fig:lindendiff} provides a direct visualization of how the evaluation of Ingleton's inequality evolves under successive transformations by gates in $\{H_i,\ T_i,\ C_{i,j} \}$. Since $H$ and $T$ are local unitaries, the Ingleton difference is fixed under all combinations of $H$ and $T$. Nevertheless, the preceding application of $H$ and $T$ prepare the system, by arranging local basis structure, so that entanglement can be altered in the appropriate ways required to achieve violation. Each change to the Ingleton difference occurs via a $CNOT$ gate, though importantly not every $CNOT$ application will modify this difference. 

\begin{figure}
    \centering
    \begin{overpic}[width=.8\linewidth]{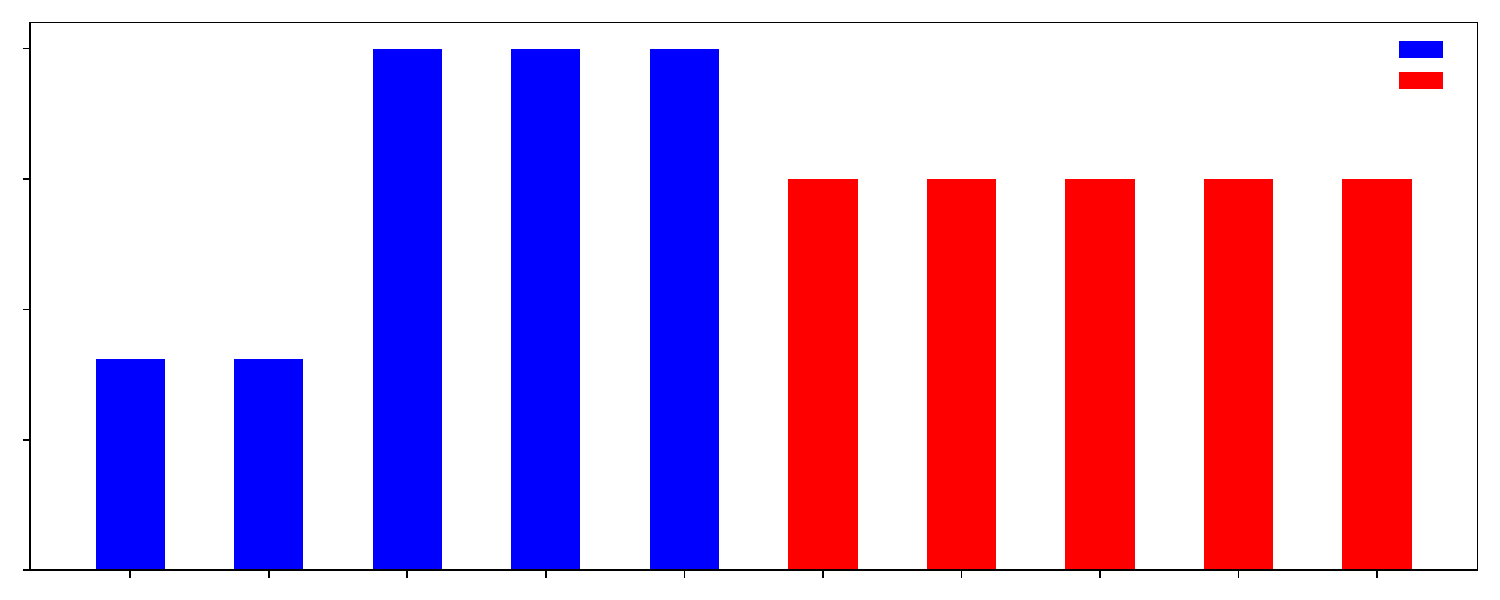}
    \put(7, -2) {A}
    \put(16, -2) {B}
    \put(22, -2) {ABC}
    \put(32, -2) {ABD}
    \put(43, -2) {CD}
    \put(51, -2) {AB}
    \put(60, -2) {AC}
    \put(69, -2) {AD}
    \put(79, -2) {BC}
    \put(89, -2) {BD}
    \put(-4.0, 1.5) {0.0}
    \put(-5, 10) { 0.5}
    \put(-5, 19) { 1.0}
    \put(-5, 28) { 1.5}
    \put(-5, 36) { 2.0}
    
   
    \put(87,36){ \tiny RHS}
    \put(87,34){ \tiny LHS}
  \end{overpic}
    \caption{Entropy vector components, specifically those appearing in Ingleton's inequality in Eq.~\eqref{Ingleton}, for the violating state $\ket{\psi}_{ABCDR}$. Subsystem $CD$, appearing on the right-hand side of Eq.~\eqref{Ingleton}, maximizes its entanglement entropy when the last $CNOT$ gate in Figure~\ref{fig:lindencircuit} is applied, exceeding all other $2$-party regions and yielding violation.}
    \label{fig:individualEntropies}
\end{figure}
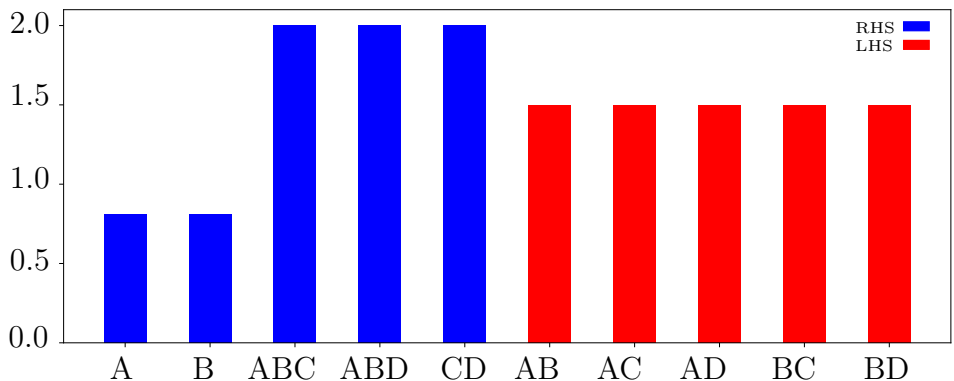

Once the state $\ket{\psi_{ABCDR}}$ is prepared, we analyze the components of $\vec{S}_{\rho_{ABCD}}$ that appear in the Ingleton instance given by Eq.\ \eqref{Ingleton}. For the $4$-qubit subsystem $\rho_{ABCD}$, on which Eq.\ \eqref{Ingleton} fails, we plot the subsystem entropies contributing violation of Ingleton's inequality in Figure~\ref{fig:individualEntropies}.
Subsystem $\rho_{ABCD}$ contains six pairs of qubits, five of whose entropies appear on the left-hand side of Eq.\ \eqref{Ingleton} and one of which, subsystem $CD$, that appears on the right. When the final $CNOT$ gate of Figure~\ref{fig:contractedCircuit} is applied, the entropy of $CD$ is maximized $(S_{CD} = 2)$, and becomes larger than the entropies of all other $2$-party subsystems. 

The relationship between $2$-party entanglement entropies plays a critical role in the evaluation of Ingleton's inequality. To highlight the dependence on two-party entropies, we can rewrite~\cite{Linden2013} Eq.\ \eqref{Ingleton} using mutual information as
\begin{equation}\label{MutualInformationRewrite}
    I(B:C \mid A) + I(A:D \mid B) + R \geq 0,
\end{equation}
where $R$ is a linear combination of $2$-party entropies given by
\begin{equation}\label{R}
    R \equiv S_{BC} + S_{AD} - S_{CD} - S_{AB}.
\end{equation}
Quantities $I(B:C \mid A) \equiv S_{BA} + S_{AC} - S_{ABC} - S_A$ and $I(A:D \mid B) \equiv S_{AB} + S_{DB} - S_{ABD} - S_B$ give the correlation of $B$ and $C$ when $A$ is traced out, and correlation of $A$ and $D$ when $B$ is traced out, respectively. Expressing Ingleton this way enables a more intuitive observation of the symmetry between qubits $A,\ B,\ C,$ and $D$ in the $2$-party subsystems. The exchange of qubits $A$ and $C$ flips the sign on $R$ in Eq.\ \eqref{R}. If $\rho_{ABCD}$ is symmetric up to this qubit swap, then $R=0$ and the parity change does not affect the satisfaction of Eq.\ \eqref{MutualInformationRewrite}. However, as illustrated in Figure~\ref{fig:individualEntropies}, the entropies of the $2$-party subsystems obey the relation  
\begin{equation}\label{CDbigger}
    S_{CD} > \{S_{AB},\ S_{AC},\ S_{AD},\ S_{BC},\ S_{BD}\}, 
\end{equation}
breaking the qubit exchange symmetry that leaves $R$ fixed, and thus resulting in Ingleton violation. This feature holds for all Ingleton-violating states~\cite{Linden2013}.

Recall that Ingleton's inequality defines multiple facets of the stabilizer entropy cone~\cite{Linden2013}, and is therefore satisfied by all stabilizer states. Consequently, any failure of Ingleton's inequality necessarily signals the presence of non-stabilizerness in the state. Probing this connection between non-stabilizerness and evaluation of Ingleton's inequality, we examine the evolution of quantum magic in $\rho_{ABCD}$ throughout the Ingleton-violating circuit shown in Figure~\ref{fig:lindencircuit}. An efficiently computable magic monotone is the stabilizer Renyi entropy (SRE), defined in Eq.\ \eqref{eq:SRE1}. While a useful magic monotone for pure states, SRE is not well-defined for mixed states~\cite{haug:2025}. Therefore we instead utilize the $n$-qubit magic witness $\mathcal{W}_{\alpha}$ to compute the magic~\cite{haug:2025} of $\rho_{ABCD}$, defined
\begin{equation}\label{MagicWitness}
    \mathcal{W}_{\alpha}\left(\rho \right) \equiv \frac{1}{1-\alpha} \ln \left(2^{-n} \sum_{P \in \mathcal{P}_n}|\Tr \left( \rho P\right)|^{2\alpha} \right) - \frac{1-2\alpha}{1-\alpha}S_2\left(\rho\right),
\end{equation}
where $S_2\left(\rho\right)$ denotes the $2$-Renyi entropy $S_2\left(\rho\right) = -\ln \Tr\left( \rho^2\right)$. The witness $\mathcal{W}_\alpha$ can be expressed in terms of the SRE $\mathcal{M}_{\alpha}$ as
\begin{equation}
    \mathcal{W}_{\alpha}\left(\rho \right) = \mathcal{M}_{\alpha}\left(\rho \right) - 2S_2\left(\rho\right),
\end{equation}
and reduces to the SRE, i.e. $\mathcal{W}_{\alpha} = \mathcal{M}_{\alpha}$, for pure states.

\begin{figure}
    \centering
    \begin{overpic}[width=.9\linewidth]{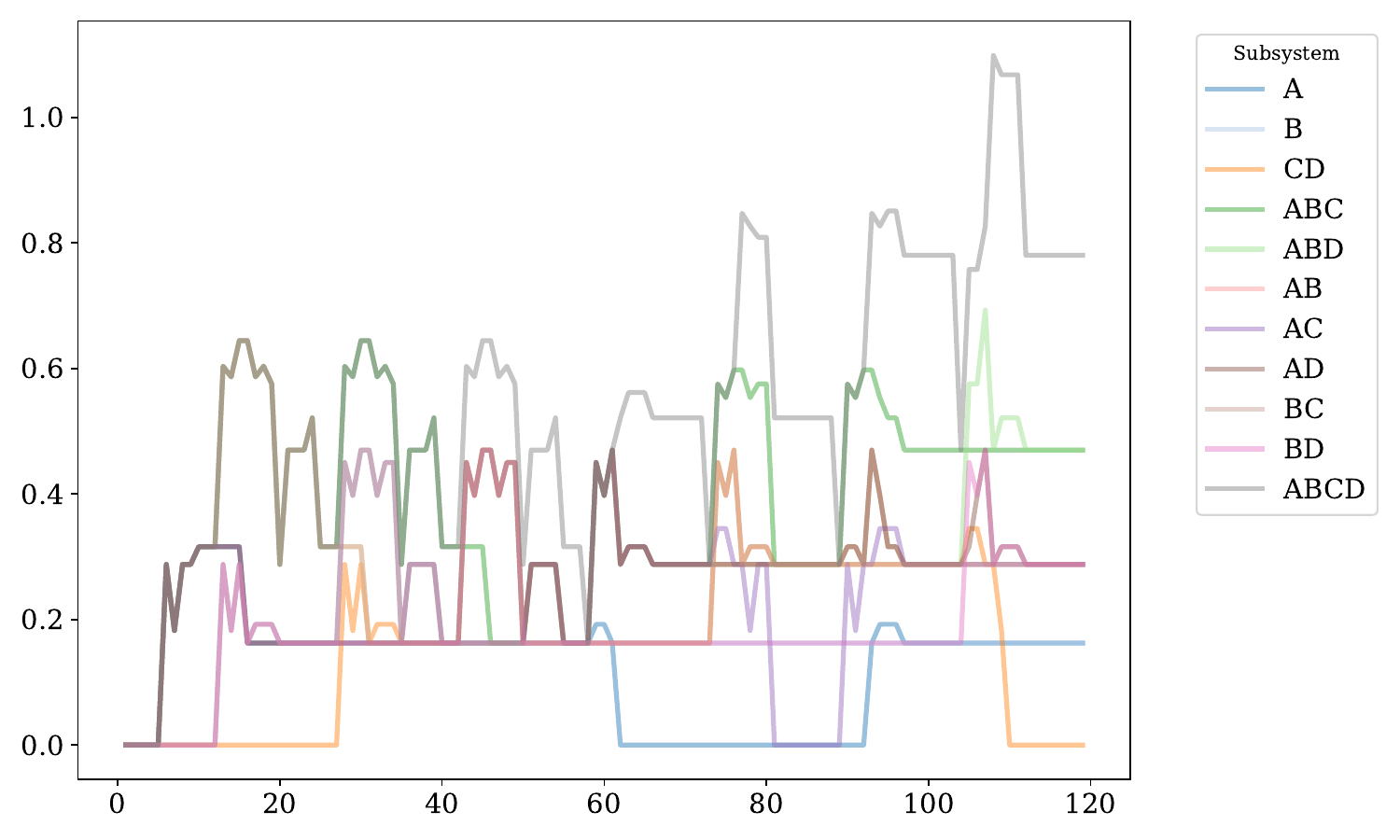}
    \put(34, -2) {Gate Number}
    \put(-5, 28) { $\mathcal{W}_{2}$}
    \put(72.4,0){\dashline{2}(0,3.9)(0,58)}
  \end{overpic}
    \caption{Evolution of the magic witness $\mathcal{W}_2$ across the circuit that prepares $\ket{\psi}_{ABCDR}$ in Eq.~\eqref{PureState}. A vertical dashed line indicates the point of Ingleton violation. As the system evolves toward Ingleton violation, the total magic $\mathcal{W}_2(\rho_{ABCD})$ steadily increases. $\mathcal{W}_2$ fluctuates for different subsystems as $CNOT$ gates redistribute the magic.}
    \label{fig:SREevolution}
\end{figure}

As the system $\rho_{ABCD}$ evolves towards Ingleton violation under the circuit shown in Figure~\ref{fig:lindencircuit}, we evaluate the magic of each subsystem $I \subseteq \rho_{ABCD}$ using the witness $\mathcal{W}_2\left(\rho_{I}\right)$. The evolution of $\mathcal{W}_2\left(\rho_{I}\right)$, after the application of each gate, is presented in Figure~\ref{fig:SREevolution}, for each subsystem $I \subseteq \rho_{ABCD}$ appearing in Ingleton's inequality.
Since magic is invariant%
\footnote{The total magic of a state cannot vary under Clifford unitaries. However, existing magic can be shared from one subsystem to another with entangling operations, e.g. the $CNOT$ gate, without changing the overall magic of the state.} %
under Clifford group action, only $T$ gates can introduce magic into the state and increase the value of $\mathcal{W}_2\left(\rho_{ABCD}\right)$. In contrast, the $CNOT$ gate can only redistribute existing magic, transferring it between different subsystems. As $\rho_{ABCD}$ evolves toward Ingleton violation, the total magic of $\rho_{ABCD}$ increases steadily, though not monotonically. A similar trend is observed in subsystems $ACD$ and $ABD$, the two $3$-party entropies appearing on the right-hand side of Eq.\ \eqref{Ingleton}. Notably, the intervals in Figure~\ref{fig:SREevolution} where the total magic $\mathcal{W}_2\left(\rho_{ABCD}\right)$ remains high, while $\mathcal{W}_2$ fluctuates across smaller subsystems indicates a restructuring of the magic distribution in the state, driven by $CNOT$.

\begin{figure}[h]
    \centering
    \includegraphics[width=0.8\linewidth]{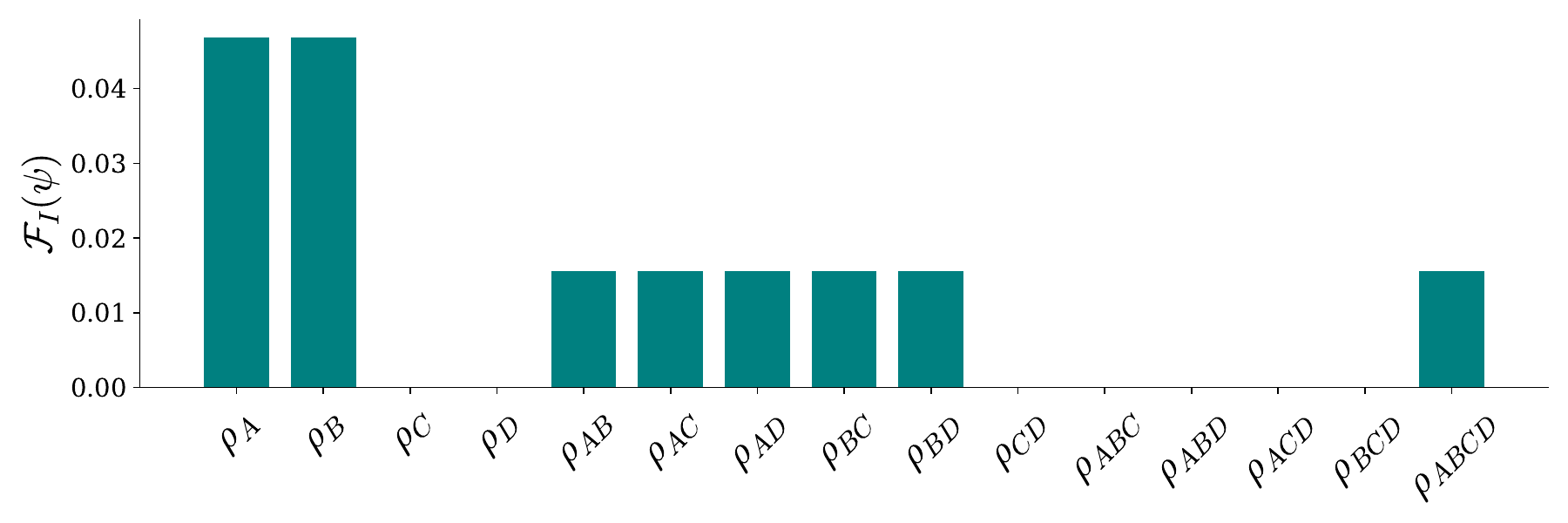}
    \caption{Non-flatness $\mathcal{F}_I$, evaluated at violation, for the bipartitions of $\rho_{ABCD}$ appearing in Ingleton's inequality. The quantity $\mathcal{F}_I$ lower bounds the non-local magic shared across each subsystem. Notably, all $2$-party subsystems on the left-hand side of Eq.\ \eqref{Ingleton} have non-zero $\mathcal{F}_I$, while subsystem $CD$, appearing on the right-hand side, has no non-flatness at violation. This contrast reveals an asymmetric distribution of minimum non-local magic among $2$-party subsystems contributing to Ingleton violation.}
    \label{fig:NonFlatness}
\end{figure}

Beyond simply possessing quantum magic, the form and distribution of this magic significantly influences the the evaluation of Ingleton's inequality. States that violation Ingleton's inequality possess non-local magic $\mathcal{M}^{NL}$, i.e. magic which is embedded in the entanglement structure of the state and cannot be removed by local unitary action. States with non-local magic require a non-trivial combination of both entanglement and magic, making them not only a powerful resource for quantum computation, but also a valuable tool for probing the role of magic in entropy inequalities. As seen in Eq.\ \eqref{eq:NLMagic}, the direct calculation of $\mathcal{M}^{NL}$ is computationally expensive, requiring an optimization over all local unitaries that act on each factor of the bipartite Hilbert space. Accordingly, we instead evaluate the non-flatness of the entanglement spectrum, which provides a lower bound on the non-local magic. For a subsystem $I \subseteq \rho_{ABCD}$, the non-flatness $\mathcal{F}_I$ is defined as 
\begin{equation}\label{eq:anti-flatness}
    \mathcal{F}_I = Tr (\rho_I^{3})- Tr^2(\rho_I^2),
\end{equation}
where $\mathcal{F}_I = 0$ corresponds to a flat spectrum. States with maximal entanglement have, by definition, a flat entanglement spectrum, as do all classically simulable systems such as stabilizer states. A non-zero value of $\mathcal{F}_I$ indicates the presence of non-stabilizerness, and a non-zero amount of non-local magic in the state, shared across the Hilbert space partition $\mathcal{H} = \Hil_I \otimes \Hil_{\overline{I}}$. Figure~\ref{fig:NonFlatness} shows the non-flatness for $\rho_{ABCD}$, across all possible bipartitions of $\Hil$, at failure of Ingleton's inequality.
We observe from Figure~\ref{fig:NonFlatness} that a minimal amount of non-local magic exists in specific subsystems of $\rho_{ABCD}$. Notably, all $2$-party subsystems with entropies appears on the left-hand side of Ingleton's inequality, as in Eq.\ \eqref{Ingleton}, possess some non-zero amount of non-local magic. Subsystem $CD$ on the other hand, the sole $2$-party subsystem appearing on the right-hand side of Eq.\ \eqref{Ingleton}, has no minimum non-local magic at violation. Recall that subsystem $CD$ necessarily possesses the largest amount of entanglement entropy, at violation, among all $2$-party subsystems. This phenomena suggests, as we will substantiate in Section~\ref{StatsSection}, an inverse relationship between entanglement and non-local magic as a state moves towards violation of Ingleton's inequality. 

In this section, we constructed and analyzed a quantum circuit that prepares, from the computational basis, a state which violates Ingleton's inequality. Decomposing this circuit into Hadamard, $T$, and $CNOT$ quantum gates, we analyzed the evolution of the state's entropy vector, quantum magic, and non-local magic as the state transitioned from the interior to the exterior of the Ingleton entropy cone. We highlighted the central role of $2$-party entanglement entropies for Ingleton evaluation, and demonstrated that violation occurred when the entropy of subsystem $CD$, the only $2$-party entropy appearing on the right-hand side of Eq.\ \eqref{Ingleton}, exceeds all other $2$-party entropies, thereby breaking the qubit exchange symmetry that preserves the inequality. Moreover, we showed that Ingleton violation necessitates the existence of non-stabilizerness in the state, as quantified by the magic witness $\mathcal{W}_\alpha$, and tracked the evolution of $\mathcal{W}_\alpha$ in each subsystem the circuit progressed and violation was achieved. Finally, we identified the spectrum of minimal non-local magic for our Ingleton-violating state, lower-bounded by the non-flatness of the entanglement spectrum across all possible bipartitions of the Hilbert space. Collectively, these results offer a correspondence between Ingleton violation and the interplay between entanglement and magic across distinct subsystems.

\subsection{Probing the Edge of the Ingleton Entropy Cone}\label{EdgeOfConeSection}

When studying the transition between entropy cones, regions of particular interest are the the edges, or extremal facets, of a cone. These regions hold particular interest since small perturbations about a state which saturates a facet inequality, i.e. with entropy vector that lives on one of these edges, can easily drive the state to satisfaction or violation of the entropy inequality defining that facet. In this section, we utilize our reinforcement learning algorithm to perturb around states which saturate Ingleton's inequality, generating new violating states with varying degrees of violation. We evaluate the entanglement and magic structure of these states, comparing their resource properties to the Hilbert space distance from a saturating state, and the amount by which the state violates Ingleton's inequality.

We first initialize a $6$-qubit register in state that non-trivially%
\footnote{A product state, with entropy vector consisting of all zero entries, will \textit{trivially} saturate any entropy inequality. We therefore refer to saturation by an entangled state as \textit{non-trivial}.}
saturates Ingleton's inequality. The state $\ket{\psi}_{Satur.}$, defined
\begin{equation}
\label{sat_state}
\begin{split}
\ket{\psi}_{Satur.} &\equiv
\frac{1}{4} \ket{000000} + \frac{1}{4} \ket{000001} + \frac{1}{\sqrt{32}} (1 + i) \ket{001110} - \frac{1}{\sqrt{32}} (1 + i) \ket{001111}\\
& \quad + \frac{1}{4} \ket{010000} + \frac{1}{4} \ket{010001} + \frac{1}{\sqrt{32}} (1 + i) \ket{011110} - \frac{1}{\sqrt{32}} (1 + i) \ket{011111}\\
& \quad + \frac{1}{4} (1 - i) \ket{101010} + \frac{1}{4} (1 + i) \ket{101011}+ \frac{1}{4} (1 - i) \ket{111000}\\
& \quad+ \frac{1}{4} (1 + i) \ket{111001},
\end{split}
\end{equation}
saturates numerous instances of Ingleton's inequality, each generated by a qubit exchange on the instance given in Eq.\ \eqref{Ingleton}. Moreover, the state $\ket{\psi}_{Satur.}$ is one of the states encountered along the circuit in Figure~\ref{fig:lindencircuit}, used to prepare $\rho_{ABCD}$ from Eq.~\eqref{eq:lindenstate}. We apply our reinforcement learning protocol, beginning from $\ket{\psi}_{Satur.}$, and observe the transition out of the Ingleton entropy cone, to different Ingleton-violating states $\ket{\psi_{f_i}}$. For each $\ket{\psi_{f_i}}$ identified using our machine learning algorithm, beginning from $\ket{\psi}_{Satur.}$, we quantify its distinction from the initial violating state $\ket{\psi_{ABCDR}}$ in Eq.\ \eqref{PureState} using the trace distance, shown in Table \ref{distancemeasures}. 
\begin{table}[h]
\centering
\begin{tabularx}{0.9\textwidth}{ 
    >{\centering\arraybackslash}X |
    >{\centering\arraybackslash}X |
    >{\centering\arraybackslash}X }
$\ket{\psi_{f_i}}$ & $\sqrt{1-|\langle\psi_{f_i}|\psi_{ABCDR}\rangle|^2}$ & Ingleton Difference\\
\hline
1  & 0.999 & 0.123\\
2  & 0.992 & 0.062\\
3  & 0.993 & 0.025\\
4  & 0.994 & 0.123\\ 
\caption{Each Ingleton-violating state $\ket{\psi_{f_i}}$ is generated by perturbing around a Ingleton-saturating state using reinforcement learning. The second column gives the trace distance of each $\ket{\psi_{f_i}}$ from the state $\ket{\psi}_{ABCDR}$, from Eq.\ \eqref{PureState}. The amount by which each $\ket{\psi_{f_i}}$ violates Ingleton's inequality is given in the third column.}
\label{distancemeasures}
\end{tabularx}
\end{table}
Each state $\ket{\psi_{f_i}}$ in the table has a trace distance close to $1$ from $\ket{\psi_{ABCDR}}$, indicating that the violating states identified by perturbing near the edge of the Ingleton entropy cone are highly distinguishable from the initial violator $\ket{\psi_{ABCDR}}$. Accordingly, the circuits preparing each $\ket{\psi_{f_i}}$ are likewise distinct from that in Figure~\ref{fig:lindencircuit}, and are shown in Appendix~\ref{MoreIngletonCircuit}.

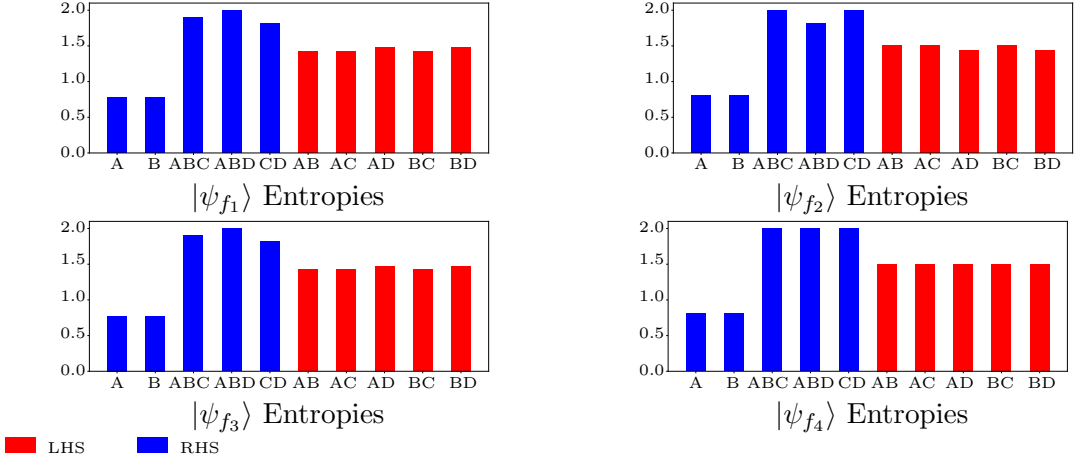
\begin{figure}
\begin{minipage}{\textwidth} 
    \centering
    \begin{subfigure}{.45\textwidth}
    \centering
      \begin{overpic}[width=.8\linewidth, clip=false]{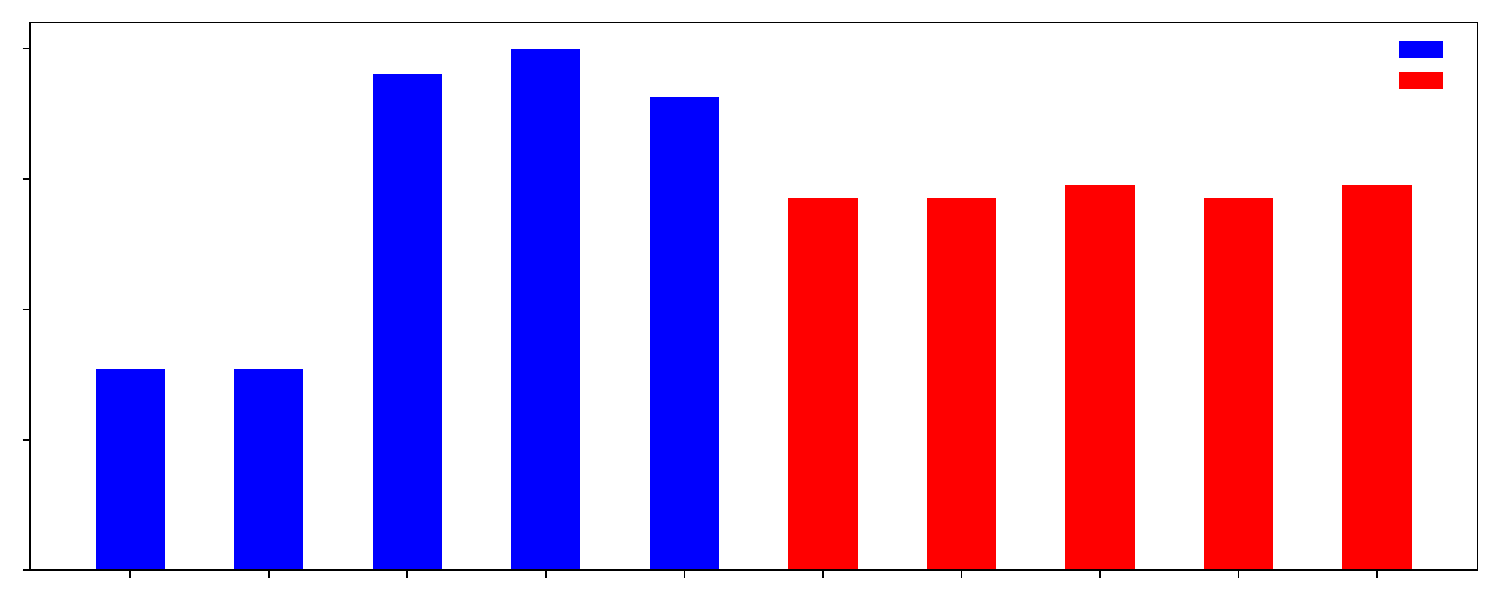}
        \put(7, -2) {\tiny A}
        \put(16, -2) {\tiny B}
        \put(21, -2) {\tiny ABC}
        \put(32, -2) {\tiny ABD}
        \put(43, -2) {\tiny CD}
        \put(51, -2) {\tiny AB}
        \put(60, -2) {\tiny AC}
        \put(69, -2) {\tiny AD}
        \put(79, -2) {\tiny BC}
        \put(89, -2) {\tiny BD}
        \put(-5, 1) {\tiny 0.0}
        \put(-5, 10) {\tiny 0.5}
        \put(-5, 19) {\tiny 1.0}
        \put(-5, 28) {\tiny 1.5}
        \put(-5, 36) {\tiny 2.0}
        \put(93,33.2){\color{white}\rule{2.5mm}{2.7mm}}
      \end{overpic}
      \caption*{$\ket{\psi_{f_1}}$ Entropies}
      \label{fig:ev_sub2}
    \end{subfigure}
\hspace{0.04\textwidth}
    \begin{subfigure}{.45\textwidth}
    \centering
      \begin{overpic}[width=.8\linewidth]{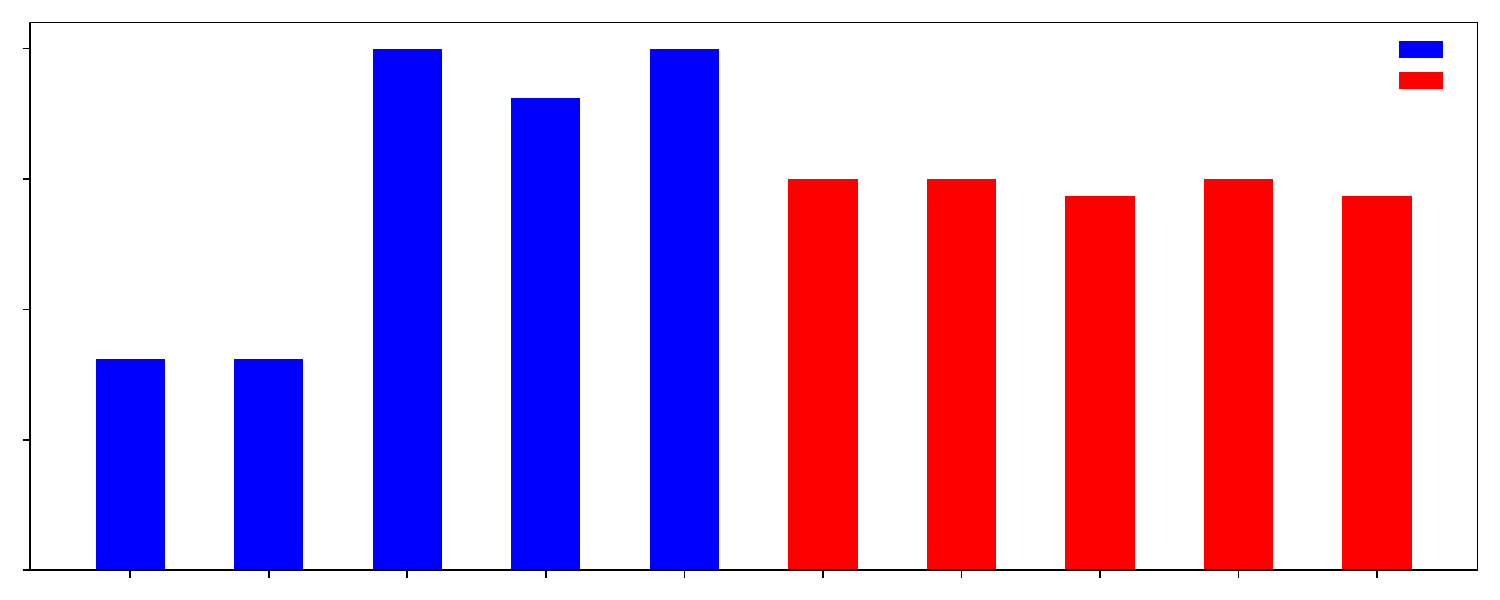}
        \put(7, -2) {\tiny A}
        \put(16, -2) {\tiny B}
        \put(21, -2) {\tiny ABC}
        \put(32, -2) {\tiny ABD}
        \put(43, -2) {\tiny CD}
        \put(51, -2) {\tiny AB}
        \put(60, -2) {\tiny AC}
        \put(69, -2) {\tiny AD}
        \put(79, -2) {\tiny BC}
        \put(89, -2) {\tiny BD}
        \put(-5, 1) {\tiny 0.0}
        \put(-5, 10) {\tiny 0.5}
        \put(-5, 19) {\tiny 1.0}
        \put(-5, 28) {\tiny 1.5}
        \put(-5, 36) {\tiny 2.0}
        \put(93,33.2){\color{white}\rule{2.5mm}{2.7mm}}
      \end{overpic}
      \caption*{$\ket{\psi_{f_2}}$ Entropies}
      \label{fig:ev_sub3}
    \end{subfigure}%
\end{minipage}

\begin{minipage}{\textwidth}
    \centering
    \begin{subfigure}{.45\textwidth}
    \centering
      \begin{overpic}[width=.8\linewidth]{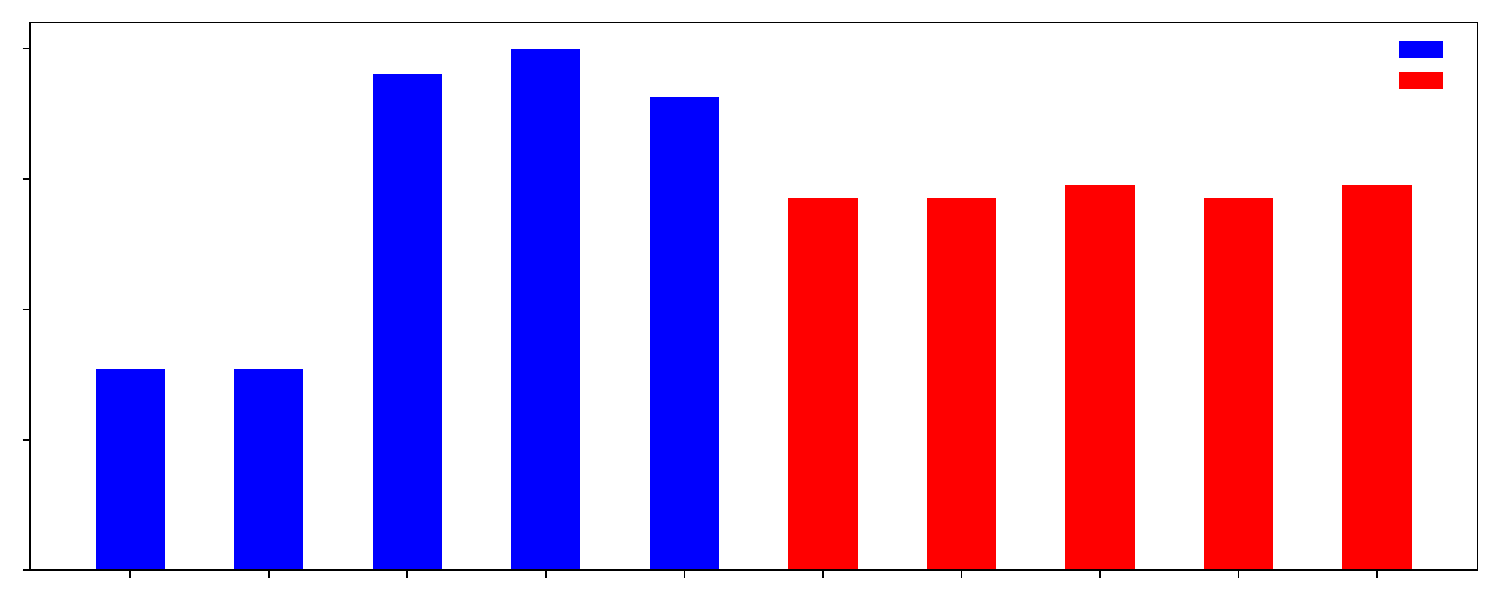}
        \put(7, -2) {\tiny A}
        \put(16, -2) {\tiny B}
        \put(21, -2) {\tiny ABC}
        \put(32, -2) {\tiny ABD}
        \put(43, -2) {\tiny CD}
        \put(51, -2) {\tiny AB}
        \put(60, -2) {\tiny AC}
        \put(69, -2) {\tiny AD}
        \put(79, -2) {\tiny BC}
        \put(89, -2) {\tiny BD}
        \put(-5, 1) {\tiny 0.0}
        \put(-5, 10) {\tiny 0.5}
        \put(-5, 19) {\tiny 1.0}
        \put(-5, 28) {\tiny 1.5}
        \put(-5, 36) {\tiny 2.0}
        \put(93,33.2){\color{white}\rule{2.5mm}{2.7mm}}
      \end{overpic}
      \caption*{$\ket{\psi_{f_3}}$ Entropies}
      \label{fig:ev_sub4}
    \end{subfigure}
\hspace{0.04\textwidth}
    \begin{subfigure}{.45\textwidth}
    \centering
      {\begin{overpic}[width=.8\linewidth]{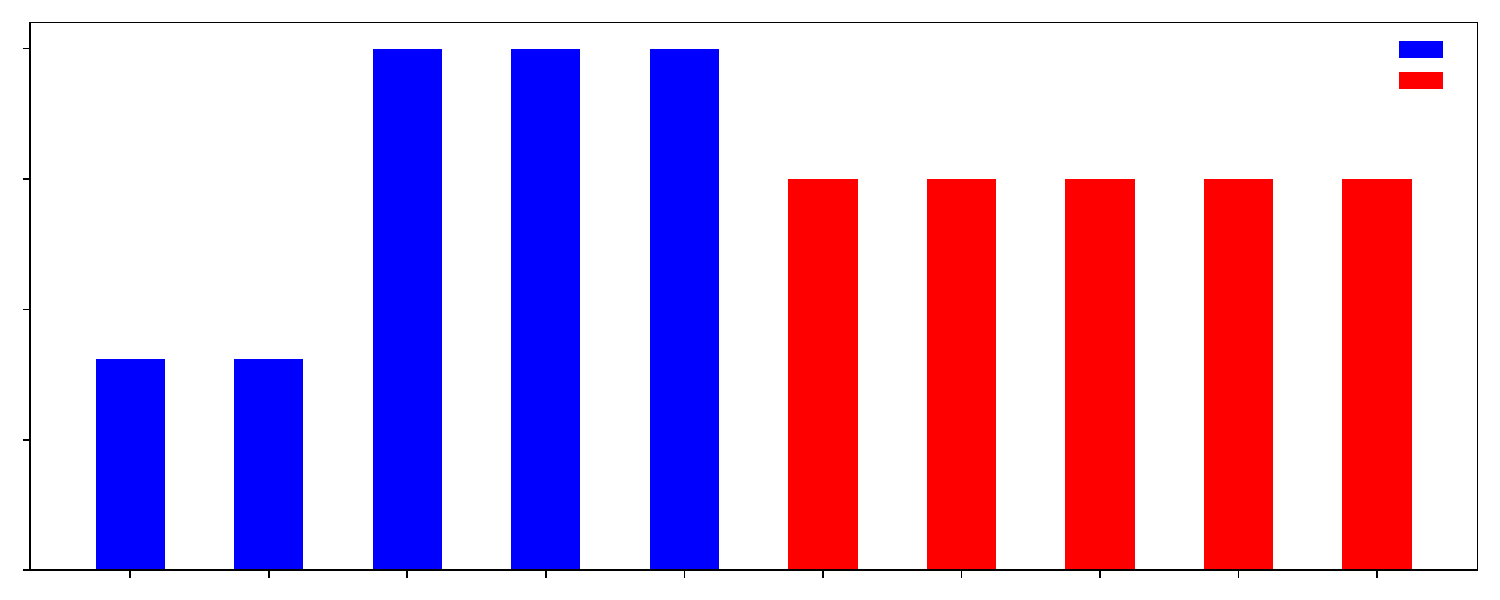}
          \put(7, -2) {\tiny A}
          \put(16, -2) {\tiny B}
          \put(21, -2) {\tiny \rotatebox{0}{ABC}}
          \put(32, -2) {\tiny ABD}
          \put(43, -2) {\tiny CD}
          \put(51, -2) {\tiny AB}
          \put(60, -2) {\tiny AC}
          \put(69, -2) {\tiny AD}
          \put(79, -2) {\tiny BC}
          \put(89, -2) {\tiny BD}
    
          \put(-5, 1)  {\tiny 0.0}
          \put(-5, 10) {\tiny 0.5}
          \put(-5, 19) {\tiny 1.0}
          \put(-5, 28) {\tiny 1.5}
          \put(-5, 36) {\tiny 2.0}
    
          \put(93,33.2){\color{white}\rule{2.5mm}{2.7mm}}   
      \end{overpic}}
      \label{fig:sfig5}
      \caption*{$\ket{\psi_{f_4}}$ Entropies}
    \end{subfigure}
\end{minipage}

{\tiny
    \color{red}\rule{4mm}{2mm}\; \color{black}LHS
    \hspace{2em}
    \color{blue}\rule{4mm}{2mm}\; \color{black}RHS
}
\caption{Entanglement entropies, for terms appearing in Ingleton's inequality, of states $\ket{\psi_{f_i}}$ from Table~\ref{distancemeasures}. Similar features, e.g. $S_{CD}$ the largest $2$-party entropy, are present, while other variations to the entanglement profile at violation are also observed.}
\label{fig:ev_perturbingstates}
\end{figure}

For each $\ket{\psi_{f_i}}$ in Table~\ref{distancemeasures}, we likewise detail the entropy vector. Figure~\ref{fig:ev_perturbingstates} gives the subsystem entropies involved in the evaluation of Ingleton's inequality in Eq.\ \eqref{Ingleton}, for each $\ket{\psi_{f_i}}$ identified by the agent.
Although each state $\ket{\psi_{f_i}}$ is highly distinguishable from $\ket{\psi_{ABCDR}}$, we nevertheless observe similar features in their entanglement entropy spectrum. As shown in Figure~\ref{fig:individualEntropies}, and necessarily required for Ingleton violation, the entropy of $S_{CD}$ exceeds all other $2$-party entanglement entropies. Interestingly, the entropy vector of $\ket{\psi_{f_i}}$ is identical to that of $\ket{\psi_{ABCDR}}$ on the subsystems appearing in Eq.\ \eqref{Ingleton}, despite the states being separated by a trace distance of $0.994$. In contrast to the entanglement profile of $\ket{\psi_{ABCDR}}$, states $\ket{\psi_{f_1}}$ and $\ket{\psi_{f_3}}$ have $3$-party entanglement entropies $S_{ABC}$ and  $S_{ABD}$ larger than $S_{CD}$, while $\ket{\psi_{f_2}}$ has $S_{ABC}$ and  $S_{ABD}$ less than $S_{CD}$. These variations demonstrate that, while $S_{CD}$ must be greater than all other $2$-party entropies to produce the symmetry breaking that leads to Ingleton violation, it need not equal the $3$-party entropies.

\begin{figure}
    \centering 
    \begin{subfigure}{0.49\textwidth}
        \centering
        \begin{overpic}[width=\linewidth]{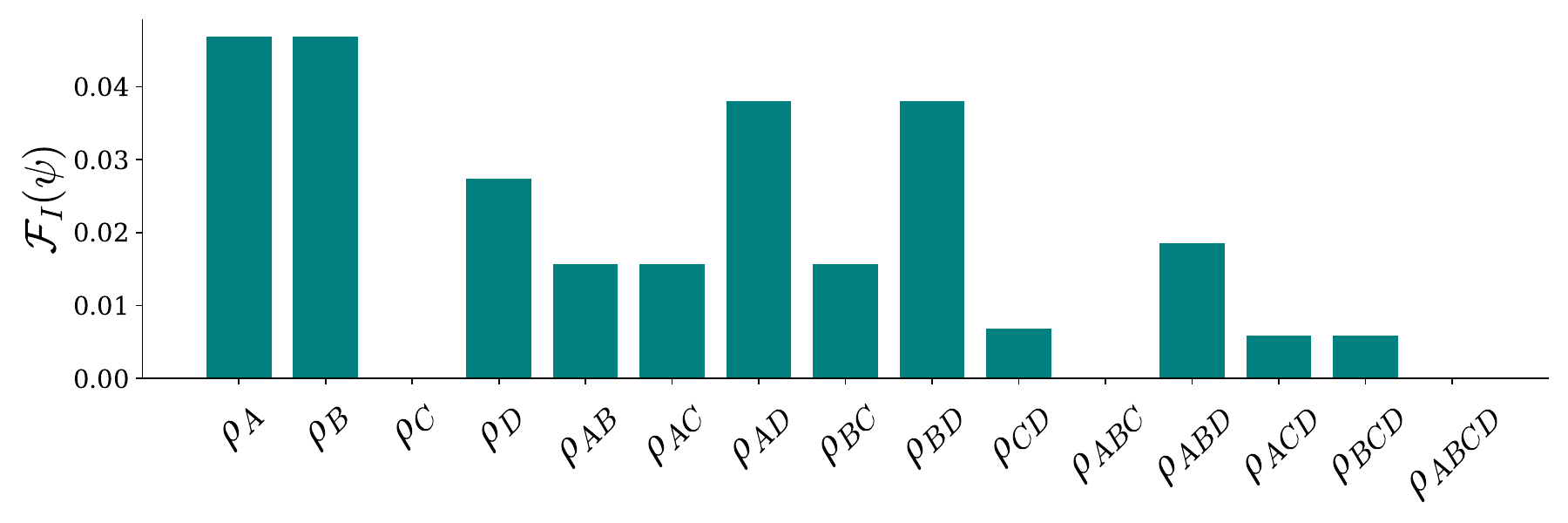}
            \put (7,-1.8) {}
        \end{overpic}
        \caption*{Spectrum non-flatness of $\ket{\psi_{f_1}}$}
    \end{subfigure}
    \begin{subfigure}{0.48\textwidth}
        \centering
        \begin{overpic}[width=\linewidth]{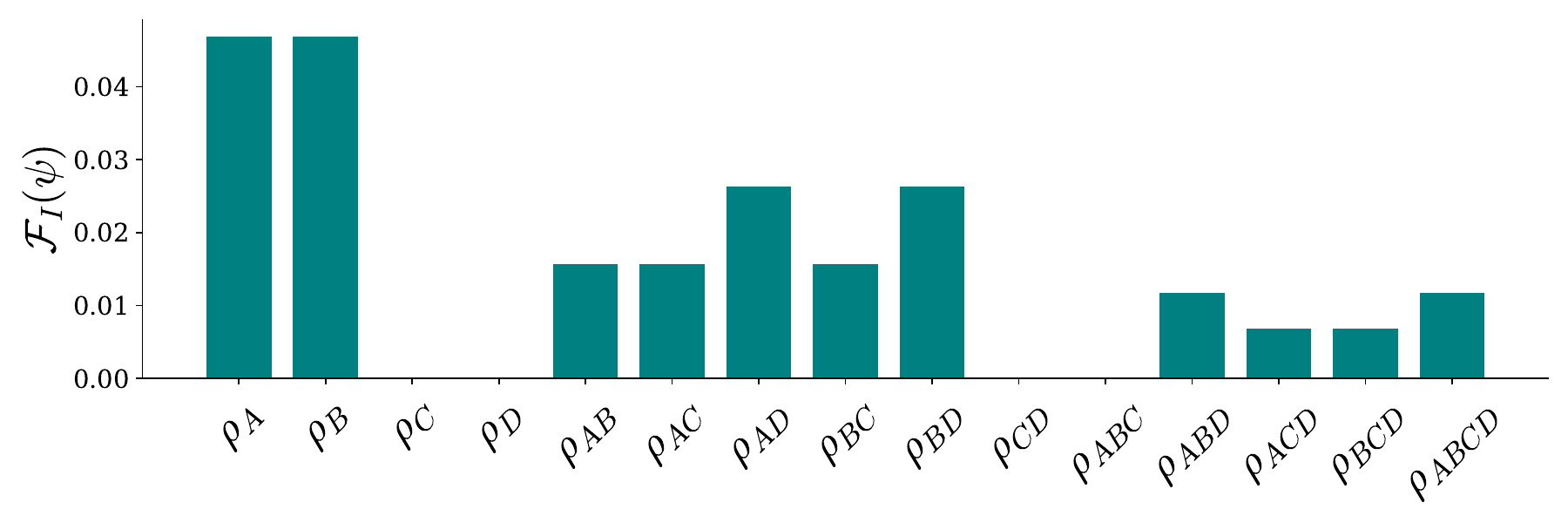}
            \put (7,-1.8) {}
        \end{overpic}
        \caption*{Spectrum non-flatness of $\ket{\psi_{f_2}}$}
    \end{subfigure}
    \\[1em]
    \begin{subfigure}{0.48\textwidth}
        \centering
        \begin{overpic}[width=\linewidth]{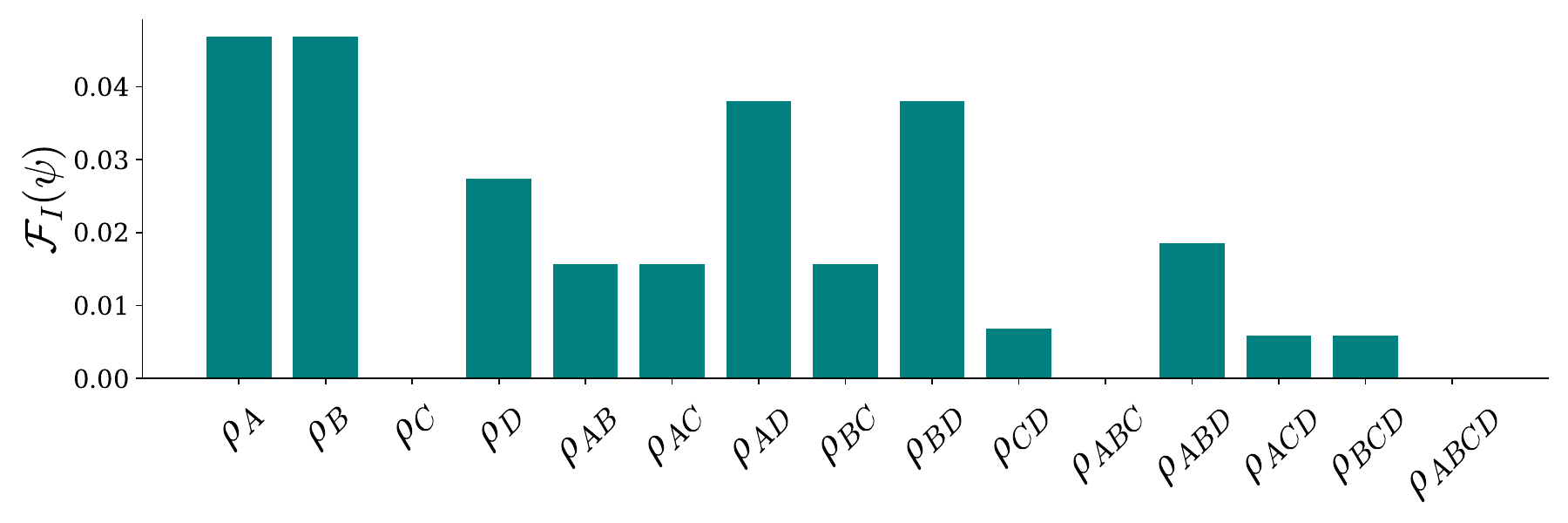}
            \put (7,-1.8) {}
        \end{overpic}
        \caption*{Spectrum non-flatness of $\ket{\psi_{f_3}}$}
    \end{subfigure}
    \begin{subfigure}{0.48\textwidth}
        \centering
        \begin{overpic}[width=\linewidth]{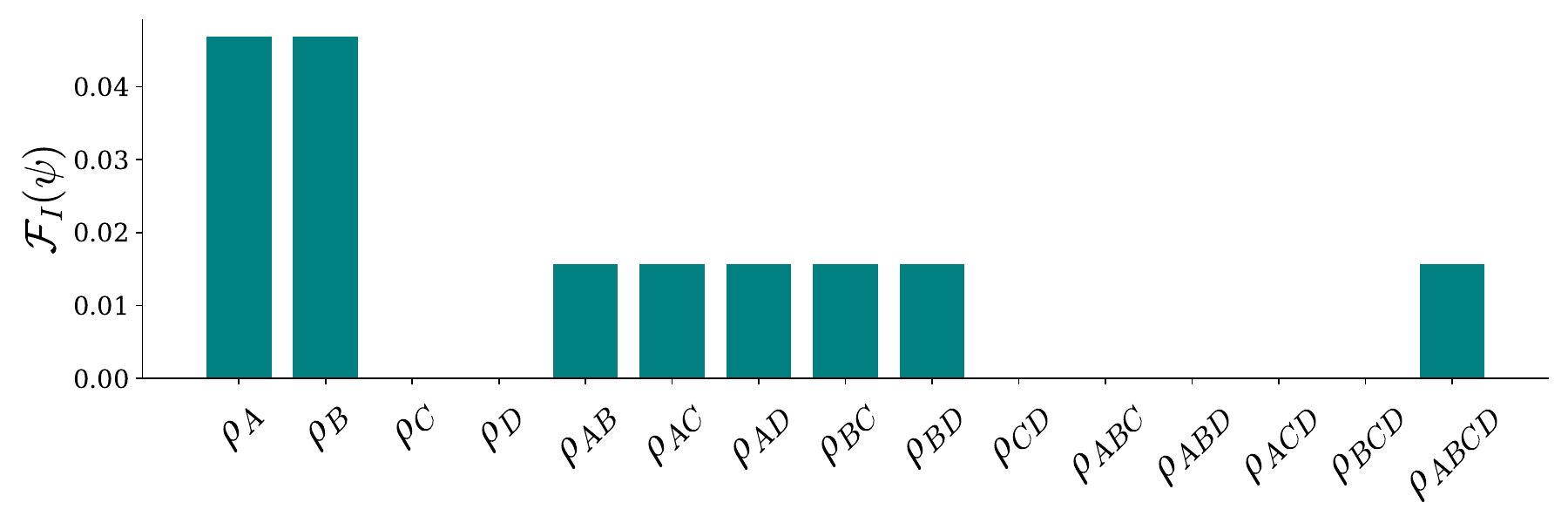}
            \put (7,-1.8) {}
        \end{overpic}
        \caption*{Spectrum non-flatness of $\ket{\psi_{f_4}}$}
    \end{subfigure}
    \caption{Entanglement spectrum non-flatness $\mathcal{F}_I$ for states $\ket{\psi_{f_i}}$. A non-zero value of $\mathcal{F}_I$ appears for $\rho_{CD}$ and $\rho_{D}$ in $\ket{\psi_{f_1}}$ and $\ket{\psi_{f_3}}$ indicating the presence of non-local magic. Additionally, states $\ket{\psi_{f_1}},\ \ket{\psi_{f_2}},$ and $\ket{\psi_{f_3}}$ witness the emergence of non-local magic in each $3$-party subsystem.}
    \label{fig:flatness_perturb}
\end{figure}

For each state $\ket{\psi_{f_i}}$ we additionally compute the non-flatness of the entanglement spectrum, shown in Figure~\ref{fig:flatness_perturb}, across each Hilbert space bipartition appearing Ingleton's inequality. For a subsystem $A \subseteq \ket{\psi_{f_i}}$, we calculate the non-flatness $\mathcal{F}_I$ using Eq.\ \eqref{eq:anti-flatness}, which serves as a lower-bound for the non-local magic shared across that partition.
Figure~\ref{fig:flatness_perturb} reveals variation in the entanglement spectrum flatness across the states $\ket{\psi_{f_i}}$. For states $\ket{\psi_{f_1}}$ and $\ket{\psi_{f_3}}$ the values $\mathcal{F}\left(\rho_{CD}\right)$ and $\mathcal{F}\left(\rho_{D}\right)$ are non-zero, indicating a non-zero amount of non-local magic shared between these subsystems and their complement regions. As in the previous analysis, subsystem $CD$ consistently exhibits the smallest non-flatness value of any $2$-party subsystem. Moreover, the states $\ket{\psi_{f_1}},\ \ket{\psi_{f_2}},$ and $\ket{\psi_{f_3}}$ possess non-local magic in each $3$-party subsystem. States $\ket{\psi_{f_1}},\ \ket{\psi_{f_2}},$ and $\ket{\psi_{f_3}}$ are also those for which $S_{CD}$ does not equal $S_{ABC}$ and $S_{ABD}$, highlighting how deviations in the entanglement structure yield an emergence of non-local magic. 

In this section we employed our reinforcement-learning framework to perturb around a state $\ket{\psi}_{Satur.}$ that non-trivially saturates multiple instances of Ingleton's inequality, placing its entropy vector on an edge of the Ingleton entropy cone. Starting from $\ket{\psi}_{Satur.}$, the reinforcement agent identified $4$ new Ingleton-violating states, each distinct from $\ket{ABCDR}$ in Section~\ref{ResourceEvolutionSection} and separated from $\ket{ABCDR}$ by substantial Hilbert space distance. We analyzed the entanglement entropy and minimum non-local magic profiles for these newly discovered states, and revealed shared features, such as the $2$-party dominance of $S_{CD}$, along with notable differences in how $S_{CD}$ relates to $3$-party entropies and how  non-local magic emerges across various subsystems. These results demonstrate that controlled perturbations near extremal facets of the Ingleton entropy cone can generate a broad and varied family of violating states, a concept we rigorously develop in Section~\ref{RaritySection}, with diverse quantum resource signatures at violation. In the next section, we introduce an optimization scheme for generating arbitrary sets of Ingleton-violating states, and perform a comprehensive statistical analysis of their resource profiles as well as the rarity of Ingleton violation in the Hilbert space.

\section{Generation and Characterization of Ingleton-Violating States}\label{StatsSection}

Ingleton's inequality constrains numerous entropy cones, including the stabilizer and holographic cones~\cite{Linden2013,Bao2015,HernandezCuenca2019,Bao2020}, and also characterizes certain states with complex entanglement structures and non-stabilizer resources. Accordingly, identifying Ingleton-violating states is a problem of both theoretical and practical interest, with applications ranging from an improved understanding of entropy cone nesting to benchmarking quantum advantage and classical hardness~\cite{Pollack:2024bnj}. However, as we will demonstrate, Ingleton-violating states are extremely rare in the Hilbert space. Addressing this challenge, we introduce a computational framework that systematically explores the optimization landscape of Ingleton's inequality and generates arbitrary numbers of Ingleton-violating states, with customizable violation magnitudes. 

Our computational protocol formulates the problem of Ingleton violation as a classical optimization over the Ingleton gap, defined simply as the left-hand side of Eq.\ \eqref{Ingleton} minus the right-hand side, which we minimize using two complementary optimization algorithms. Using our results, we infer a theoretical upper-bound for violations of Ingleton's inequality, and analyze the evolution of entanglement and magic as maximal violation is approached. We conduct a comprehensive statistical analysis to examine the stability of Ingleton violation, as well as the rarity of Ingleton-violating states, constructing a detailed quantum resource profile for Ingleton-violating states. This framework not only enables the controlled production of states which violate Ingleton's inequality, but provides an understanding of the quantum resource structure that leads to violation.

\subsection{An Optimization Protocol for Generating Ingleton-Violating States}\label{OptimizationSection}

States that violate Ingleton's inequality are exceptionally rare, with only a few explicit examples~\cite{Linden2013,Fong2008Ingleton} reported prior to this work. In this section, we introduce a computational framework for generating arbitrary numbers of Ingleton-violating states. We show how the search for Ingleton violators can be formulated as a classical optimization problem, and efficiently implemented using gradient-free evolutionary algorithms. While this protocol is initially designed to identify states that maximally violate Ingleton's inequality, the techniques can be readily adapted to produce states which violate Ingleton's, or any other inequality, by a controlled, arbitrary amount.

Ingleton's inequality, as defined in Eq.\ \eqref{Ingleton}, is a $4$-party entropy inequality, though parties can be relabeled following qubit exchange and each party can itself consist of multiple qubits. Therefore, when evaluating Ingleton's inequality on an $n$-party state, with $n \geq 4$, we must consider all instances of the inequality arising from subsystem relabeling, e.g. $A \leftrightarrow C$ which produces the instance
\begin{equation}
   S_{AB} + S_{AC} + S_{BC} + S_{BD} + S_{CD}  \geq S_{B} + S_{C} + S_{ABC} + S_{BCD} + S_{AD}.
\end{equation}
Furthermore, for $n > 4$ qubits, we must consider all \textit{lifts} of Eq.\ \eqref{Ingleton} to higher qubit parties, e.g. $A \rightarrow AE$ which gives the instance 
\begin{equation}
   S_{ABE} + S_{ACE} + S_{ADE} + S_{BC} + S_{BD} \geq S_{AE} + S_{B} + S_{ABCE} + S_{ABDE} + S_{CD}.
\end{equation}
The set of all permuted and lifted forms of Eq.\ \eqref{Ingleton} defines the complete set $\mathcal{I}$ of Ingleton inequalities for an $n$-qubit state, with cardinalities shown in Table \ref{tab:symmetry_group-caridnality}.
\begin{table}[h]
    \centering
    \begin{tabular}{|c|c c c c c c|}
        \hline
         $\mathbf{n}$ & 4 & 5 & 6 & 7 & 8 & $\ldots$\\ 
         \hline
         $\mathbf{|\mathcal{I}|}$ & 3 & 90 & 780 & 6090 &  39270 & $\ldots$\\
         \hline
    \end{tabular}
    \caption{Cardinality $|\mathcal{I}|$ of the full set of Ingleton inequalities, including instances generated by qubit exchange and higher qubit lifts, for an $n$-qubit system.}
    \label{tab:symmetry_group-caridnality}
\end{table}
An Ingleton-violating state is considered to be a state which fails any instance of Ingleton's inequality represented in Table~\ref{tab:symmetry_group-caridnality}.

We formulate our search for states that violate Ingleton's inequality as an optimization problem over the Ingleton ``gap'', denoted $g_I$ for a specific instance $I$ of Ingleton's inequality and defined%
\footnote{The Ingleton gap is simply the additive inverse of the Ingleton difference in Eq.\ \eqref{IngletonDifference}, however, we introduce this new terminology to simplify the language distinguishing the two quantities.} %
simply as the left-hand side of the inequality minus the right-hand side. For example, given the canonical statement of Ingleton's inequality $I_{0}$ in Eq.\ \eqref{IngletonDifference}, the Ingleton gap for a state $\psi$ is calculated
\begin{equation}\label{IngletonGap}
   g_{I_{0}}\left(\psi \right) = \left(S_{AB} + S_{AC} + S_{AD} + S_{BC} + S_{BD}\right) - \left(S_{A} + S_{B} + S_{ABC} + S_{ABD} + S_{CD}\right).
\end{equation}
We design our optimizer to minimize the cost $\mathcal{C}$, defined as the minimum Ingleton gap evaluated over all symmetry orbits and lifts in $\mathcal{I}$, explicitly
\begin{equation}\label{CostFunction}
   \mathcal{C} \equiv \min_{I\in \mathcal{I}} \{g_I\},
\end{equation}
with $g_I$ computed for all states encountered in the evolution. The cost function in Eq.\ \eqref{CostFunction} enables an exhaustive search over all instances of Ingleton's inequality, identifying states which maximally violate the inequality. However, given that the cardinality of $\mathcal{I}$ grows combinatorically, as shown in Table~\ref{tab:symmetry_group-caridnality}, this method is not efficient at large scales. Fortunately, since Ingleton's inequality is a $4$-party entropy inequality, we maximally need $4$ ancillary qubits to purify any Ingleton-violating $4$-qubit mixed state. Following Theorem~\ref{MinQubitTheorem}, we therefore apply our optimization protocol beginning at $6$ qubits, and restrict our search to pure systems of $6,\ 7,$ and $8$ qubits. Moreover, if the goal is to identify a state which violates a specific instance of Ingleton's inequality, our optimization protocol can be performed over solely the specified instance, significantly reducing the computational overhead and enabling an additional level of customization to the search. 

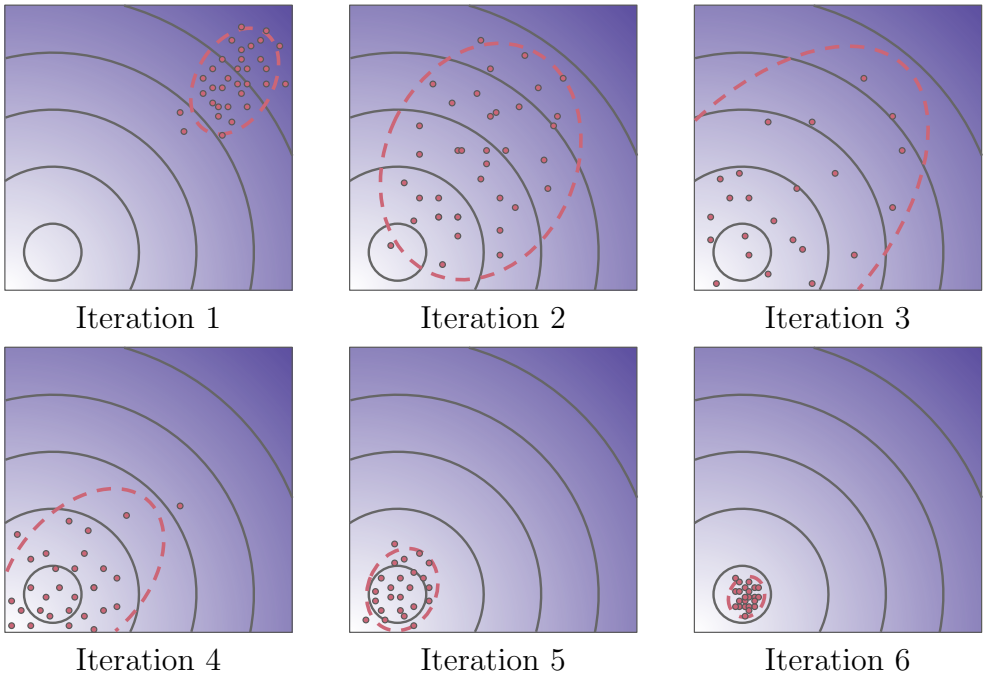
\begin{figure}[h]
    \centering
    \begin{overpic}[width=\linewidth, grid=False]{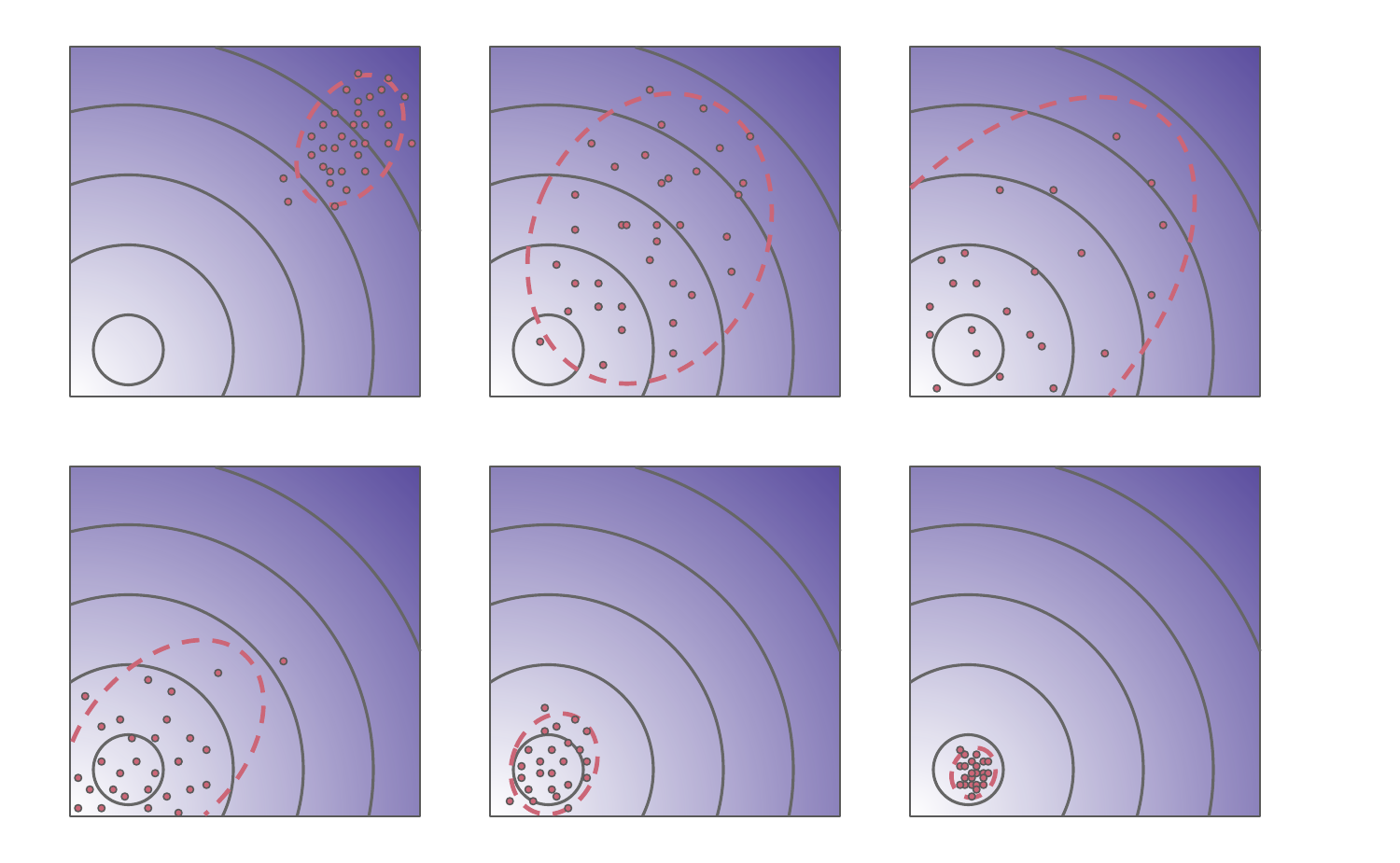}
        \put (17.5,30) {\makebox[0pt]{\Centerstack{Iteration 1}}}
        \put (47.5,30) {\makebox[0pt]{\Centerstack{Iteration 2}}}
        \put (77.5,30) {\makebox[0pt]{\Centerstack{Iteration 3}}}
        \put (17.5,0) {\makebox[0pt]{\Centerstack{Iteration 4}}}
        \put (47.5,0) {\makebox[0pt]{\Centerstack{Iteration 5}}}
        \put (77.5,0) {\makebox[0pt]{\Centerstack{Iteration 6}}}
    \end{overpic}
    \caption{Qualitative description of CMA-Evolution Strategy (CMA-ES). The algorithm reaches the optimum solution by evaluating cost at sampled points (shown in red) from a multi-variate normal distribution (dashed ellipse). The mean and standard deviation of the distribution are shifted towards the points where evaluation yields a lower cost, until the optimizer converges to a minima.}
    \label{fig:CMAES-evol}
\end{figure}

We next select an appropriate optimization algorithm to efficiently explore this high-dimensional landscape, a choice which we demonstrate reveals important insights about the structure and distribution of Ingleton-violating states in the Hilbert space. We find that conventional gradient-based approaches that rely on computing the Jacobian or the Hessian, e.g. gradient descent, the Broyden-Fletcher-Goldfarb-Shanno (BFGS) algorithm, and Powell's method, either fail to find a global minima for $\mathcal{C}$, or exhibit far slower convergence than non-gradient based methods. This behavior suggests that the optimization landscape associated with Ingleton's inequality is highly irregular and non-convex \cite{Boston2020ViolationsOT}. Accordingly, we adopt two gradient-free optimization strategies, the Covariance Matrix Adaptation Evolution Strategy (CMA-ES) \cite{Hansen:2001} and the Constrained Optimization BY Linear Approximation (COBYLA) method \cite{Powell:1994xno}. In the CMA-ES approach, each new candidate point is sampled from a multivariate normal distribution, and the cost function is evaluated at each point. After each iteration, the mean and variance of the distribution are updated according to the cost function evaluations, until the algorithm eventually converges to a solution. A qualitative illustration of the CMA-ES process is shown in Figure \ref{fig:CMAES-evol}. 

The COBYLA method alternatively operates by building local linear approximations to the objective and constraint functions, and performing trust-region steps to improve the candidate solution until a convergence criterion is satisfied. It is naturally suited to highly irregular or non-analytic landscapes because it does not require derivatives of either the objective or the constraint functions \cite{MALAN2013148}. Both CMA-ES and COBYLA can accommodate equality and inequality constraints on variables, which for our search corresponds to a normalization requirement for input vectors. CMA-ES conducts a global exploration, is thereby more robust to noise and multimodality, and typically requires more evaluations; whereas COBYLA converges more quickly in smoother, lower-dimensional settings, but is more sensitive to local minima and geometric constraint. These complementary strengths make derivative-free optimization methods an appealing alternative to reinforcement learning based black-box strategies for complex optimization tasks. In this work, we independently make use of both COBYLA and CMA-ES, depending on the situation. The former is employed to search over states in the Hilbert space and to track the evolution of resources at each iteration, while the latter is used specifically to identify violators and to substantiate the results related to the maximum violation, which we discuss in the next subsection.

The computational generation of Ingleton-violating states begins by applying the map $f: \mathcal{H}^{2^n} \rightarrow \mathbb{R}^{2^{n+1}}$, which takes every pure state $\ket{\psi} \in \mathcal{H}^{2^n}$ to a normalized real vector $x \in \mathbb{R}^{2^{n+1}}$. These normalized vectors serve as inputs to the CMA-ES and COBYLA optimizers. Notably, the map $f$ preserves global phase distinction, i.e. the states $\ket{\psi}$ and $e^{\iota\phi}\ket{\psi}$ in $\mathcal{H}^{2^n}$ are mapped to distinct vectors in $\mathbb{R}^{2^{n+1}}$. An initial vector $x$ is randomly generated and passed to the CMA-ES or COBYLA optimizer, which iteratively explores the search space until the cost function $\mathcal{C}(x)$ converges. The state $x^*$ that maximally violates Ingleton's inequality is defined as
\begin{equation}\label{ConvergenceCriteria}
    x^\ast \equiv \argmin_{x \in \mathbb{R}^{2^{n+1}}} \{\mathcal{C}(x) \},
\end{equation}
with $\mathcal{C}(x)$ given as in Eq.\ \eqref{CostFunction}. Once identified, the state $x^*$ is mapped back to $\mathcal{H}^{2^n}$ using the inverse function $f^{-1}$, and both the quantum state and violation magnitude are returned to the user. While Eq.\ \eqref{ConvergenceCriteria} ensures convergence of the optimizer to a maximally-violating state, this criterion can be easily adjusted, either by introducing an additive offset to $\mathcal{C}(x)$ or by terminating the search once a target violation is reached, to generate Ingleton violations of arbitrary magnitude.

We developed a computational protocol to generate Ingleton-violating states by framing the problem as an optimization over the Ingleton gap, defined as the difference between the left-hand and right-hand side of Eq.\ \eqref{Ingleton}. Our approach considered violations of any Ingleton instance, including those arising from qubit exchange and lifts to higher-qubit systems. We utilized two gradient-free evolutionary optimizers, namely CMA-ES and COBYLA, since gradient-based schemes proved largely ineffective, likely due to the highly irregular and non-convex landscape of the cost function. Mapping states $\ket{\psi} \in \mathcal{H}^{2^n}$ to normalized vectors $x \in \mathbb{R}^{2^{n+1}}$, each optimizer iteratively minimizes the cost function, and identifies states which maximally violate Ingleton's inequality, returning the final state and violation amount to the user. In the following sections, we generate ensembles of Ingleton-violating states, study the evolution of quantum resources as states approach maximal violation, and perform a statistical analysis of the distribution, rarity, and relative positions of Ingleton-violating states in the Hilbert space.

\subsection{Maximal Violation Convergence and Quantum Resource Evolution}\label{ConvergenceSubsection}

It is straightforward to verify that the cost function $\mathcal{C}$ in Eq.\ \eqref{CostFunction} is globally non-convex, therefore convergence to the true global minimum is not guaranteed. This feature motivates the natural questions: to what extent can we verify that an optimizer converges to the global minimum, and how reliably can we estimate the theoretical maximum violation of Ingleton's inequality? In this section, we address both questions by developing an empirical method to probe the structure of the solution landscape, providing evidence of near-global optimality and establishing a practical bound for maximum Ingleton violation. We further analyze the evolution of entanglement entropy and quantum magic as this maximum violation is approached, and discuss the physical implications of a quantum resource structure that yields Ingleton violation.

In convex optimization, any well-suited algorithm will converge to the same global minimum regardless of the initialization point $x_0$. However, for non-convex problems the optimization landscape typically contains multiple local minima, and thereby the solution will strongly depend on the choice of $x_0$ \cite{Barber_Ha_2018}. Formally, given a function $g: \mathbb{R}^{2^{n+1}} \rightarrow \mathbb{R}$, an optimization algorithm $\mathcal{A}$ generates a sequence $\{x_k\}_{k \in \mathbb{N}}$ according to the recursion relation
\begin{equation}\label{Sequence}
    x_{k+1} = \mathcal{A}(x_k,g).
\end{equation}
Given Eq.\ \eqref{Sequence}, the final (optimal) solution $x^*$ is then defined as the limit of this sequence, explicitly
\begin{equation}\label{SequenceSolution}
    x^\ast(x_0) = \lim_{k\rightarrow\infty}x_k,
\end{equation}
where $x_0$ again denotes the initial input state. In practice, the sequence in Eq.\ \eqref{Sequence} truncates after some maximum number of iterations have been performed, typically when the cost function plateaus to the optimal value.

\begin{figure}[b]
    \centering
    \includegraphics[width=\linewidth]{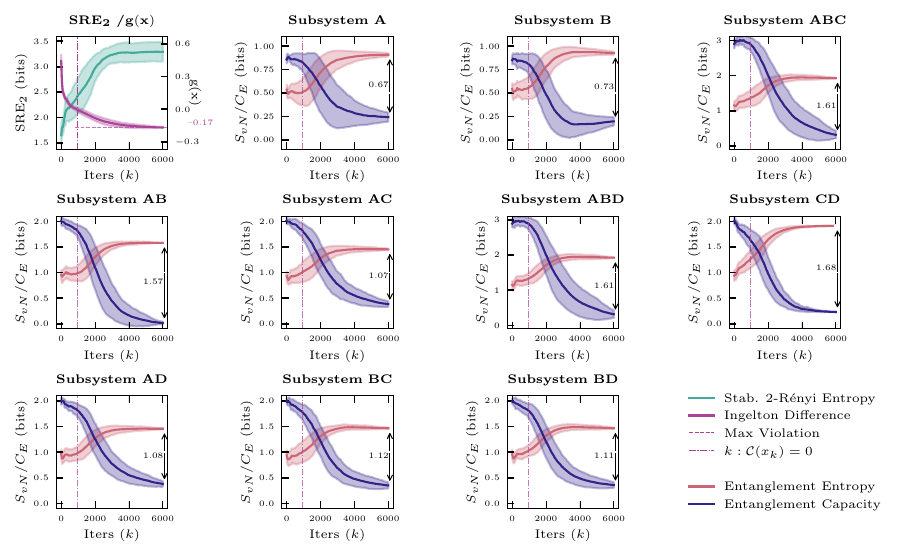}
    \caption{Evolution of entanglement and magic in $6$-qubit state as system approaches maximum violation of Ingleton's inequality. Entanglement entropy $S_{vN}$ (red curve) increases monotonically towards violation, while entanglement capacity $C_E$ (blue curve) sharply drops during optimization. Total magic (green curve) rises and saturates. The dashed vertical line marks the point of Ingleton saturation. The shaded region gives the standard deviation from sampling a population of solutions with different $x_0$.}
    \label{fig:cobyla_convergence}
\end{figure}

We use Eqs.\ \eqref{Sequence} and \eqref{SequenceSolution} to infer both the existence and the value of a maximum violation for the Ingleton's instance given by Eq.\ \eqref{Ingleton}. We repeat the optimization protocol detailed in Section~\ref{OptimizationSection} approximately $10^4$ times, to better ensure that the determined maximum violation is not merely a local minimum. For each repetition we initialize the optimizer with a new state $x_0$, sampled uniformly from the unit hyper-sphere $S^{2^{n+1}}$ embedded in $\mathbb{R}^{2^{n+1}+1}$ \cite{Marsaglia_1972,10.1145/377939.377946}. We perform this procedure for $6,\ 7,$ and $8$-qubit systems, each time maximizing the violation of Eq.\ \eqref{Ingleton}, finding in every case that the Ingleton gap $g\left (x^\ast(x_0) \right)$ reliably converges to the same minimum value
\begin{equation}\label{MaxViolation}
    g_I\left (x^\ast(x_0) \right) \approx -0.1699,
\end{equation}
regardless of the choice of initialization $x_0$ and Ingleton instance $I\in \mathcal{I}$. Figure~\ref{fig:cobyla_convergence} illustrates a representative convergence profile for a $4$-qubit subsystem $\rho_{ABCD}$, prepared on a $6$-qubit pure state.
\begin{figure}
     \centering
     \begin{subfigure}[c]{0.45\textwidth}
        \centering
        \vskip 0pt 
        {\includegraphics[width=\textwidth]{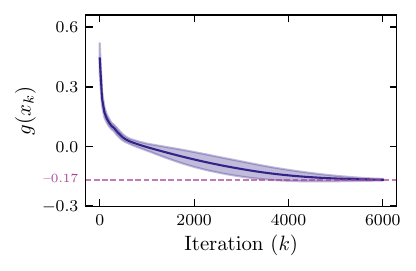}}
        \label{fig:evolution_analysis-cost}
     \end{subfigure}
     \hfill
     \begin{subfigure}[c]{0.45\textwidth}
        \centering
        \vskip 0pt 
        {\includegraphics[width=\textwidth]{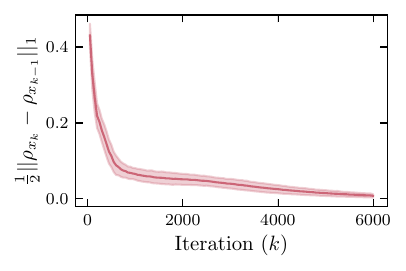}}
        \label{fig:evolution_analysis-dist}
     \end{subfigure}
     \caption{Asymptotic convergence of sequence, in Eq.\ \eqref{SequenceSolution}, as \(k\rightarrow \infty\). Left image shows convergence of the Ingleton gap $g_{I}$, for a $6$-qubit system, to the minimum value, i.e. maximum violation, shown by the dashed line. Right image shows the trace distance approaches zero between \(\rho_{x_k}\) and \(\rho_{x_{k-1}}\), where \(\rho_x \equiv \ketbra{f^{-1}(x)}{f^{-1}(x)}\). The shaded regions indicate standard deviation from the mean.}
     \label{fig:MaxGapPlots}
 \end{figure}

The observed convergence at $6,\ 7,$ and $8$ qubits to the same maximal violation given in Eq.\ \eqref{MaxViolation} indicates that the gap function $g_I(x)$ already attains its full range at $6$ qubits, assuming single-qubit parties in Eq.\ \eqref{Ingleton}. Consequently, $6$ qubits is sufficient to realize maximal violation of the Ingleton instance in  Eq.\ \eqref{Ingleton}. This observation aligns with the intuition that, while Ingleton's inequality is a $4$-party inequality, the largest entropy term in Eq.\ \eqref{Ingleton} involves $3$ qubits, e.g. $S_{ABC}$, and can be purified using an additional $3$ qubits. However, this reasoning is ultimately incomplete, since subsystem entropies are also constrained by numerous additional inequalities, such as subadditivity and strong-subadditivity \cite{Robinson:347009, LiebRuskai1973}, and potentially others \cite{Bao2015} depending on the state. Accordingly, dimensionality arguments must be applied carefully when inferring the minimal system size required for maximal violation.

For the same $6$-qubit realization of maximal violation in Figure~\ref{fig:MaxGapPlots}, we also examine how entanglement and magic evolve as the optimizer approaches maximal Ingleton violation, highlighting the interplay between these two resources. Figure~\ref{fig:cobyla_convergence} displays the evolution of entanglement entropy and the capacity of entanglement, a lower bound on non-local magic, for each subsystem appearing in the Ingleton instance in Eq.\ \eqref{Ingleton}. The figure additionally shows the evolution of overall magic, quantified using the stabilizer $2$-Renyi entropy, for the full system $\rho_{ABCD}$ throughout the optimization.
As the optimizer drives the system towards maximal Ingleton violation, the total magic of $\rho_{ABCD}$ increases steadily, as shown in the top-left panel of Figure~\ref{fig:cobyla_convergence}, highlighting the feature that Ingleton-violating states occupy regions of the Hilbert space with high magic content. Interestingly however, we observe a pronounced collapse of entanglement capacity as the system approaches maximal violation, revealing an inverse relationship between minimal non-local magic and increasing Ingleton violation. Notably, each plot in Figure~\ref{fig:cobyla_convergence} exhibits a sharp information-theoretic phase transition occurring when the entanglement entropy $S_{vN}$ crosses entanglement capacity $C_{E}$ near initial violation of Ingleton's inequality. To this end, the positivity of a difference between entanglement entropy and entanglement capacity serves as a proxy for failure of Ingleton's inequality. These features signal a reorganization of multipartite correlations, i.e. entanglement entropy and non-local magic, required to achieve a (maximal) violating configuration

As the Ingleton violation increases, a widening gap emerges between $S_{vN}$ and $C_{E}$, eventually stabilizing near the maximal-violating state.  The largest such gap occurs in subsystem $CD$ which, as discussed following Figure~\ref{fig:individualEntropies}, plays a critical role in the evaluation of Ingleton's inequality with an entropy that maximizes at violation. The tripartite subsystems $ABC$ and $ABD$ each show a gradual increase in entanglement entropy, saturating at a submaximal value $S_{vN} \approx 2$. In contract, their respective entanglement capacities decay to nearly zero, indicating that while these regions retain substantial entanglement, they rapidly lose the ability to generate additional entanglement. The universal collapse of entanglement capacity across all subsystems reflects an increasing fragility of subsystem entanglement~\cite{DeBoer:2018kvc} as the system nears maximal violation, i.e. as the multipartite correlations become highly constrained the state approaching an extremal point in the resource landscape.

\begin{figure}[h]
    \centering
    \includegraphics[width = 0.7\linewidth]{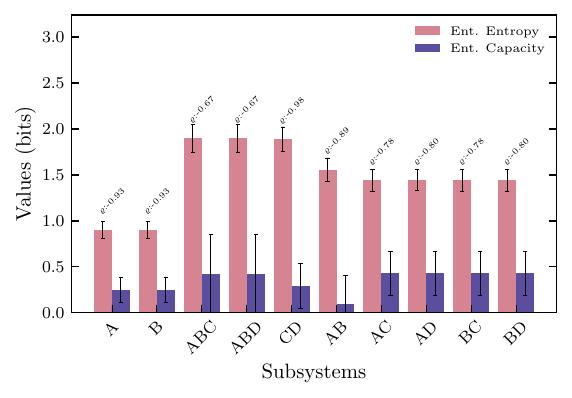}
     \caption{Histogram showing average entanglement entropy and entanglement capacity for $\sim 10^4$ maximum Ingleton violators, with error bars giving standard deviation. The two resources are found to be anti-correlated, with magnitude given by the Pearson correlation coefficient $\varrho$. Subsystem $CD$ possesses the strongest anticorrelation.}
     \label{fig:ResourceCorrelation}
 \end{figure}

We further investigate the relationship between entanglement entropy and capacity of entanglement by performing a correlation analysis over an ensemble, of size $\sim 10^4$, of maximal Ingleton-violating states. For each state in the ensemble, we compute $S_{vN}$ and $C_E$ on every subsystem appearing in Ingleton's inequality, as given in Eq.\ \eqref{Ingleton}. Figure~\ref{fig:ResourceCorrelation} plots the average entanglement entropy of each subsystem, as well as the average entanglement capacity. The Pearson coefficient $\varrho$, which measures the strength and direction of a linear relationship between the two parameters, is indicated above each column.
A consistent pattern emerges across the ensemble of maximal violators, in which entanglement entropy and entanglement capacity are strongly anticorrelated. For each subsystem appearing in Ingleton's inequality, maximal violation coincides with an increase in $S_{vN}$ and a corresponding decrease in $C_E$, yielding a lower minimal amount of non-local magic. The Pearson coefficient $\varrho \in [-1,\ 1]$ quantifies this relationship, and the values $\varrho_{CD}  = -0.98$ and $\varrho_A = \varrho_B = -0.93$ indicate a strong, almost perfectly monotonic, inverse dependence between the two resources. Notably, subsystems $A,\ B,$ and $CD$ all appear on the right-hand side of Eq.\ \eqref{Ingleton}, and contribute dominantly to the entanglement entropies that produce violations of Ingleton's inequality. The $3$-party subsystems $ABC$ and $ABD$ possess the largest overall of entanglement entropies, but their entanglement capacities remain suppressed, reinforcing the behavior seen in Figure~\ref{fig:cobyla_convergence} that the system maintains substantial, yet highly constrained, entanglement. Following the analyses in Figures~\ref{fig:cobyla_convergence} and \ref{fig:ResourceCorrelation}, we therefore conjecture that Ingleton-violating states have high entanglement, high total magic, and low non-local magic. 

In this section we introduced an optimization protocol for generating states that violate Ingleton's inequality, providing numerous examples in Appendix~\ref{MoreIngletonStates} and substantially augmenting the set of known Ingleton-violating states reported in the literature. Moreover, this optimization protocol enables the identification of arbitrarily many Ingleton-violating states, with controllable violation magnitudes. Repeatedly initializing the optimizer on randomly-selected $6,\ 7,$ and $8$-qubit states, we observed consistent convergence to the same minimum Ingleton gap $g_{Min.} \approx -0.1699$, indicating the existence of a maximum violation amount for Ingleton's inequality and confirming that $6$ qubits were sufficient to realize the maximum violation of Eq.\ \eqref{Ingleton}. We analyzed the convergence behavior of the optimizers as the system evolved towards maximum violation, and tracked the evolution of different quantum resources. Specifically, we tracked the total magic of the violating system $\rho_{ABCD}$, as well as the entanglement entropy and entanglement capacity, a lower-bound on non-local magic, for each subsystem involved in the evaluation of Eq.\ \eqref{Ingleton}. Our results showed that maximum Ingleton violation coincides with a steady increase in total magic, but a pronounced collapse of entanglement capacity, and thereby minimum non-local magic. A statistical analysis further revealed a strong anticorrelation between entanglement entropy and entanglement capacity, across all relevant subsystems in maximal Ingleton-violating states, indicating a characteristic resource profile of high entanglement, high overall magic, and low non-local magic. Together, these findings demonstrated that Ingleton-violating states occupy extremal regions of the Hilbert space, marked by high entanglement and high magic, but with fragile entanglement structure and potentially very little non-local magic. In the next section we examine the stability of Ingleton violation, and assess the rarity of Ingleton-violating states in the Hilbert space.

\subsection{Stability Analysis and Rarity of Ingleton Violation}\label{RaritySection}

Having established the minimum Ingleton gap in Eq.\ \eqref{MaxViolation}, which corresponds to the maximal violation of Ingleton's inequality in Eq.\ \eqref{Ingleton}, a natural next step is to investigate both the distribution of maximally violating states and the stability of maximally violating solutions. The question of solution stabilizer, in the context of optimization~\cite{stability_in_optimization}, addresses the sensitivity of obtained solutions to perturbations in the problem's formulation, input data, or choice of algorithms. Conducting this stabilizer analysis can therefore provide insight into the uniqueness of local minima, as well as the suitability of the chosen optimization algorithm for exploring the entropy vector landscape. Moreover, these details help quantify the rarity of Ingleton-violating states in the Hilbert space, and reveal statistical correlations between specific quantum resources at maximal violation.

We begin our stability analysis by measuring the distance between optimal solutions $x^\ast(x_0)$ and $x^\ast(y_0)$, obtained from initial states $x_0$ and 
$y_0$ randomly sampled from $S^{2^{n+1}}$ with $x_0\neq y_0$. Distances are quantified using two complementary measures, each evaluated for solutions $x^*$ that minimize the Ingleton gap. First, the $2$-norm, often called the Euclidean norm, is defined
\begin{equation}\label{EucNorm}
   D_{E}(x_0,y_0) \equiv ||x^\ast(x_0) -x^\ast(y_0)||_2,
\end{equation}
and provides a canonical vector norm for complex vectors. Alternatively, the Schatten $1$-norm, commonly referred to as the trace distance~\cite{Nielsen_Chuang_2012}, is defined for operators as
\begin{equation}\label{TraceNorm}
   D_{T}(x_0,y_0) \equiv ||\rho_{x^\ast(x_0)} - \rho_{x^\ast(y_0)}||_1,
\end{equation}
where $\rho_{x^\ast(x_0)}$ is the density matrix associated to the wavefunction of the solution vector
\begin{equation}
   \rho_{x^\ast(x_0)} \equiv \ket{f^{-1}(x^\ast(x_0))}\bra{f^{-1}(x^\ast(x_0))}.
\end{equation}
While the Euclidean norm $D_{E}(x_0,y_0)$ captures geometric separation of solution vectors, the trace distance $D_{T}(x_0,y_0)$ quantifies the distinguishability of the corresponding states. 

\begin{figure}[h]
    \centering
    \includegraphics[width=.7\linewidth]{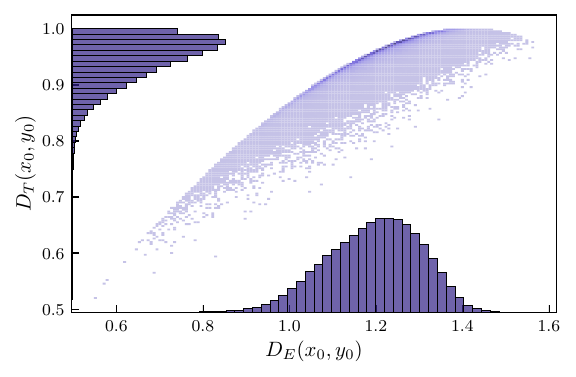}
     \caption{Scatterplot of Euclidean and trace distance for final states obtained by optimization, each starting from a random initial state $x_0$. Histograms on each axis depict the density of the scatter points, indicating that optimized output states are distinct, and tend to be far-separated from each other.}
     \label{fig:distancePlot}
 \end{figure}

For optimal solutions $x^\ast(x_0)$ and $x^\ast(y_0)$, derived respectively by applying the optimizer to randomly-generated states $x_0$ and $y_0$, Figure~\ref{fig:distancePlot} illustrates the distance between these solutions using the distance measures defined in Eqs.\ \eqref{EucNorm} and \eqref{TraceNorm}. The Euclidean distance $D_{E}(x_0,y_0)$ is plotted on the horizontal axis, while the trace distance $D_{T}(x_0,y_0)$ is shown on the vertical axis. A histogram accompanies each axis, displaying the distribution of distances according to each measure. Figure~\ref{fig:distancePlot} is generated from $10^{6}$ randomly sampled pairs $\left(x_0,\ y_0\right)$.
Every point in Figure~\ref{fig:distancePlot} has a distance $D_E,\ D_T> 0.5$ \cite{PhysRevA.108.012409}, indicating that distinct initial states $x_0$ and $y_0$ converge to unique solutions $x^\ast(x_0)$ and $x^\ast(y_0)$ under optimization. Furthermore, the distribution of distances shows that, while optimal solutions achieve the same maximal violation given by Eq.\ \eqref{MaxViolation}, they remain widely separated in the Hilbert space. This observation suggests that Ingleton-violating states are sparse and far apart.

\begin{figure}[h]
     \centering
     \includegraphics[width=.7\linewidth]{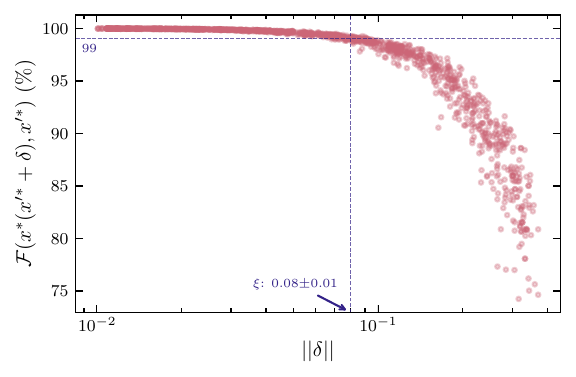}
     \caption{Plot illustrating the robustness of optimization and the stability radius, $\xi$. Each point in the scatter plot represents the fidelity, $\mathcal{F}$ between a maximally-violating state $x^{\prime\ast}$ and solution obtained by optimizing a perturbed state $x^{\prime\ast} + \delta$. A vertical line at $\xi$ marks the perturbation threshold beyond which $\mathcal{F}$ drops below $99\%$ (horizontal line).}
     \label{fig:fidelityPlot}
 \end{figure}

Furthering our understanding of the Ingleton optimization landscape, we perform a stability analysis with respect to perturbations in the input vector. Specifically, we examine the trace distance between a solution vector $x^{\prime\ast} \equiv x^\ast(x_0^\prime)$ and the solution vector obtained by starting the optimization from a $\delta$-perturbed input  $x^\ast(x^{\prime\ast}+\delta)$, where $x^\ast(x^{\prime \ast}) = x^{\prime \ast}$. The optimization is considered robust to perturbation $\delta$ if the resulting distance remains bounded within a constant multiple $C$ of $||\delta||$, explicitly 
\begin{equation}
    \text{dist.}( x^\ast(x_0 + \delta),\ x^\ast(x_0)) \leq C ||\delta||, \quad \forall \, ||\delta|| < \xi,
\end{equation}
where $\xi$ is defined to be the stability radius, i.e. the largest perturbation for which the optimizer output remains stable. We select the constant $C$ such that the stability radius corresponds to a fidelity
\begin{equation}
    \mathcal{F}\left(x^\ast(x^{\prime\ast} + \delta),\ x^{\prime\ast}\right) \equiv 1 - D_T\left(x^\ast(x^{\prime\ast} + \delta),\ x^{\prime\ast} \right)^2,
\end{equation}
of $99\%$ or higher between the perturbed state $x^\ast(x^{\prime\ast} + \delta)$ and the reference state $x^{\prime\ast}$, indicated by the dashed horizontal line in Figure~\ref{fig:fidelityPlot}. We calculate the stability radius to be $\xi = 0.08\pm 0.01$, as depicted by the dashed vertical line in Figure~\ref{fig:fidelityPlot}. 
This analysis empirically reaffirms the claim that distinct maximal violators of Ingleton's inequality are far-separated across the Hilbert space. Consequently, unique Ingleton-violating states can be reliably generated by performing the optimization algorithm $\mathcal{A}$, from Section~\ref{OptimizationSection}, beginning from initial vectors separated by a distance $\xi$.

\begin{figure}
    \centering
    \includegraphics[width = 0.7\linewidth]{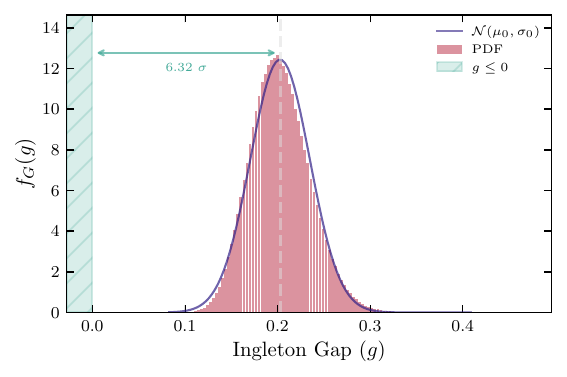}
     \caption{Probability density function of Ingleton gap $g$ (red histogram) for pure $8$-qubit Haar random states. The distribution converges to a normal distribution $\mathcal{N}$ (blue curve), with a KL divergence of $0.2$, for large sample size $N > 10^6$. The shaded green region represents Ingleton violation, located $6\sigma$ from mean, making such states exceedingly rare.}
     \label{fig:IngDist}
 \end{figure}

The analysis above suggests that violations of Ingleton's inequality occur within a bounded neighborhood of the maximal violating states. Moreover, although such states can be identified using optimization, both generic Ingleton-violating states, and maximal violators in particular, appear to be exceptionally rare in the Hilbert space. Formalizing this observation, and quantifying the rarity of Ingleton-violating states, we examine the Ingleton gap for Haar-random sampled states. Specifically, we randomly select a unitary $U$ from the Haar measure on a $2^8$-dimensional Hilbert space, and apply $U$ to the state $\ket{+}^{\otimes 8}$. For each resulting state $U\ket{+}^{\otimes 8}$, we compute the entropy vector and corresponding Ingleton gap $g$ over all instances, as in Eq.\ \eqref{IngletonGap}, for the subsystem $\rho_{ABCD}$ defined on the first $4$ qubits of $U\ket{+}^{\otimes 8}$. The distribution $f_G(g)$ of the Ingleton gap, illustrated in Figure~\ref{fig:IngDist}, has a mean $\mu_0 \approx 0.2026$ and standard deviation $\sigma_0 \approx 0.0321$, and closely resembles a normal distribution $\mathcal{N}(\mu = \mu_0,\ \sigma = \sigma_0)$ with a Kullback-Leibler divergence of $0.02$.
Recall Ingleton's inequality is saturated when the Ingleton gap vanishes, i.e. $g=0$, and violated when $g<0$. Despite sampling over $10^6$ Haar-random states, we observe no violations of Ingleton's inequality. This absence of Ingleton violators is consistent with $f_G(g)$ in Figure~\ref{fig:IngDist}, which indicates that Ingleton-violating states are expected to occur more than $6\sigma$ away from the mean, placing them deep in the tail of the distribution and rendering their appearance in random sampling highly unlikely.

In this section we analyzed the stability and rarity of Ingleton-violating states within the Hilbert space. First, we performed a stability analysis by measuring the distances between optimal (maximal-violating) states obtained from different randomly-generated input states. Using both Euclidean and trace distance measures, we demonstrated that distinct input states converge to distinct output states, indicating that clusters of Ingleton-violating states are far-separated in the Hilbert space. We then examined the robustness of maximal Ingleton violations, establishing a finite and relatively small radius within which small perturbations of the input reliably reproduce the same violating solution. The existence and magnitude of this stability radius confirms that each maximally-violating state is surrounded by a sharply bounded neighborhood of attraction. Finally, we analyzed a distribution of the Ingleton gap across on millions of Haar-random states and showed that violations occurs far in the statistical tail of the distribution, over $6$ standard deviations from the mean gap value, emphasizing the exceptional rarity of Ingleton-violating states. Together, these results provide a comprehensive characterization of both the stability and scarcity of Ingleton-violating states. We now conclude this manuscript by discussing these findings and outlining directions for future work.

\section{Discussion}\label{DiscussionSection}

In this work, we present a hybrid machine learning and optimization framework to study the emergence and dynamics of entanglement entropy and quantum magic as a system evolves towards violation of a chosen entropy inequality. Using a reinforcement learning algorithm, trained on a reward function defined by the target inequality, we systematically explore the Hilbert space using a preselected gate set to identify states that violate the inequality. After each gate application, the state's entropy vector and magic are evaluated to determine subsequent gates. Focusing on Ingleton's inequality, the protocol generates new sets of violating states, with tunable violation amounts, significantly extending the limited examples previously known.

We first prove that a minimum of $6$ qubits are required to violate Ingleton's inequality, and analyze the interplay between entanglement and magic underscoring Ingleton violation. Tracking the evolution of quantum resources across Ingleton-violating circuits, we reveal how entropy vectors, magic, and non-local magic evolve as states exit the Ingleton entropy cone. The reinforcement learning agent effectively learns these transitions by traversing the entanglement/magic landscape, uncovering constructive paths through the Hilbert space that integrate geometric, information-theoretic, and resource considerations. This approach provides a generic framework for probing the dynamics of entropy vectors and better understanding how non-classical correlations enable entropy inequality violation and constrain quantum information processing. While our analysis focuses on Ingleton's inequality in this work, the methods are sufficiently general to study any linear entropy constraint, and evolution under any set of quantum gates. Collectively, these techniques introduce a new algorithmic strategy for exploring the geometry of entropy cones, and navigating the landscape of quantum resources through the lens of entropy inequalities. 

Beyond our circuit-based approach, we develop a numerical optimization scheme to directly identify Ingleton-violating states in the Hilbert space. Employing the evolution optimizers Covariance Matrix Adaptation Evolution Strategy (CMA-ES) and Constrained Optimization BY Linear Approximation (COBYLA), we empirically determine a maximal violation of Ingleton's inequality, which we explicitly verify up through $8$-qubits, and track quantum resource evolution as states approach and saturate this bound. Our optimization method generates arbitrarily many Ingleton-violating states, with customizable degrees of violation. We perform a stability analysis and statistical characterization of Ingleton violating states, demonstrating that they lie deep in the tail of the Haar-random distribution and are confined to sharply-defined, isolated regions, $\epsilon$-neighborhoods, of the Hilbert space. Using CMA-ES and COBYLA we demonstrate consistent convergence to the same maximal violation amount across random initializations, reaffirming the robustness and isolation of maximal-violating states. These findings rigorously quantify the ``non-generic'' nature of Ingleton violation, and provide a deterministic strategy to construct broad families of maximally-violating states, previously unknown in the literature. 

The machine learning and classical optimization techniques introduced in this work can be readily adapted to probe the landscape of entanglement and magic, as well as any computable quantum resource, offering a data-driven approach for exploring quantum complexity. While entanglement and magic each provide independent advantages for specific tasks, it is the combination of these resources that drives quantum states beyond the reach of efficient classical simulation. For example, states with maximal entanglement and no magic, e.g. stabilizer states, are efficiently classically simulable~\cite{aaronson2004improved}. Conversely, states with high magic and minimal entanglement, e.g. matrix product states (MPS) at low bond dimension, are likewise classically tractable using tensor network methods~\cite{PhysRevB.110.045101, PhysRevB.109.174207}. Classical simulation becomes challenging in the regime where both entanglement and magic exist in significant  \cite{PhysRevLett.116.250501}. Ingleton-violating states inhabit in such regions of the Hilbert space, making them suitable test cases for probing the boundary between classically simulable and manifestly quantum computation. The efficient identification of Ingleton-violating states, with custom degree of violation, together with the circuits that prepare them, can improve navigation of the entanglement/magic phase space. This includes addressing questions such as: Is it more resource-efficient to generate entanglement first, then inject magic, or to begin with high magic states and distribute entanglement among them? Moreover, entropy inequality-based methods provide a novel framework for exploring trajectories through the quantum resource landscape of the Hilbert space. Incorporating information-theoretic constraints, such approaches reveal structural features of states and circuits which ultimately govern their utility for performing specific computational tasks.

The techniques established in this work provide a versatile toolkit for engineering quantum circuits with custom information-theoretic properties. In particular, this framework enables the precise and selective manipulation of entropy vectors within specific subsets of a quantum register. For example, one can design circuits which strictly preserve a chosen entropic constraint~\cite{Hayden_Jozsa_Petz_Winter_2004}, e.g. MMI, on a designated $3$-party subset of qubits, while simultaneously inducing a controlled violation of Ingleton's inequality on a different $4$-party subsystem. This ability to selectively enforce or violate linear inequalities allows for the creation of multipartite systems with complex information-theoretic structure within a single global state. Beyond a theoretical interest, such systems could offer a testbed for benchmarking future optimization strategies aimed at probing quantum correlations and exploring the interplay between classical and quantum information resources.

Another natural direction is to systematically quantify the complexity of preparing quantum states~\cite{10.5555/2011686.2011688, Seddon_Campbell_2019} with desired entanglement entropy or magic profiles, given access to a specific gate set. Understanding the minimal resources, both in terms of circuit depth and gate type, as well as non-stabilizer resources, needed to realize particular entropy vectors can provide insight into the practicality of preparing specific multipartite systems on different hardware platforms. Beyond static entanglement measures, this framework can be extended to probe dynamical properties, such as operator entanglement~\cite{PhysRevLett.126.030601, Andreadakis:2025mfw} and the spreading of entanglement and magic, by promoting the circuit to a quantum channel. This extension would offer a more complete understanding of how unitary evolution distributes quantum information and non-classical resources across a multipartite system, offering a foundation for the controlled manipulation of complex information-theoretic structures in many-body quantum states.

MMI and Ingleton's inequality partially characterize the holographic entropy cone, the convex hull of entropy vectors corresponding to states with a smooth classical dual geometric in AdS/CFT~\cite{HernandezCuenca2019, Bao2020a, Bao2015, Czech2021}. Since our protocol can describe controlled violation, and equivalently satisfaction, of linear entropy inequalities, we could naturally consider the possibility of defining a class of ``holography-preserving'' quantum circuits. In other words, we could characterize the class of quantum circuits, composed of a chosen gate set, which keep a state's entropy vector inside the holographic entropy cone. For a holographic CFT prepared on a system of qubits, evolution by a holography-preserving circuit would, at least, guarantee the state's entanglement structure remains compatible with a smooth geometric representation in AdS. A deeper understanding of how unitary transformations of a quantum state correspond to geometric transformations of the dual spacetime could enable the simulation of gravitational processes, those that induce changes in the dual geometry, using quantum circuits that preserve the full set of holographic entropy inequalities.

Methods developed in this work can also be applied beyond the Ingleton setting, particularly to quantum resource theory in chemical and condensed matter systems. A growing body of literature has emphasized that entanglement, non-classical correlations, and resource conversion structures play a central role across quantum chemistry and strongly correlated materials \cite{Ding:2025dac, Aliverti-Piuri:2024dub, Boguslawski:2013fzb, Duperrouzel:2015gas, Perez-Obiol:2023wdz}. In this broader context, the entropy vector formalism and the tools introduced here for tracking the evolution of entanglement and magic under constrained circuit dynamics offer a natural bridge to resource-theoretic approaches in electronic-structure theory. These techniques could enable a more unified way of quantifying how non-classicality, manifesting as entanglement or magic emerges, propagates, and transforms across physically relevant models. 

In classical information theory, Ingleton's inequality is satisfied by any network that admits a linear coding solution~\cite{Ingleton:1971,Boston2020ViolationsOT,Dougherty:2005}, and its violation signals that some non-linear processing is necessary to achieve the full communication capacity of the network. As quantum networks begin to emerge~\cite{Hahn_2019,Hayashibook:2007,Coffman2000,Kobayashi:2011}, characterizing Ingleton-violating entropy vectors, i.e. multipartite entanglement patterns that lie outside some linear regime, may offer useful insights into the distinction between linearly and non-linearly preparable entanglement structures. The efficient and controlled distribution of entanglement is a fundamental process underlying quantum networks~\cite{Hein2003,WalterGrossEisert2016,HeinDurEisertBriegel2006,Negrin:2024tyj}, enabling capabilities such as entanglement swapping and secure key distribution~\cite{Bennett:2014rmv,Ekert,Briegel_1998,Pirandola2018,Gottesman:2002gg,Hughes:2025jwa}, teleportation-based gates~\cite{Bennett1993,Raussendorf2001,Wehner2018}, and enhanced metrology using distributed quantum sensing~\cite{ Huelga1997,Higgins2007,Zhuang2018}. In this context, Ingleton's inequality provides a diagnostic for identifying entanglement patterns whose distributed preparation may require additional computational resources, parallel coordination between nodes, or even some non-stabilizer operation within nodes to achieve. While a complete operational understanding of Ingleton's inequality in quantum networks remains unexplored, continued study may help clarify the limitations of linear coding techniques in such architectures, and provide insight into the generation and distribution of certain non-stabilizer resources across distant nodes.

\acknowledgments

The authors thank Ning Bao, J. Shepard Bryan IV, Bartek Czech, Jesus Fuentes, Tobias Haug, Cynthia Keeler, Scott Nie, Jason Pollack, Martin Savage, and Dylan Van Allen for helpful discussions. This work is supported by the Department of Energy (DOE) Office of Science (SC) Grant No DOE DE-FOA-0003432. This work is also supported by Grant No GBMF12976 of the Gordon and Betty Moore Foundation. This material is based upon work supported by the U.S. Department of Energy, Office of Science, Office of Advanced Scientific Computing Research, Department of Energy Computational Science Graduate Fellowship under Award Number DE-SC0024386.

\newpage

\begin{appendix}
\addtocontents{toc}{\protect\setcounter{tocdepth}{1}}

\section{Additional Circuits and Magic Evolution}\label{MoreIngletonCircuit}

These are the circuits corresponding to Violators 2-4 in section~\ref{ResourceEvolutionSection}. The Q-Learning algorithm was initialized at the state given by Eq.~\eqref{sat_state} and evolved until Ingleton violation was reached.  

\begin{figure}[h]
    \centering
    \begin{subfigure}[b]{0.9\textwidth}
        \centering
        \includegraphics[width=\linewidth]{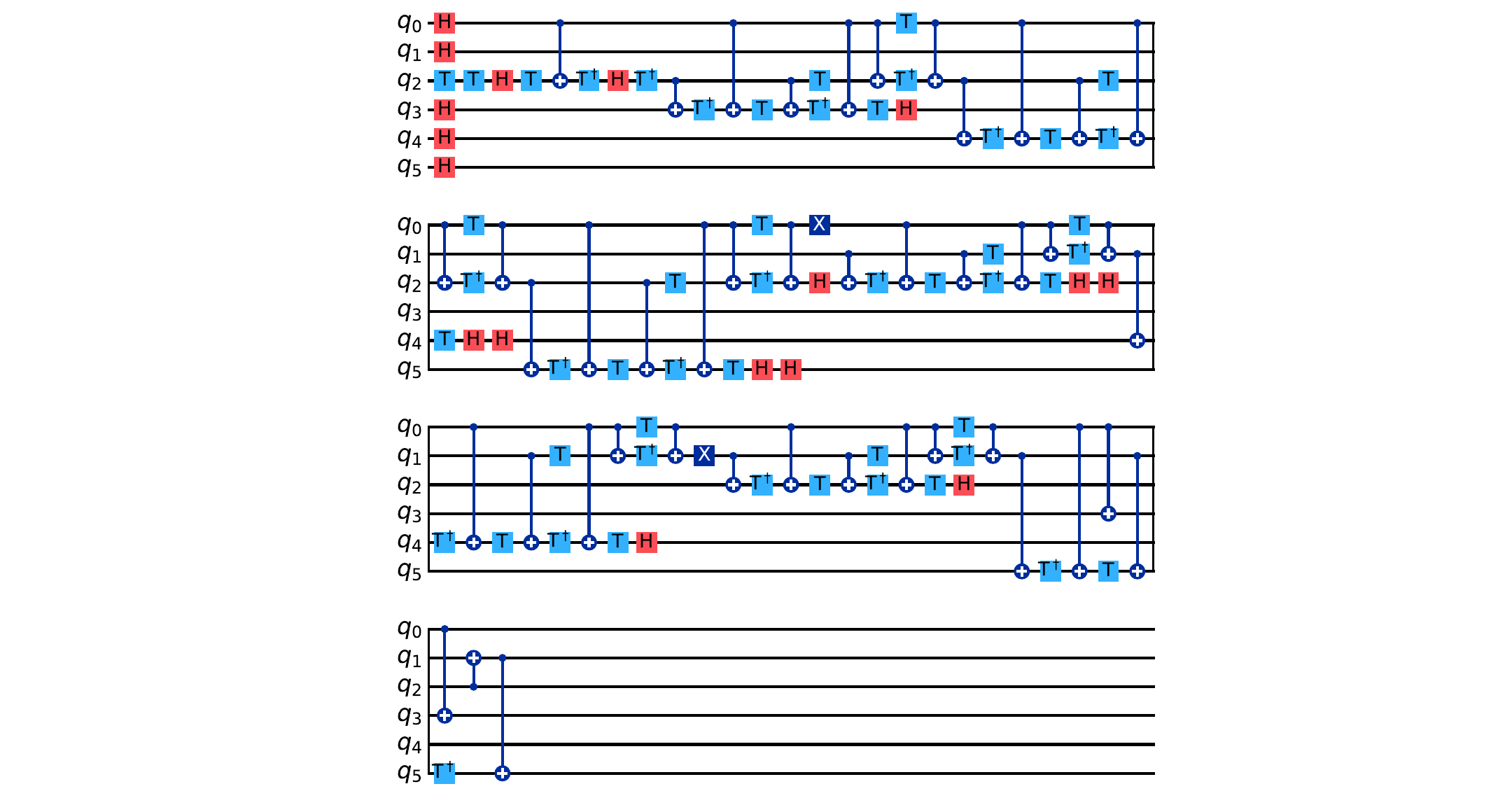}
        \caption{Violator 2}
    \end{subfigure}
    \begin{subfigure}[b]{0.9\textwidth}
        \centering
        \includegraphics[width=\linewidth]{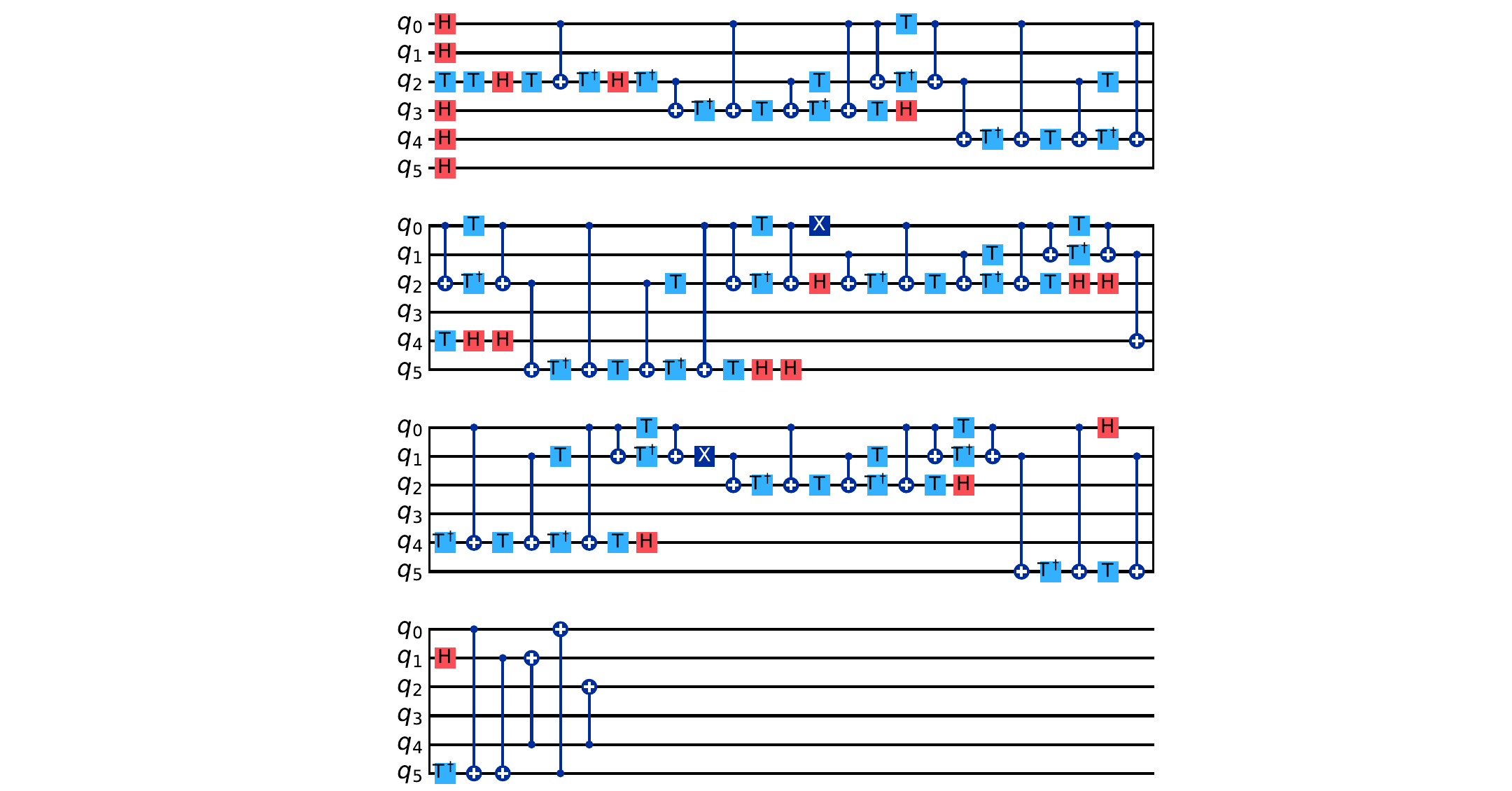}
        \caption{Violator 3}
    \end{subfigure}

    \vspace{1em} 

    \caption{Violators 2-3 found by perturbing Eq.~\eqref{sat_state} and QL paradigm in Figure~\ref{fig:ql_workflow}.}
    \label{fig:QL_circs}
\end{figure}

\begin{figure}[h]
    \centering

    \begin{subfigure}[b]{0.9\textwidth}
        \centering
        \includegraphics[width=\linewidth]{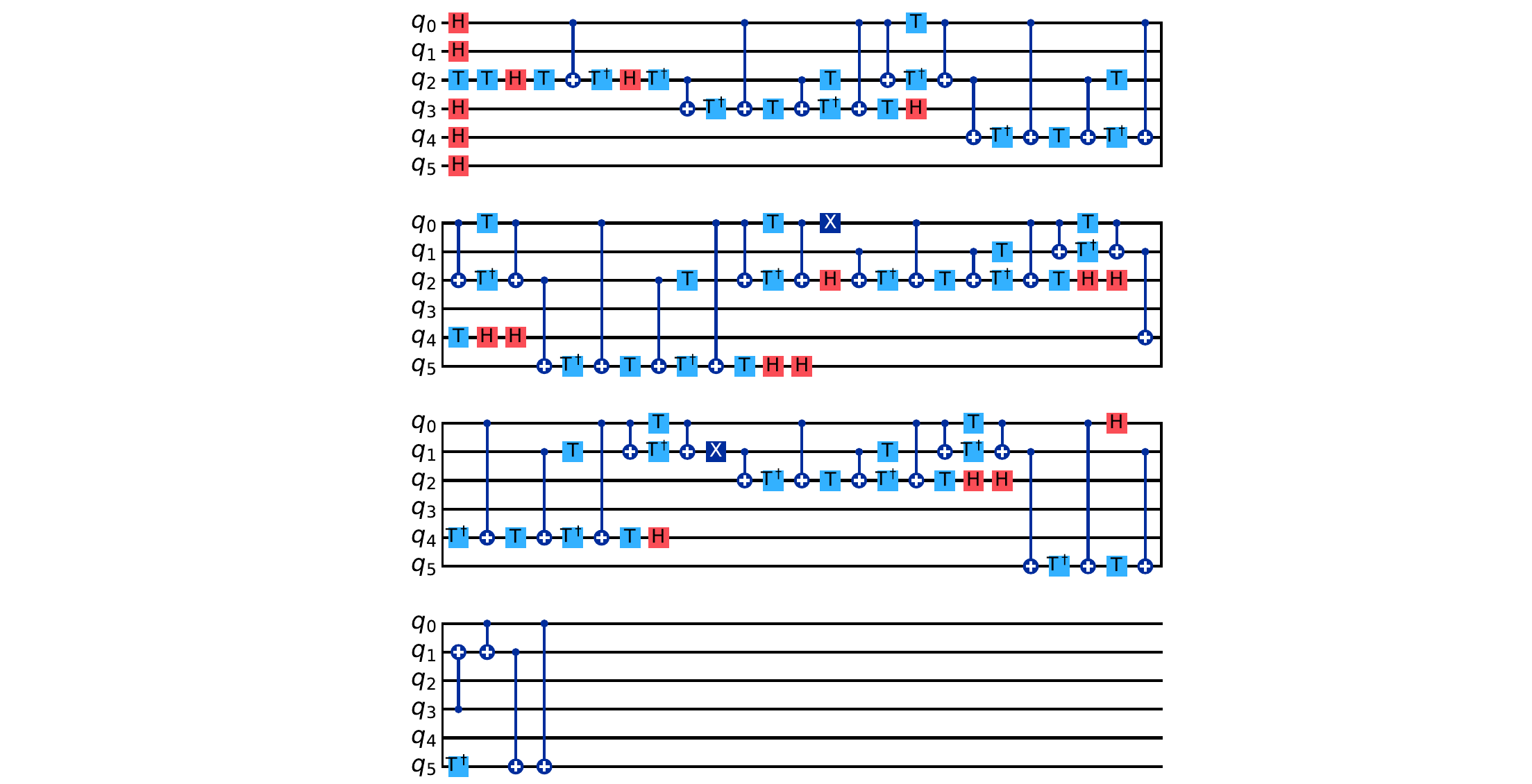}
        \caption{Violator 4}
    \end{subfigure}
    \begin{subfigure}[b]{0.9\textwidth}
        \centering
        \includegraphics[width=\linewidth]{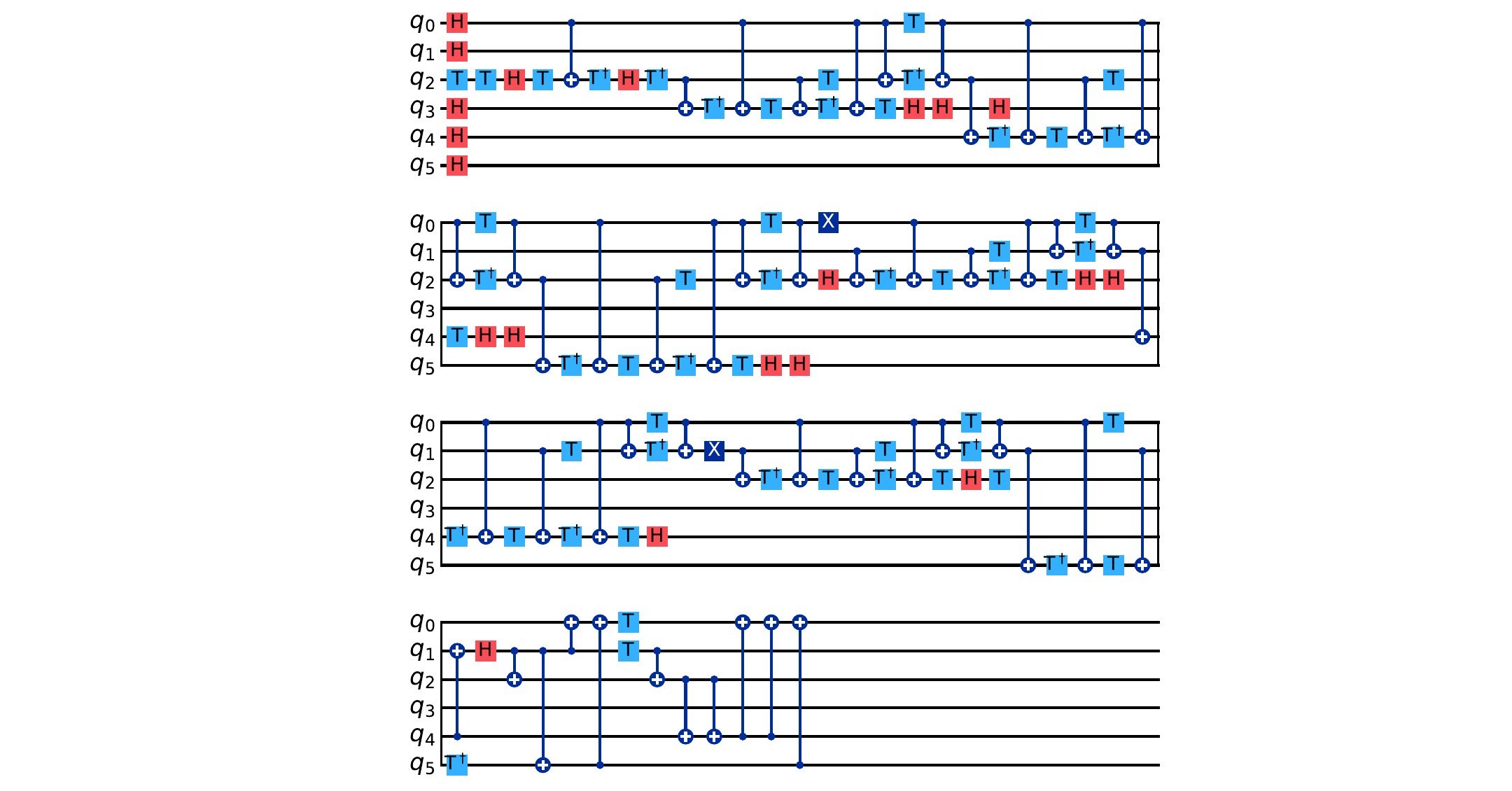}
        \caption{Violator 5}
    \end{subfigure}

    \caption{Violators 4-5 found by perturbing \eqref{sat_state} and QL paradigm \ref{fig:ql_workflow}.}
    \label{fig:QL_circs1}
\end{figure}

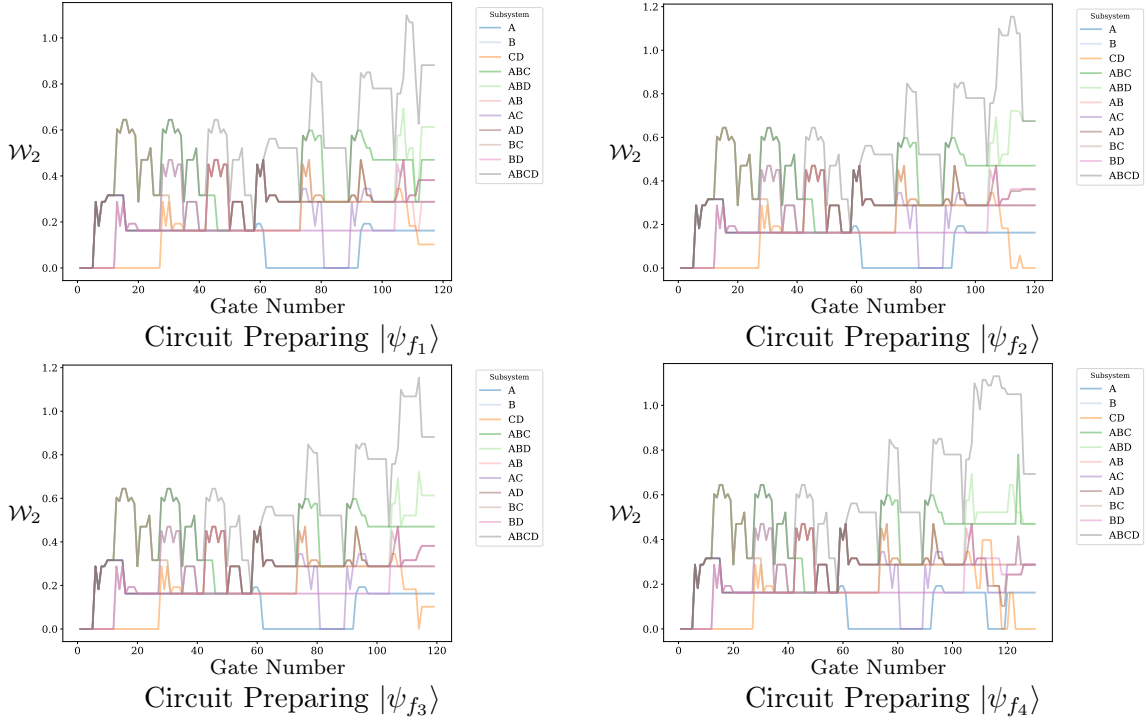
\begin{figure}
    \begin{minipage}{\textwidth}
    \centering
    \begin{subfigure}{0.45\textwidth}
    \centering
        \begin{overpic}[width=\linewidth]{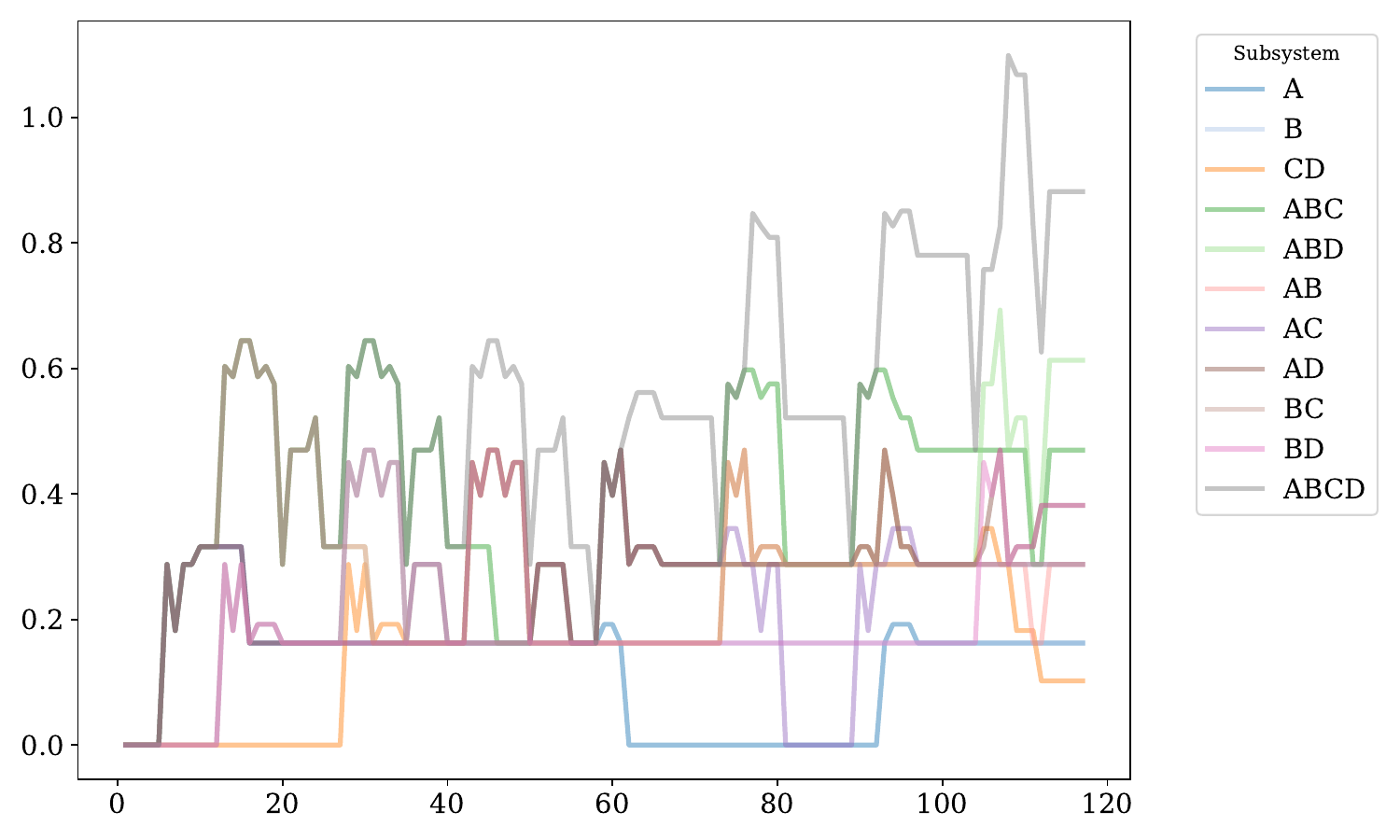}
            \put(34, -2) {\scriptsize Gate Number}
            \put(-7, 28) { \scriptsize $\mathcal{W}_{2}$}
        \end{overpic}
        \caption*{Circuit Preparing $\ket{\psi_{f_1}}$}
        \label{fig:sre2}
    \end{subfigure}
    \hspace{2em}
    \begin{subfigure}{0.45\textwidth}
    \centering
        \begin{overpic}[width=\linewidth]{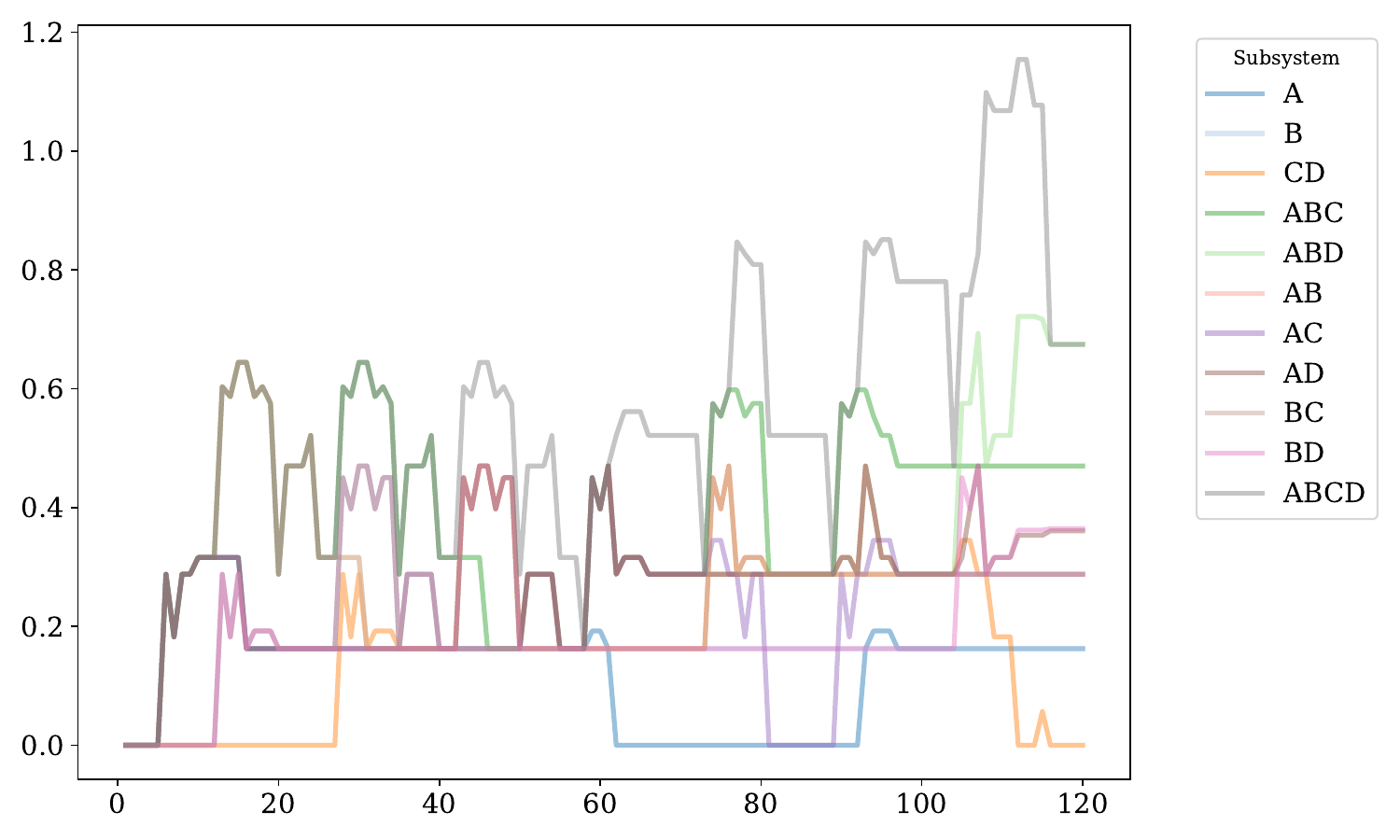}
            \put(34, -2) {\scriptsize Gate Number}
            \put(-7, 28) { \scriptsize $\mathcal{W}_{2}$}
        \end{overpic}
        \caption*{Circuit Preparing $\ket{\psi_{f_2}}$}
        \label{fig:sre3}
    \end{subfigure}
    \end{minipage}
   
    \begin{minipage}{\textwidth}
    \centering
    \begin{subfigure}{0.45\textwidth}
    \centering
        \begin{overpic}[width=\linewidth]{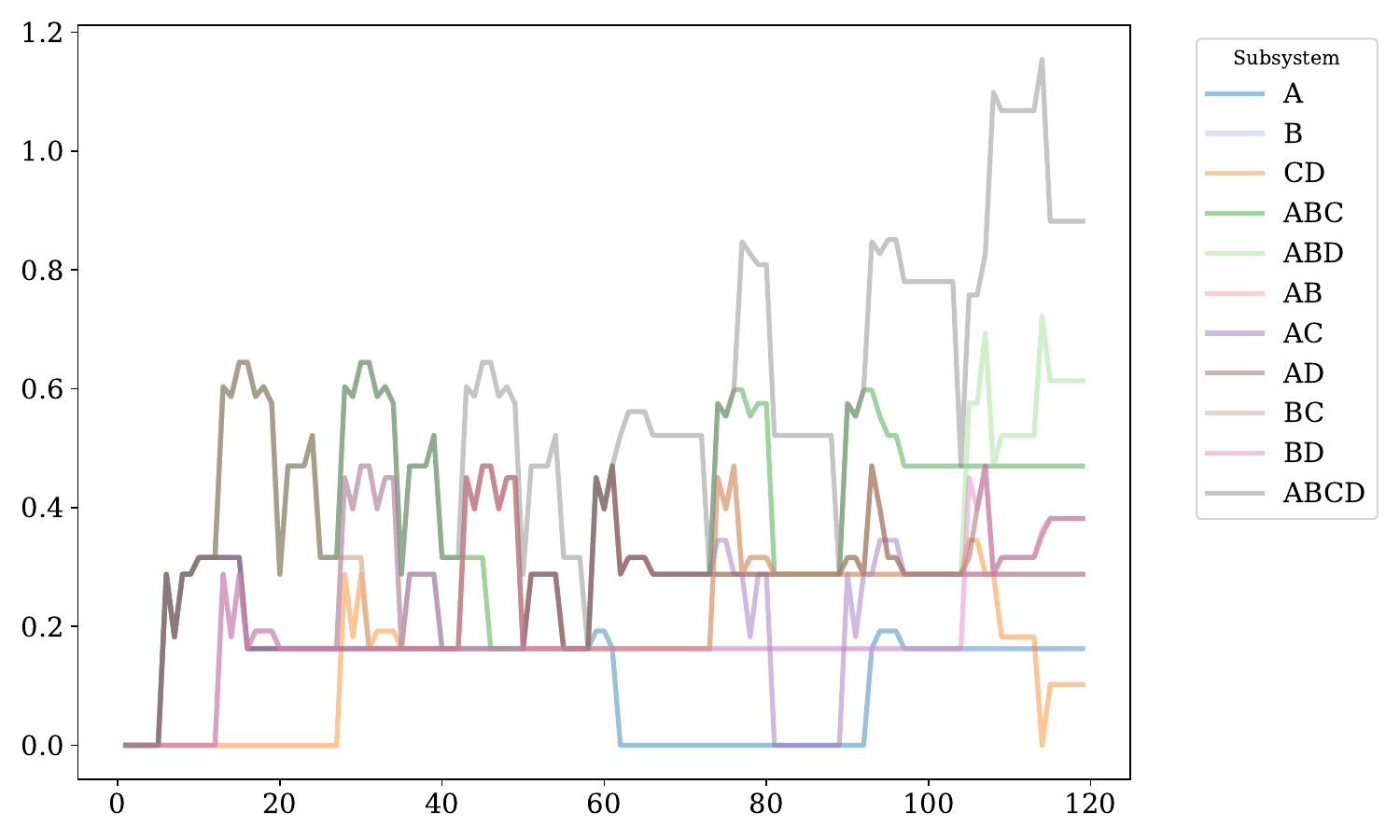}
            \put(34, -2) {\scriptsize Gate Number}
            \put(-7, 28) { \scriptsize $\mathcal{W}_{2}$}
        \end{overpic}
        \caption*{Circuit Preparing $\ket{\psi_{f_3}}$}
        \label{fig:sre4}
    \end{subfigure}
    \hspace{2em}
    \begin{subfigure}{0.45\textwidth}
    \centering
        \begin{overpic}[width=\linewidth]{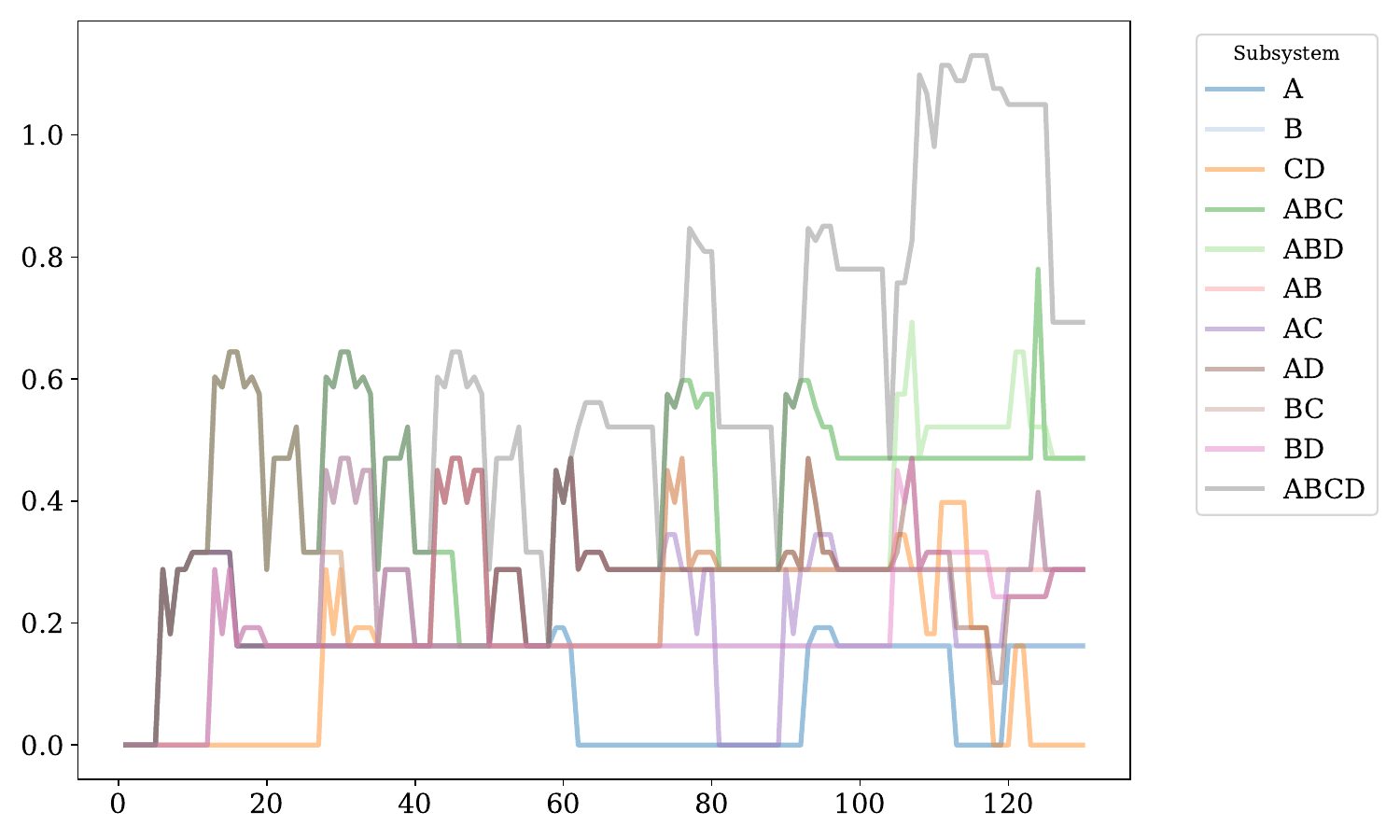}
            \put(34, -2) {\scriptsize Gate Number}
            \put(-7, 28) { \scriptsize $\mathcal{W}_{2}$}
        \end{overpic}
        \caption*{Circuit Preparing $\ket{\psi_{f_4}}$}
        \label{fig:sre5}
    \end{subfigure}
    \end{minipage}
    \caption{Evolution of the non-local magic witness for mixed states obtained from Ingleton-violating circuits identified via Q-learning–based perturbations of the known violator. In circuits 2 and 4, subsystem CD exhibits a positive $\mathcal{W}_{2}$ value at the point of violation, in contrast to the reference (literature) circuit where $\mathcal{W}_{2}(CD) = 0$.}
    \label{fig:sre_perturbing_states}
\end{figure}

For each circuit in Appendix~\ref{MoreIngletonCircuit} that perturbs about a state on the edge of the Ingleton entropy cone, we track the overall magic of the system as the entropy vector is driven out of the Ingleton entropy cone. Figure~\ref{fig:sre_perturbing_states} shows the evolution of $\mathcal{W}_2$, the magic witness defined in Eq.\ \eqref{MagicWitness}, across each circuit that prepares a state $\ket{\psi}_{f_i}$, beginning from the initial state $\ket{0}^{\otimes 6}$.
In each circuit, the overall magic of subsystem $ABCD \subset \ket{\psi}_{f_i}$ increases throughout the evolution, as the system is driven towards Ingleton violation. 

\newpage

\section{Examples of Ingleton-Violating States}\label{MoreIngletonStates}

Below we include a few explicitly examples of Ingleton-violating states, with various violation amounts. A significantly larger sample list is publicly available at \cite{Khumalo_Quantum_Resource_Dynamics}.

\input{ingleton_states.tex}

Please find the complete codebase, data and all other generated examples of violations at \cite{Khumalo_Quantum_Resource_Dynamics}. Additional functions to compute entropy vectors and evaluate entropy inequalities available at \cite{githubStab,githubCayley,fuentes_munizzi2025mmi}.
\end{appendix}

\bibliographystyle{JHEP}
\bibliography{Ingleton}

\end{document}

%% file: ingleton_states.tex
\tiny
\begin{longtable}{l r X}
\toprule
\# & $\mathcal{C}(x)$ & $x$ \\
\midrule
1 & -0.013800 & [$(0.1654+0.1221i)$, $(0.0395+0.0232i)$, $(0.1554+0.0616i)$, $(0.0344-0.0227i)$, $(0.0839+0.1969i)$, $(0.1397+0.1293i)$, $(0.0248+0.1645i)$, $(0.114+0.0867i)$, $(0.0716+0.0485i)$, $(-0.0081-0.0009i)$, $(0.0833-0.0047i)$, $(0.016-0.0251i)$, $(0.0245+0.1413i)$, $(0.0394+0.0849i)$, $(0.0395+0.0826i)$, $(0.065+0.0538i)$, $(0.1626+0.0993i)$, $(0.1512+0.0826i)$, $(0.0788+0.0872i)$, $(0.0912+0.0569i)$, $(0.0873-0.0117i)$, $(0.1327+0.0942i)$, $(0.0115-0.0049i)$, $(0.0392+0.0781i)$, $(0.1558+0.2106i)$, $(0.0403+0.1969i)$, $(0.0401+0.0691i)$, $(0.0267+0.0415i)$, $(0.1547+0.0782i)$, $(0.0999+0.1834i)$, $(0.0102+0.0081i)$, $(0.0191+0.0466i)$, $(0.0036+0.0012i)$, $(0.0188+0.0294i)$, $(0.0079+0.0328i)$, $(0.0129+0.0553i)$, $(0.0986+0.0429i)$, $(0.0512+0.0263i)$, $(0.1215+0.0655i)$, $(0.0535+0.0601i)$, $(0.158-0.0022i)$, $(0.1083+0.0962i)$, $(0.074-0.0214i)$, $(0.0655+0.0393i)$, $(0.1991+0.0899i)$, $(0.1055+0.1415i)$, $(0.0911+0.0808i)$, $(0.0206+0.0902i)$, $(0.1883+0.0537i)$, $(0.0906+0.0824i)$, $(0.1761+0.0691i)$, $(0.076+0.08i)$, $(0.1119+0.023i)$, $(0.1226+0.094i)$, $(0.0879+0.0516i)$, $(0.1036+0.1045i)$, $(0.0652+0.0551i)$, $(0.0146+0.0614i)$, $(0.0961+0.0537i)$, $(0.0363+0.0663i)$, $(-0.0173+0.0128i)$, $(-0.0186+0.0028i)$, $(0.0029+0.0308i)$, $(0.0028+0.0297i)$] \\
2 & -0.169300 & [$(0.0353+0.0162i)$, $(0.1359+0.0062i)$, $(0.0641+0.0823i)$, $(-0.0679+0.0832i)$, $(0.1532+0.0311i)$, $(0.1053+0.0882i)$, $(0.0305+0.0323i)$, $(0.1043+0.0315i)$, $(0.1136-0.009i)$, $(0.0438+0.025i)$, $(-0.0243+0.1106i)$, $(0.0544+0.132i)$, $(0.0985+0.0909i)$, $(0.1383+0.0951i)$, $(0.1528+0.022i)$, $(0.1008+0.0577i)$, $(0.0766+0.159i)$, $(-0.0139+0.0268i)$, $(0.0181-0.1236i)$, $(0.0586+0.0826i)$, $(0.1042+0.001i)$, $(0.1895+0.1108i)$, $(-0.02+0.0739i)$, $(-0.0873-0.0576i)$, $(-0.0752-0.0158i)$, $(-0.0263+0.1072i)$, $(0.1303+0.1564i)$, $(0.1078+0.0138i)$, $(0.0833+0.0324i)$, $(0.025-0.0286i)$, $(0.0316+0.0694i)$, $(0.0614+0.1973i)$, $(-0.0672+0.0652i)$, $(-0.0549+0.08i)$, $(0.1382+0.0084i)$, $(0.1226-0.0022i)$, $(-0.0042+0.0005i)$, $(-0.0254-0.0073i)$, $(0.0806+0.0622i)$, $(0.0985+0.1004i)$, $(0.0431+0.0958i)$, $(0.0083+0.0948i)$, $(0.0282-0.0574i)$, $(0.0593-0.0278i)$, $(0.1154-0.042i)$, $(0.1483-0.0171i)$, $(-0.0608+0.0821i)$, $(-0.0944+0.0642i)$, $(0.1105+0.1196i)$, $(0.0222-0.1042i)$, $(-0.0201-0.0938i)$, $(-0.0165+0.2199i)$, $(-0.0504-0.0445i)$, $(0.0482+0.0888i)$, $(0.1714+0.1668i)$, $(0.1049+0.006i)$, $(0.0615-0.0554i)$, $(0.1425+0.1664i)$, $(-0.0247+0.1421i)$, $(-0.0406-0.1243i)$, $(0.148+0.1112i)$, $(0.0618+0.0231i)$, $(0.0234-0.0762i)$, $(0.0815+0.0979i)$] \\
3 & -0.168300 & [$(0.0778-0.0261i)$, $(-0.0389+0.0914i)$, $(-0.0128+0.117i)$, $(0.0877+0.0084i)$, $(0.2012+0.1165i)$, $(0.0407-0.0294i)$, $(0.151+0.0288i)$, $(-0.0254+0.0378i)$, $(-0.0601+0.0572i)$, $(0.1104+0.0827i)$, $(0.024-0.1283i)$, $(-0.0798+0.1563i)$, $(-0.0093-0.0667i)$, $(0.077+0.1935i)$, $(-0.0095+0.0531i)$, $(0.1006+0.0804i)$, $(-0.0299+0.0371i)$, $(0.1081+0.0944i)$, $(-0.0627+0.15i)$, $(0.1677+0.0083i)$, $(0.0639+0.0373i)$, $(0.1208+0.1092i)$, $(0.0434-0.0334i)$, $(0.0367+0.1491i)$, $(0.0514+0.1515i)$, $(0.1002-0.0689i)$, $(0.0681-0.0548i)$, $(-0.0524+0.0665i)$, $(0.1758-0.1053i)$, $(-0.0765+0.1031i)$, $(0.1475+0.0381i)$, $(-0.0168-0.0063i)$, $(0.1586+0.0145i)$, $(-0.0705+0.0121i)$, $(0.0623+0.1238i)$, $(0.0401-0.0408i)$, $(0.0326-0.0014i)$, $(0.0773+0.1527i)$, $(0.0434-0.058i)$, $(-0.0121+0.1619i)$, $(0.0687+0.1692i)$, $(0.071-0.0789i)$, $(0.1037+0.0126i)$, $(-0.0638-0.0006i)$, $(0.0326+0.0327i)$, $(0.0656+0.0915i)$, $(-0.0094+0.1098i)$, $(0.1078+0.0305i)$, $(-0.0256-0.1443i)$, $(0.0739+0.1841i)$, $(0.072+0.0233i)$, $(0.0547+0.0031i)$, $(0.1679-0.002i)$, $(0.1208-0.0536i)$, $(0.0716-0.0532i)$, $(0.0949+0.0602i)$, $(0.0871+0.019i)$, $(0.0463-0.0187i)$, $(-0.0275-0.1481i)$, $(0.008+0.1753i)$, $(0.0835-0.0558i)$, $(0.0893+0.0257i)$, $(0.1369+0.0612i)$, $(0.0653-0.0764i)$] \\
4 & -0.031400 & [$(0.0553-0.0075i)$, $(0.1798+0.0578i)$, $(0.2109-0.036i)$, $(0.1368-0.062i)$, $(0.1513+0.0691i)$, $(0.198+0.0558i)$, $(0.0197+0.0323i)$, $(0.1722+0.0596i)$, $(0.0849+0.1477i)$, $(0.1211+0.1405i)$, $(0.053-0.0082i)$, $(0.0862+0.0921i)$, $(0.0435+0.0625i)$, $(0.0578+0.2039i)$, $(0.0389+0.0998i)$, $(0.08+0.1321i)$, $(0.1816+0.0898i)$, $(0.1357+0.0444i)$, $(0.0333+0.0656i)$, $(0.088+0.0514i)$, $(-0.0521+0.1263i)$, $(0.0672+0.1669i)$, $(0.0776+0.0625i)$, $(0.0716+0.0577i)$, $(0.1596+0.0073i)$, $(0.1153+0.0211i)$, $(0.1063+0.0365i)$, $(0.0923+0.0148i)$, $(0.0188+0.0473i)$, $(0.0691+0.0436i)$, $(0.016-0.004i)$, $(0.0499+0.0288i)$, $(0.0781+0.0764i)$, $(0.2134-0.0046i)$, $(0.0068+0.0451i)$, $(0.1268+0.0402i)$, $(0.0783-0.0396i)$, $(0.123+0.0341i)$, $(0.0467+0.0083i)$, $(0.1011+0.0028i)$, $(0.0576+0.0708i)$, $(0.1278+0.077i)$, $(0.03+0.0814i)$, $(0.1332+0.0722i)$, $(0.0514+0.0296i)$, $(0.0856+0.082i)$, $(0.1158+0.0126i)$, $(0.1089+0.0149i)$, $(-0.0076+0.1605i)$, $(0.1428+0.0622i)$, $(0.019+0.0853i)$, $(0.1018+0.0096i)$, $(0.0921+0.148i)$, $(0.1341+0.0161i)$, $(0.059+0.1045i)$, $(0.0929+0.0124i)$, $(0.0513-0.0163i)$, $(0.0023+0.0399i)$, $(0.0283-0.0192i)$, $(0.0256+0.0338i)$, $(0.1255+0.0407i)$, $(0.0373+0.0081i)$, $(0.1094+0.0495i)$, $(0.033+0.0018i)$] \\
5 & -0.167900 & [$(0.0575+0.0319i)$, $(-0.1243-0.0452i)$, $(0.015+0.0419i)$, $(0.0403-0.0003i)$, $(-0.0408-0.005i)$, $(0.1274+0.104i)$, $(0.042+0.0348i)$, $(-0.0587+0.0133i)$, $(0.0007+0.046i)$, $(0.035+0.0142i)$, $(0.041+0.1035i)$, $(-0.0228-0.0343i)$, $(0.0207+0.0422i)$, $(-0.0388+0.013i)$, $(0.0053+0.0485i)$, $(0.0228+0.1305i)$, $(0.135-0.0302i)$, $(-0.2188+0.0709i)$, $(0.1454+0.0086i)$, $(-0.0561-0.0147i)$, $(-0.0003+0.0458i)$, $(0.1486-0.0501i)$, $(0.1611+0.0231i)$, $(-0.1993-0.0103i)$, $(0.0691+0.0477i)$, $(0.1458+0.048i)$, $(0.196+0.0635i)$, $(0.1011+0.103i)$, $(0.089+0.065i)$, $(-0.0303+0.0165i)$, $(0.1003+0.1129i)$, $(0.1705+0.0561i)$, $(0.0259-0.0152i)$, $(-0.0259+0.0449i)$, $(0.0357+0.0078i)$, $(0.0081-0.0234i)$, $(-0.0203+0.0033i)$, $(0.0983+0.0164i)$, $(-0.0189-0.0203i)$, $(0.1179+0.103i)$, $(0.0703+0.0238i)$, $(-0.0371-0.0846i)$, $(0.1494+0.0167i)$, $(-0.1314-0.1117i)$, $(0.0178-0.0219i)$, $(0.043+0.0934i)$, $(0.0096-0.0301i)$, $(0.1533+0.1429i)$, $(0.0759+0.0226i)$, $(-0.0675+0.0421i)$, $(0.0037+0.0443i)$, $(0.2892+0.121i)$, $(-0.0586+0.0119i)$, $(0.2501+0.1188i)$, $(0.0694+0.0561i)$, $(-0.0151+0.039i)$, $(0.0397+0.0952i)$, $(0.0802+0.0863i)$, $(0.1567+0.1891i)$, $(-0.026+0.0733i)$, $(0.0724+0.1306i)$, $(-0.0977-0.1046i)$, $(0.0718+0.1982i)$, $(0.0114-0.0321i)$] \\
6 & -0.169000 & [$(0.0502+0.0245i)$, $(0.0038+0.091i)$, $(0.0302+0.0334i)$, $(0.0294+0.0479i)$, $(-0.0677-0.0105i)$, $(0.1329+0.091i)$, $(-0.0499-0.0256i)$, $(0.1201+0.0792i)$, $(0.1152+0.1353i)$, $(-0.0584-0.0263i)$, $(0.1222+0.0566i)$, $(-0.0518+0.0233i)$, $(0.0256+0.0571i)$, $(0.046+0.0211i)$, $(0.0348+0.0785i)$, $(0.0298-0.0254i)$, $(0.0475+0.1135i)$, $(0.0485+0.0337i)$, $(0.1926+0.161i)$, $(-0.0987-0.0333i)$, $(0.0894+0.0821i)$, $(0.0223+0.0355i)$, $(-0.0824-0.0025i)$, $(0.198+0.0471i)$, $(0.1554+0.0942i)$, $(-0.0372+0.0267i)$, $(-0.0276+0.0134i)$, $(0.1437+0.0341i)$, $(-0.1298-0.0309i)$, $(0.2626+0.0645i)$, $(0.0863-0.0035i)$, $(0.0326+0.043i)$, $(0.06+0.0412i)$, $(0.0387+0.1309i)$, $(0.0963+0.0908i)$, $(0.0009+0.0398i)$, $(0.1482+0.1633i)$, $(-0.0644-0.0062i)$, $(0.0944+0.0388i)$, $(0.0095+0.0631i)$, $(-0.019+0.0657i)$, $(0.1418+0.0469i)$, $(-0.0683-0.0057i)$, $(0.1958+0.0627i)$, $(-0.0045+0.0424i)$, $(0.127+0.0391i)$, $(0.0835+0.0933i)$, $(0.0218-0.0331i)$, $(-0.1154+0.0069i)$, $(0.2267+0.144i)$, $(0.0531+0.0615i)$, $(0.0605+0.0682i)$, $(0.1329+0.1072i)$, $(-0.0124+0.0544i)$, $(-0.0341+0.053i)$, $(0.1544+0.0274i)$, $(0.1567+0.0745i)$, $(-0.0211+0.0698i)$, $(-0.0273+0.043i)$, $(0.1547+0.0222i)$, $(0.0824+0.0631i)$, $(0.0581+0.0333i)$, $(0.2482-0.0047i)$, $(-0.1057+0.086i)$] \\
7 & -0.167600 & [$(0.0091+0.0185i)$, $(0.1001+0.1151i)$, $(0.0968+0.1022i)$, $(-0.017+0.0166i)$, $(0.1145+0.0699i)$, $(-0.0219+0.0372i)$, $(-0.0134-0.0349i)$, $(0.1378+0.098i)$, $(0.07+0.1646i)$, $(0.0289+0.0018i)$, $(0.0055+0.0686i)$, $(0.127+0.0554i)$, $(0.0118+0.0276i)$, $(0.1323+0.081i)$, $(0.1173+0.0873i)$, $(-0.0051+0.0104i)$, $(0.125+0.1037i)$, $(-0.0549+0.0157i)$, $(0.0777+0.1693i)$, $(-0.0207-0.091i)$, $(0.0637+0.0672i)$, $(0.0169+0.0227i)$, $(0.1098-0.0462i)$, $(-0.0098+0.1386i)$, $(-0.0281+0.1283i)$, $(0.1192-0.0524i)$, $(0.0279+0.0284i)$, $(0.0783+0.0477i)$, $(0.013-0.0782i)$, $(0.1178+0.1682i)$, $(-0.0346+0.0225i)$, $(0.1416+0.0227i)$, $(0.0834+0.0022i)$, $(0.0094+0.1822i)$, $(0.0229+0.1329i)$, $(0.0503-0.0081i)$, $(-0.0424+0.2388i)$, $(0.0918-0.1371i)$, $(0.0206+0.0559i)$, $(0.0725+0.0506i)$, $(0.0784+0.0785i)$, $(0.0277+0.1126i)$, $(0.1171-0.0666i)$, $(0.0227+0.2447i)$, $(0.0472+0.0544i)$, $(0.0731+0.0989i)$, $(0.0227+0.1882i)$, $(0.0649-0.0876i)$, $(-0.0017-0.0454i)$, $(0.0968+0.1635i)$, $(-0.0753+0.0599i)$, $(0.1535+0.0049i)$, $(-0.0042+0.0424i)$, $(0.0796+0.0557i)$, $(0.0666-0.0743i)$, $(0.031+0.1815i)$, $(-0.0103+0.1775i)$, $(0.0562-0.0901i)$, $(0.0662+0.0617i)$, $(0.0049+0.0326i)$, $(0.1466+0.0443i)$, $(-0.0807+0.0866i)$, $(0.0821+0.1262i)$, $(-0.0308-0.0418i)$] \\
8 & -0.168900 & [$(0.0538+0.032i)$, $(0.0231+0.1358i)$, $(0.0756+0.1129i)$, $(0.1244-0.0413i)$, $(0.0437+0.1447i)$, $(-0.0305-0.0014i)$, $(0.1269+0.0598i)$, $(-0.1685+0.0877i)$, $(0.0028+0.1739i)$, $(0.1983-0.1797i)$, $(0.0033-0.027i)$, $(0.0493+0.1386i)$, $(0.0627+0.0028i)$, $(-0.1323+0.0763i)$, $(-0.0136+0.1479i)$, $(0.1614-0.0885i)$, $(0.1681+0.0231i)$, $(0.0361+0.0113i)$, $(0.1684+0.0093i)$, $(0.078-0.048i)$, $(0.0889+0.1058i)$, $(0.0814+0.1359i)$, $(0.1382+0.0653i)$, $(0.0717+0.1226i)$, $(0.0843+0.0205i)$, $(0.0434-0.0117i)$, $(0.1037-0.0764i)$, $(-0.0489+0.0842i)$, $(0.1142+0.0208i)$, $(-0.0845+0.0634i)$, $(0.0741+0.0257i)$, $(0.0397+0.069i)$, $(0.0068-0.0416i)$, $(0.105+0.1861i)$, $(0.0591-0.0312i)$, $(0.0605+0.082i)$, $(0.06+0.1574i)$, $(0.1069-0.0923i)$, $(0.0658+0.0985i)$, $(0.1062-0.0174i)$, $(0.0946-0.0258i)$, $(-0.0924+0.0206i)$, $(-0.0417-0.0568i)$, $(0.1225+0.1681i)$, $(0.0344+0.0012i)$, $(-0.0115+0.088i)$, $(0.1203+0.0662i)$, $(-0.0631-0.1095i)$, $(0.0922-0.0274i)$, $(-0.0534+0.046i)$, $(0.1112-0.0756i)$, $(-0.1042+0.0765i)$, $(0.0725+0.0513i)$, $(-0.0133+0.0038i)$, $(0.0547+0.0343i)$, $(0.0433+0.0313i)$, $(0.0259+0.0134i)$, $(0.0384+0.0008i)$, $(-0.0129+0.0524i)$, $(0.1516-0.0488i)$, $(-0.0877+0.0329i)$, $(0.2072+0.1127i)$, $(-0.021+0.0467i)$, $(0.1335+0.0214i)$] \\
9 & -0.168500 & [$(0.2325+0.0631i)$, $(-0.0761+0.0975i)$, $(-0.0865+0.0709i)$, $(0.1311-0.0381i)$, $(-0.1601-0.0031i)$, $(0.009+0.1142i)$, $(0.1811+0.0359i)$, $(0.1145+0.0751i)$, $(-0.067+0.0507i)$, $(0.2479+0.0199i)$, $(0.1169-0.0279i)$, $(0.015+0.0996i)$, $(0.142+0.0668i)$, $(0.0262+0.0596i)$, $(0.0254+0.1022i)$, $(0.197+0.1151i)$, $(0.1735+0.0294i)$, $(0.0924-0.1093i)$, $(-0.0612+0.0468i)$, $(0.067+0.0978i)$, $(0.0006-0.0232i)$, $(0.0913+0.0803i)$, $(0.0736+0.0459i)$, $(0.0382+0.031i)$, $(-0.0196-0.0612i)$, $(0.1336-0.029i)$, $(0.0964+0.0255i)$, $(0.0411+0.048i)$, $(-0.0069+0.1035i)$, $(-0.0877+0.0999i)$, $(0.0622-0.035i)$, $(0.1777-0.0534i)$, $(0.1372+0.0171i)$, $(-0.0177+0.0834i)$, $(-0.0368+0.0419i)$, $(0.0633-0.0503i)$, $(0.0598+0.0234i)$, $(0.2646+0.0541i)$, $(0.0213+0.0343i)$, $(-0.0593+0.0964i)$, $(-0.1397+0.0979i)$, $(-0.0433+0.1087i)$, $(0.1046-0.1068i)$, $(0.0839-0.0901i)$, $(0.0578+0.0103i)$, $(-0.0207+0.0049i)$, $(0.0456+0.0348i)$, $(0.1532+0.0147i)$, $(0.0148+0.0184i)$, $(-0.1032+0.0188i)$, $(0.0484+0.0046i)$, $(0.163-0.0373i)$, $(-0.0242+0.0006i)$, $(0.0396+0.0472i)$, $(0.1048+0.0464i)$, $(0.1235+0.067i)$, $(-0.0406+0.0357i)$, $(0.0515+0.042i)$, $(0.0914-0.0348i)$, $(0.0902-0.0191i)$, $(0.1339+0.0828i)$, $(0.158+0.1059i)$, $(0.018+0.0433i)$, $(0.0864+0.056i)$] \\
10 & -0.004300 & [$(0.0097+0.0392i)$, $(-0.0471+0.0175i)$, $(0.0861+0.101i)$, $(0.046+0.04i)$, $(0.1926+0.0732i)$, $(0.1145-0.0029i)$, $(0.129+0.0669i)$, $(0.042+0.0122i)$, $(0.0065+0.0268i)$, $(0.0037+0.0267i)$, $(0.0393+0.1238i)$, $(0.0718+0.104i)$, $(0.1047+0.1608i)$, $(0.1362+0.1169i)$, $(0.0472+0.1168i)$, $(0.0633+0.0991i)$, $(0.0938-0.0092i)$, $(0.1201+0.0057i)$, $(0.0792+0.0657i)$, $(0.1423+0.11i)$, $(0.0994+0.0503i)$, $(0.2063+0.0947i)$, $(0.0995+0.0005i)$, $(0.179+0.0442i)$, $(0.0564-0.0036i)$, $(0.0449+0.0252i)$, $(-0.0037+0.0461i)$, $(-0.001+0.1149i)$, $(0.0266+0.0796i)$, $(0.0688+0.164i)$, $(0.0514+0.0323i)$, $(0.0652+0.1011i)$, $(0.0303+0.0734i)$, $(0.0624+0.0394i)$, $(0.0461+0.0696i)$, $(0.0736+0.0265i)$, $(0.053+0.1193i)$, $(0.091+0.065i)$, $(0.045+0.1423i)$, $(0.0958+0.0894i)$, $(0.0319+0.0497i)$, $(0.0384+0.1097i)$, $(0.0552+0.0566i)$, $(0.0628+0.1047i)$, $(0.0975+0.0619i)$, $(0.1374+0.1197i)$, $(0.0894+0.0728i)$, $(0.1314+0.1487i)$, $(0.0694+0.0621i)$, $(0.1597+0.0635i)$, $(0.0265+0.0264i)$, $(0.1078+0.0521i)$, $(0.0649-0.0138i)$, $(0.1613-0.0393i)$, $(0.1113+0.0088i)$, $(0.2264-0.0284i)$, $(0.0355+0.1684i)$, $(0.092+0.0834i)$, $(0.0019+0.1154i)$, $(0.0447+0.0348i)$, $(0.0598+0.1102i)$, $(0.0624-0.008i)$, $(0.0906+0.1629i)$, $(0.1113+0.0262i)$] \\

\bottomrule
\end{longtable}